\documentclass[onecolumn,nofootinbib,superscriptaddress,notitlepage,amsmath,amssymb,fleqn,aps,prd,preprintnumbers,10pt]{revtex4-1}
\usepackage{xcolor,braket,graphicx,array,xspace,wasysym,placeins,hyperref,todonotes,comment,cancel}
\usepackage[all]{hypcap}

\newcommand{\eps}{\varepsilon}

\newcommand{\dilog}{\operatorname{Li_2}}

\newcommand{\MCatNLO}{M\protect\scalebox{0.8}{C}@N\protect\scalebox{0.8}{LO}\xspace}
\newcommand{\SMCatNLO}{S-M\protect\scalebox{0.8}{C}@N\protect\scalebox{0.8}{LO}\xspace}
\newcommand{\MEPSatNLO}{M\protect\scalebox{0.8}{E}P\scalebox{0.8}{S}@N\protect\scalebox{0.8}{LO}\xspace}
\newcommand{\MEPS}{M\protect\scalebox{0.8}{E}P\scalebox{0.8}{S}\xspace}

\newcommand{\Alaric}{A\scalebox{0.8}{LARIC}\xspace}
\newcommand{\Sherpa}{S\scalebox{0.8}{HERPA}\xspace}
\newcommand\LHC{LHC\xspace}
\newcommand\ATLAS{A\protect\scalebox{0.8}{TLAS}\xspace}
\newcommand\CMS{C\protect\scalebox{0.8}{MS}\xspace}
\newcommand\ALICE{A\protect\scalebox{0.8}{LICE}\xspace}
\newcommand\LHCb{L\scalebox{0.8}{HC}B\xspace}
\newcommand\FCC{FCC\xspace}
\newcommand\CERN{CERN\xspace}
\newcommand{\Rivet}{R\scalebox{0.8}{IVET}\xspace}
\newcommand\ALEPH{A\scalebox{0.8}{LEPH}\xspace}

\newcommand\OPAL{O\scalebox{0.8}{PAL}\xspace}
\newcommand\JADE{J\scalebox{0.8}{ADE}\xspace}

\hypersetup{hidelinks,
  pdfauthor={Stefan Hoeche,Frank Krauss,Peter Meinzinger,Daniel Reichelt},
  pdftitle={Recoil-Safe Subtraction, Matching and Merging in e+e- to hadrons},
  pdfkeywords={QCD, Perturbation Theory}
}

\begin{document}
\preprint{FERMILAB-PUB-25-0479-T, IPPP-25-49, ZU-TH 49/25, CERN-TH-2025-141, MCNET-25-18}
\title{Recoil-Safe Subtraction, Matching and Merging in \texorpdfstring{e\textsuperscript{+}e\textsuperscript{--}\textrightarrow}{e+e- to} hadrons}
\author{Stefan H{\"o}che}
\affiliation{Fermi National Accelerator Laboratory, Batavia, IL, 60510}
\author{Frank~Krauss}
\affiliation{Institute for Particle Physics Phenomenology, Durham University, Durham DH1 3LE, UK}
\author{Peter Meinzinger}
\affiliation{Physik-Institut, Universit{\"a}t Z{\"u}rich, CH-8057 Z{\"u}rich, Switzerland}
\author{Daniel Reichelt}
\affiliation{Theoretical Physics Department, CERN, CH-1211 Geneva, Switzerland}

\begin{abstract}
We present the first next-to-leading order matched and multi-jet merged predictions
based on the \Alaric parton shower. The components needed for infrared subtraction in the
\SMCatNLO algorithm are computed analytically for the case of color singlet decays to hadronic
final states and validated against existing approaches for up to $e^-e^+\to$~5~jets.
Phenomenological results for $e^-e^+\to$ hadrons at the $Z$ pole are obtained with up to
five jets at next-to-leading order precision, for the first time using an evolution algorithm
with NLL-preserving kinematics mapping.
\end{abstract}

\maketitle

\section{Introduction}
For more than four decades, parton-shower event generators have played an important role
in deciphering the nature of elementary particles and their interactions~\cite{Buckley:2011ms,Campbell:2022qmc}.
They provide a generic method to simulate the complete hadronic final states measured
in collider experiments such as \ATLAS, \CMS, \LHCb or \ALICE at the Large Hadron Collider (\LHC)
at \CERN~\cite{ATLAS:2008xda,CMS:2008xjf,LHCb:2008vvz,ALICE:2008ngc}. They are also essential
for the planning of future experiments such as those at a potential Future Circular Collider
(\FCC)~\cite{FCC:2025lpp,FCC:2025uan,FCC:2025jtd}.
Due to the high energies involved in these measurements, the correct description of QCD
radiative effects plays a vital role. They are implemented in the simulation by parton shower
algorithms, which have been a topic of intense research activity for many years~\cite{%
  Webber:1983if,Bengtsson:1986gz,Bengtsson:1986et,Marchesini:1987cf,Andersson:1989ki,Webber:1986mc,
  Bengtsson:1986hr,Collins:1987cp,Knowles:1987cu,Knowles:1988vs,Knowles:1988hu,vanBeekveld:2022ukn,
  Nagy:2005aa,Nagy:2006kb,Schumann:2007mg,Giele:2007di,Platzer:2009jq,Hoche:2015sya,Fischer:2016vfv,
  Cabouat:2017rzi,Nagy:2012bt,Platzer:2012np,Nagy:2014mqa,Nagy:2015hwa,Platzer:2018pmd,Isaacson:2018zdi,
  Nagy:2019rwb,Nagy:2019pjp,Forshaw:2019ver,Hoche:2020pxj,DeAngelis:2020rvq,Holguin:2020joq,
  Gustafson:1992uh,Hamilton:2020rcu,Dasgupta:2018nvj,Dasgupta:2020fwr,Hamilton:2021dyz,
  Herren:2022jej,Assi:2023rbu,Hoche:2024dee,Preuss:2024vyu,Hoche:2017iem,Dulat:2018vuy,
  Gellersen:2021eci,FerrarioRavasio:2023kyg,vanBeekveld:2024qxs}.
The matching of parton showers to next-to-leading order (NLO) calculations~\cite{Frixione:2002ik,Nason:2004rx,
  Frixione:2007vw,Alioli:2010xd,Hoche:2010pf,Hoeche:2011fd,Alwall:2014hca}, and the merging
of calculations for varying jet multiplicity~\cite{Andre:1997vh,Catani:2001cc,Lonnblad:2001iq,
  Mangano:2001xp,Krauss:2002up,Lavesson:2007uu,Hoeche:2009rj,Hamilton:2009ne,Lonnblad:2011xx,
  Lonnblad:2012ng,Platzer:2012bs,Hoche:2019ncc,Lavesson:2008ah,Gehrmann:2012yg,Hoeche:2012yf,
  Frederix:2012ps,Lonnblad:2012ix} are crucial for the success of searches for physics
beyond the Standard Model, both in the form of direct limit setting, and in the form of
precision measurements of known Standard Model quantities.

All existing fully differential matching algorithms are based on using the parton shower 
evolution kernels to create an approximate higher-order real-emission correction, which is then
used to cancel the infrared singularities of the actual NLO real-emission contribution.
When spin and color correlations are taken into account, the remainder of this subtraction
is finite in four dimensions and can therefore be integrated out with Monte-Carlo methods.
It is treated as a hard correction to the inclusive reaction, which means
that it is a new type of contribution that cannot be described by the unitary parton evolution. 
The \MCatNLO matching algorithm~\cite{Frixione:2002ik} provided the first fully differential,
process-independent technique to include this hard correction in parton-level event generators. 

This paper reports on the implementation of the \SMCatNLO matching for the \Alaric parton-shower
algorithm~\cite{Herren:2022jej,Assi:2023rbu,Hoche:2024dee}, and on the creation of a fixed-order
subtraction that mirrors the parton-shower up to first order in the strong coupling. 
This is achieved by re-purposing relevant results of massless and massive Catani-Seymour
dipole subtraction~\cite{Catani:1996vz,Catani:2002hc}, in particular the solutions to the
color algebra~\cite{Bassetto:1983mvz} and the phase-space parametrization for splittings
with identified partons and with a massive spectator. The \SMCatNLO matching method
accounts for both non-trivial color correlations and non-trivial spin correlations
across the hard process and the first splitting by reweighting the parton shower with
the relevant insertion operators on a point-by-point basis~\cite{Hoeche:2011fd,Hoeche:2012fm}.

While such matching methods are useful to correct the event simulation
to the inclusive $n$-jet rate, and to correctly include the effects of a single
hard emission, they fail to correctly model the often intricate effects of correlations
among many hard, well-separated jets accompanying the inclusive process of interest. 
Such configurations require dedicated calculations for the many-jet final state, which can be
included by multi-jet merging algorithms that exist both at leading~\cite{Andre:1997vh,Catani:2001cc,
  Lonnblad:2001iq,Mangano:2001xp,Krauss:2002up,Lavesson:2007uu,Hoeche:2009rj,Hamilton:2009ne,
  Lonnblad:2011xx,Lonnblad:2012ng,Platzer:2012bs,Hoche:2019ncc} and at next-to-leading
order~\cite{Lavesson:2008ah,Gehrmann:2012yg,Hoeche:2012yf,Frederix:2012ps,Lonnblad:2012ix}
in QCD perturbation theory. We therefore also report on the implementation of 
\MEPSatNLO~\cite{Gehrmann:2012yg,Hoeche:2012yf}, a next-to-\-leading-order accurate
merging technique, for the \Alaric parton shower. It allows us to make the first
NLO-accurate predictions using a parton shower with a NLL-safe recoil scheme for observables
sensitive to three-, four- and five-jet configurations in $e^+e^-$ annihilation into hadrons. 
These results will be particularly useful to gauge the impact of higher-order corrections versus
non-perturbative power corrections from hadronization in the region around the Sudakov shoulder
of the thrust and $C$-parameter distribution.

This manuscript is organized as follows. Section~\ref{sec:analytics} discusses the
infrared subtraction method and presents the analytic expressions for the corresponding
integrals. Section~\ref{sec:validation} contains the validation of the fixed-order
calculations, and of the matching and merging procedures. Section~\ref{sec:results} presents
the first phenomenological applications and discusses the impact of the higher-order corrections.
Finally, Sec.~\ref{sec:outlook} presents an outlook and discusses further steps towards
precision simulations for the \LHC and a potential \FCC.

\section{Analytic results}
\label{sec:analytics}
\begin{figure}[t]
    \includegraphics[width=\textwidth]{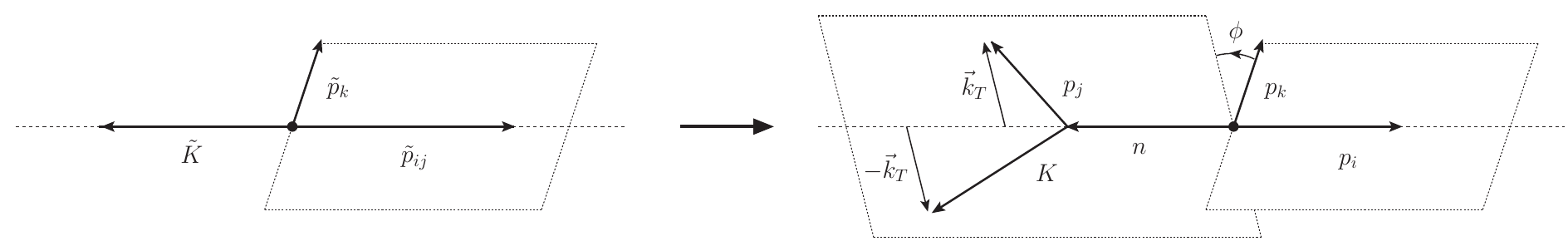}
    \caption{Momentum assignment in radiation kinematics.
    \label{fig:radiation_kinematics}}
\end{figure}
We start with the derivation of the analytic integrals that
are necessary for the construction of the fixed-order infrared
subtraction scheme corresponding to the \Alaric parton shower.  To
this end, we use the \Alaric splitting functions and the kinematic mappings
discussed in~\cite{Assi:2023rbu}; in particular we differentiate
between scalar (soft) and splitting (collinear) kinematics, and we
only address the case of massless final-state partons.

\subsection{Scalar contribution}
\label{sec:soft_integral}
The infrared structure of the real-emission corrections with
final-state partons can be described using the Catani-Seymour dipole
factorization formulae~\cite{Catani:1996vz,Catani:2002hc}:
\begin{equation}\label{eq:cs_dipole_factorization}
    \,_{m+1}\langle 1,\ldots,i,\ldots,j,\ldots,m+1|1,\ldots,i,\ldots,j,\ldots,m+1\rangle_{m+1}
    =\sum_{k\neq a}\mathcal{D}_{ij,k}(p_1,\ldots,p_{m+1})\;,
\end{equation}
where the individual dipole contribution is given by
\begin{equation}
    \mathcal{D}_{ij,k}(p_1,\ldots,p_{m+1})
    =-\frac{1}{2p_ip_j}\,_m\langle\tilde{1},\ldots,\widetilde{ij},\ldots,\widetilde{m\!+\!1}|
    \frac{{\bf T}_i{\bf T}_k}{{\bf T}_i^2}V_{ij,k}
    |\tilde{1},\ldots,\widetilde{ij},\ldots,\widetilde{m\!+\!1}\rangle_m\;.
\end{equation}
We separate the differential insertion operator $V_{ij,k}$ into a
partial fractioned scalar contribution and a pure splitting
component, which are labeled with superscripts $\,^{(s)}$ and
$\,^{(p)}$, respectively.  They can also be interpreted as a soft and
a collinear contribution~\cite{Campbell:2025lrs}
\begin{equation}\label{eq:soft_coll_sinsertion_split}
    V_{ij,k}=V_{ij,k}^{(s)}+V_{ij,k}^{(p)}\;.
\end{equation}
In complete analogy to the differential insertion operator,
$V_{ij,k}$, we define the integral of its spin average over the
one-emission phase space, $\mathbf{I}$, as a sum of soft and collinear
components:
\begin{equation}\label{eq:soft_coll_iinsertion_split}
    \mathbf{I}=\int[{\rm d}p_j]\frac{1}{2p_ip_j}\langle V_{ij,k}\rangle=
    -\frac{\alpha_s}{2\pi} \frac{1}{\Gamma(1-\eps)}\left(\frac{4\pi\mu^2}{s_{ik}}\right)^\eps
    \frac{\mathbf{T}_{ij} \mathbf{T}_{k}}{\mathbf{T}_{ij}^2}
    \left(I^{(s)}_{i,k}+I^{(p)}_{ij}\right)\;,
\end{equation}
where $s_{ik}=2\tilde{p}_i\tilde{p}_k$, and where we follow the
notation of Refs.~\cite{Catani:1996vz,Catani:2002hc}.  Note that the
spin-averaged splitting function appears in this calculation due to a
simplification of the azimuthal angle integrals over spin-dependent
contributions.  This constrains the form of the transverse vector used
in the evaluation of the splitting functions,
Eq.~\eqref{eq:massive_collinear_kernels}.

For splittings where the final-state parton $i$ is not a gluon, the
scalar components of Eq.~\eqref{eq:soft_coll_sinsertion_split}
and~\eqref{eq:soft_coll_iinsertion_split} vanish.  For all others,
they are given by the differential and integrated form of the
partial fractioned eikonal defined in the \Alaric parton-shower
algorithm~\cite{Herren:2022jej,Assi:2023rbu}.
\begin{equation}
    V_{ig,k}^{(s)}(p_i,p_j,n)=8\pi\mu^{2\eps}\alpha_s\, C_i\,
    \frac{2(p_ip_k)(p_in)}{(p_ip_j)(p_kn)+(p_kp_j)(p_in)}\;,\\
\end{equation}
where $C_i$ is the quadratic color Casimir operator of the emitting
particle, $i$.  The partial fractioned scalar insertion operators
are spin-independent, i.e., they are diagonal in the spin states of
any particle they act on.  This is due to the fact that they represent
the semi-classical component of the radiative
corrections~\cite{Gell-Mann:1954wra,Brown:1968dzy}.

The radiation kinematics are sketched in
Fig.~\ref{fig:radiation_kinematics}.  The phase space is parameterized
through the variables~\cite{Catani:1996vz,Herren:2022jej}
\begin{equation}\label{eq:def_v_z_cs}
  v=\frac{p_ip_j}{p_i\tilde{K}}
  \qquad\mathrm{and}\qquad
  z=\frac{p_i\tilde{K}}{\tilde{p}_i\tilde{K}}\;.
\end{equation}
The final-state momentum of the emitter, $\tilde{p}_i$, and the recoil momentum, $\tilde{K}$, are given by
\begin{equation}\label{eq:fi_emit_spec}
  \begin{split}
    p_i=&\;\,z\,\tilde{p}_i\;,\\
    p_j=&\;\,(1-z)\,\tilde{p}_i+v\big(\tilde{K}-(1-z+2\kappa)\,\tilde{p}_i\big)-k_\perp\;,\\
    K=&\;\tilde{K}-v\big(\tilde{K}-(1-z+2\kappa)\,\tilde{p}_i\big)+k_\perp\;,
  \end{split}
\end{equation}
with the magnitude squared of the transverse momentum defined as
\begin{equation}
  {\rm k}_\perp^2=v(1-v)(1-z)\,2\tilde{p}_i\tilde{K}-v^2\tilde{K}^2\;.
\end{equation}
For the final-state evolution of a single QCD multipole, the momentum
$\tilde{K}$ is composed of the two initial-state momenta.  In this
case, all final-state momenta are subjected to the Lorentz
transformation
\begin{equation}\label{eq:lorentz_trafo_fs}
  p_l^\mu\to\Lambda^\mu_{\;\nu}(K,\tilde{K}) \,p_l^\nu\;,
  \qquad\text{where}\qquad
  \Lambda^\mu_{\;\nu}(\tilde{K},K)=g^\mu_{\;\nu}
  -\frac{2(K+\tilde{K})^\mu(K+\tilde{K})_\nu}{(K+\tilde{K})^2}
  +\frac{2K^\mu \tilde{K}_\nu}{\tilde{K}^2}\;.
\end{equation}

The integrated scalar insertion operator corresponds to the massless
limit of Eq.~(55) in \cite{Assi:2023rbu}.  We obtain
\begin{align}
    I^{(s)}_{ik} &= \int_0^1 \mathop{\mathrm{d}z}
    \left(-\frac{\delta(1-z)}{2\eps}+\frac{z}{\left[1-z\right]_{+}}
    - 2\eps z \left[\frac{\log (1-z)}{1-z}\right]_{+} \right)\\
    &\phantom{= \int_0^1 \mathop{\mathrm{d}z} } \times 2 z^{-\eps}
    \left(\frac{np_k}{np_i}\right)^{\eps}\frac{\Gamma^2(1-\eps)}{\Gamma(1-2\eps)}
    \left(\frac{1}{\eps}+\eps \dilog\left(1-\frac{n^2 (p_ip_k)}{2 (np_i) (np_k)}\right)\right)~.
\end{align}
Note that $n$ implicitly depends on $z$. 
Using the variables
\begin{equation}
    \rho \equiv \frac{2 \tilde{p}_k \tilde{K}}{s_{ik}},\qquad
    \tau \equiv \frac{2 \tilde{p}_i \tilde{K}}{s_{ik}},\qquad
    \mu_K \equiv \frac{\tilde{K}^2}{s_{ik}}\;.
\end{equation}
we can parametrize all scalar products in terms of the squared dipole mass
$s_{ik} = 2\tilde{p}_i\tilde{p}_k$,
\begin{equation}
    2 n p_i = z \tau s_{ik},\qquad
    2 n p_k = (1-z+\rho) s_{ik},\qquad
    n^2=\big[(1-z)\tau + \mu_K\big]\,s_{ik}\;.
\end{equation}
The final result, again expanded up to $\mathcal{O}(\eps^0)$, reads
\begin{equation}\label{eq:Isoft}
    \begin{split}
        I^{(s)}_{ik} =&\; \frac{1}{\eps^2} + \frac{2}{\eps}
        + \left(6-\frac{\pi^2}{2}\right)
        + \frac{1}{2} \ln^2\left(\frac{\rho}{\tau}\right)
        + \dilog\left(1-\frac{\mu_K}{\rho \tau}\right)\\
        &-2\operatorname{Re}\left\{\dilog\left(1+\frac{1}{\rho}\right)\right\}
        - 2\Big(1+\rho+\log|\rho|\Big)\log\frac{\rho+1}{\rho} + (\rho\leftrightarrow\tau)\;,
    \end{split}
\end{equation}
where we have suppressed terms that do not contribute to the sum over
the two legs $i,k$ of a dipole due to symmetry.  Note that in the
simplest case of a color singlet decay to two jets, we would have
$\tilde{K} = -\tilde{p}_i-\tilde{p}_k$ and hence $\rho =
-1,\;\tau=-1,\;\mu_K=1$.  In this case, only the constant terms
contribute to Eq.~\eqref{eq:Isoft}, and we obtain the standard soft
integral for the massless case~\cite{Catani:1996vz,Catani:2002hc}.

\subsection{Pure splitting remainders}
\label{sec:collinear_integrals}
\begin{figure}
    \includegraphics[width=\textwidth]{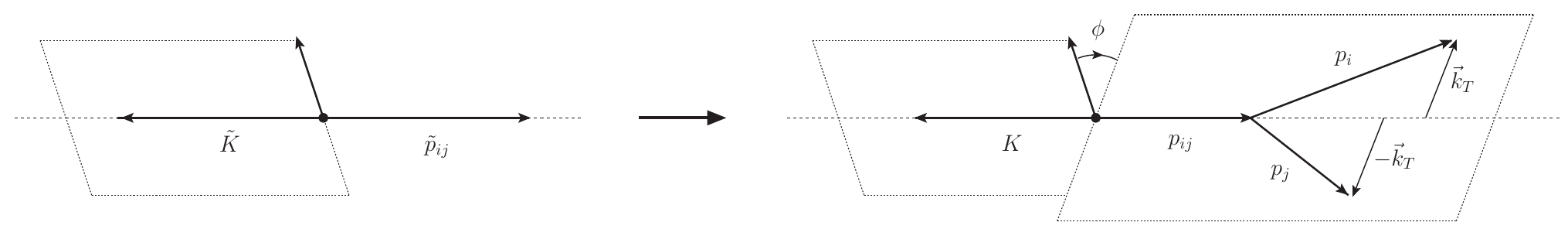}
    \caption{Momentum assignment in splitting kinematics.
    \label{fig:splitting_kinematics}}
\end{figure}
The kinematics mapping used in the computation of pure splitting
functions is sketched in Fig.~\ref{fig:splitting_kinematics}.
We define
\begin{equation}
    Q^2 = 2 \tilde{p}_i \tilde{K} + \tilde{K}^2,\qquad
    \hat{\kappa} = \frac{\tilde{K}^2}{Q^2}\;.
\end{equation}
The insertion operators are given by the spin-dependent pure
remainders of the DGLAP splitting
functions~\cite{Catani:2002hc,Campbell:2025lrs}
\begin{equation}\label{eq:massive_collinear_kernels}
    \begin{split}
        \langle s|V_{gq,k}^{\rm coll}|s' \rangle=&\;
        8\pi\mu^{2\eps}\alpha_s\, C_F\,\delta_{ss'}\,(1-\eps)\,\tilde{z}_j\;,\\
        \langle \mu|V_{q\bar{q},k}^{\rm coll}|\nu \rangle=&\;
        8\pi\mu^{2\eps}\alpha_s\, C_F\,\bigg\{
        -g^{\mu\nu}\bigg[1-\frac{2\,\eta\, z_+z_-}{1-\eps}\bigg]
        -\frac{2}{p_ip_j}\Big[\tilde{z}_i^{(m)}p_i^\mu-\tilde{z}_j^{(m)}p_j^\mu\Big]
        \Big[\tilde{z}_i^{(m)}p_i^\nu-\tilde{z}_j^{(m)}p_j^\nu\Big]\bigg\}\;,\\
        \langle \mu|V_{gg,k}^{\rm coll}|\nu \rangle=&\;
        16\pi\mu^{2\eps}\alpha_s\, C_A\,\bigg\{
        -g^{\mu\nu}\,\eta\, z_+z_-
        +\frac{1-\eps}{p_ip_j}\Big[\tilde{z}_i^{(m)}p_i^\mu-\tilde{z}_j^{(m)}p_j^\mu\Big]
        \Big[\tilde{z}_i^{(m)}p_i^\nu-\tilde{z}_j^{(m)}p_j^\nu\Big]\bigg\}\;,
    \end{split}
\end{equation}
where the transverse momentum depends on the shifted momentum fractions
\begin{equation}
    \tilde{z}_i^{(m)}=\tilde{z}_i+\frac{1}{2}\big(1-v_{ij,k}\big)\;,\qquad
    \tilde{z}_j^{(m)}=\tilde{z}_j+\frac{1}{2}\big(1-v_{ij,k}\big)\;,\qquad
    \text{with}\qquad
    v_{ij,k}=\sqrt{1-\frac{p_{ij}^2 K^2}{Q^4}}\;.
\end{equation}
In terms of the variable $y_{ij,K}=(p_ip_j)/(p_ip_j+p_{ij}K)$, the
boundaries $z_\pm$ for the integration of the light-cone momentum
fractions are given by
\begin{equation}\label{eq:cdst_z_bounds}
    z_\pm=\frac{1}{2}\pm
    \frac{\sqrt{\big[(1-y_{ij,K})(1-\hat{\kappa})+2\hat{\kappa}\big]^2-4\hat{\kappa}}}{
    2(1-y_{ij,K})(1-\hat{\kappa})}\;,
\end{equation}
and we obtain the spin-averaged forms of the pure splitting remainder functions
\begin{equation}
    \begin{split}
        \langle V_{gq,k}^{\rm coll}\rangle=&\;
        8\pi\mu^{2\eps}\alpha_s\, C_F\,(1-\eps)\,\tilde{z}_j\;,\\
        \langle V_{q\bar{q},k}^{\rm coll}\rangle=&\;
        8\pi\mu^{2\eps}\alpha_s\, C_F\,\bigg\{1-\frac{2}{1-\eps}
        \Big[\,\tilde{z}_i\tilde{z}_j-(1-\eta)z_+z_-\Big]\bigg\}\;,\\
        \langle V_{gg,k}^{\rm coll}\rangle=&\;
        16\pi\mu^{2\eps}\alpha_s\, C_A\,\bigg\{
        \tilde{z}_i\tilde{z}_j-(1-\eta) z_+z_-\bigg\}\;.
    \end{split}
\end{equation}
The pure splitting integrated counter-term is given by
\begin{equation}\label{eq:single_coll_counterterm}
  \begin{split}
    \frac{\alpha_s}{2\pi} \frac{1}{\Gamma(1-\eps)}&
    \left(\frac{4\pi\mu^2}{s_{ik}}\right)^\eps
    I^\text{(p)}_{ij}(\hat{\kappa})\equiv\int{\rm d}\Phi_{+1}
    (q;\tilde{p}_{ij},\tilde{K};p_i,p_j)\frac{1}{2p_ip_j}\langle V_{ij,k}\rangle\\
  =&\;\frac{\alpha_s}{2\pi}\,\bigg(\frac{\mu^2}{Q^2}\bigg)^\eps
    \frac{\Omega(3-2\eps)}{(4\pi)^{1-2\eps}}\,
    (1-\hat{\kappa})^{1-3\eps}
    \int{\rm d}y\,y^{-1-\eps}\,(1-y)^{1-2\eps}\,
    \big(z_+-z_-\big)^{1-2\eps}\,
  \bar{\gamma}_{ab}(\eps)\;,
  \end{split}
\end{equation}
where $y=y_{ij,K}$, and where
\begin{equation}\label{eq:coll_anom_dim}
  \bar{\gamma}_{ab}(\eps)=
  \frac{\Omega(2-2\eps)}{\Omega(3-2\eps)}
  \Big(\frac{z_+-z_-}{2}\Big)^{-1+2\eps}
  \int_{z_-}^{z_+}
  {\rm d}z\,\big((z-z_-)
  (z_+-z)\big)^{-\eps}\,\frac{\langle V_{ij,k}\rangle}{8\pi\alpha_s\mu^{2\eps}}\;.
\end{equation}
We use the techniques in~\cite{Catani:2002hc,Assi:2023rbu} to compute
the integrals as a power series in $\eps$.  In particular,
\begin{equation}\label{eq:single_coll_counterterm_ml}
  \begin{split}
  \frac{\alpha_s}{2\pi} \frac{1}{\Gamma(1-\eps)}\left(\frac{4\pi\mu^2}{s_{ik}}\right)^\eps
  I^{(p)}_{ij}(\hat{\kappa})=&\; \frac{\alpha_s}{2\pi}\bigg(\frac{\mu^2}{Q^2}\bigg)^\eps
  \frac{(4\pi)^{\eps}}{\Gamma(1-\eps)}\int{\rm d}y\,\left\{-\frac{\delta(y)}{\eps}
  \bigg(\frac{1-\sqrt{\hat{\kappa}}}{1+\sqrt{\hat{\kappa}}}\bigg)^{-\eps}+\frac{1}{[y]_+}\right\}\\
  &\qquad\qquad\times\frac{\big(\big[(1-y)(1-\hat{\kappa})+2\hat{\kappa}\big]^2-4\hat{\kappa}\big)^{1/2-\eps}}{(1-\hat{\kappa})^{1-\eps}}\,
  \frac{\Gamma(1-\eps)^2}{\Gamma(2-2\eps)}\,
  \bar{\gamma}_{ab}(\eps)\;.
  \end{split}
\end{equation}
Performing the $y$-integration and expanding the result up to $\mathcal{O}(\eps^0)$ yields
\begin{equation}
  \begin{split}
    I^{(p)}_{qq}\left(\hat{\kappa}\right) =&\; -\frac{1}{2\eps} -\frac{1}{2}\log\frac{s_{ik}}{Q^2} -1
    + \log\left(1-\hat{\kappa}\right) + \frac{1}{2}\frac{\hat{\kappa}\log\hat{\kappa}}{1-\hat{\kappa}}\;, \\
    I^{(p)}_{gg}\left(\hat{\kappa}\right) =&\; -\frac{1}{6\eps} -\frac{1}{6}\log\frac{s_{ik}}{Q^2}
    -\frac{4-\hat{\kappa}}{9(1-\hat{\kappa})}
    - \frac{\hat{\kappa}^{3/2}}{3}\frac{\arcsin\sqrt{\hat{\kappa}}-\pi/2}{(1-\hat{\kappa})^{3/2}}
    + \frac{1}{3}\log(1-\hat{\kappa})\\
    &\;-\frac{\hat{\kappa}(1-\eta)}{1-\hat{\kappa}}\bigg(\sqrt{\frac{\hat{\kappa}}{1-\hat{\kappa}}}
      \arctan\sqrt{\frac{1-\hat{\kappa}}{\hat{\kappa}}}-\frac{\log\hat{\kappa}}{2}-1\bigg)\;, \\
    I^{(p)}_{gq}\left(\hat{\kappa}\right) =&\; -\frac{2}{3\eps}-\frac{2}{3}\log\frac{s_{ik}}{Q^2}
    -\frac{16-22\hat{\kappa}}{9(1-\hat{\kappa})}
    + \frac{2\hat{\kappa}^{3/2}}{3}\frac{\arcsin\sqrt{\hat{\kappa}}-\pi/2}{(1-\hat{\kappa})^{3/2}}
    + \frac{4}{3}\log(1-\hat{\kappa}) + \frac{\hat{\kappa}\log \hat{\kappa}}{1-\hat{\kappa}}\\
    &\;+\frac{4}{3}\frac{\hat{\kappa}(1-\eta)}{1-\hat{\kappa}}\bigg(\sqrt{\frac{\hat{\kappa}}{1-\hat{\kappa}}}
      \arctan\sqrt{\frac{1-\hat{\kappa}}{\hat{\kappa}}}-\frac{\log\hat{\kappa}}{2}-1\bigg)\;.
  \end{split}
\end{equation}
Note that these results differ from the ones presented
in~\cite{Catani:2002hc} due to a different definition of the splitting
functions.  Contrary to the collinear components of Eqs.~(5.16)-(5.20)
in~\cite{Catani:2002hc}, our collinear dipole functions in
Eq.~\eqref{eq:massive_collinear_kernels} do not include the velocity
factor $1/v_{ij,k}$.  The results discussed in
Ref.~\cite{Assi:2023rbu} are obtained from the slightly more general
form presented here by setting $\eta=1$.

\FloatBarrier
\section{Numerical tests and validation}
\label{sec:validation}

To validate the implementation of results of the previous section
within the \Sherpa event generator we proceed in three steps: first, we
check the subtraction in representative processes to isolate and test
individual insertion operators.  Second, we compute integrated
cross-section for various processes to ensure the correctness of the
integrated counter-terms.  
And third, the self-consistency of the implementation is checked by
comparing the matched NLO results with pure parton-shower and
the stability of the multi-jet merging for different merging cuts. 
Throughout, we benchmark our results
against established Catani-Seymour (CS) subtraction implementation
within \Sherpa~\cite{Gleisberg:2007md}, thereby guaranteeing
consistent parameter settings and application of phase space cuts.

For each process considered in the first step, we generate phase space
points which, for a given set of emitter, emitted particle and
spectator, are rescaled into the soft and collinear limits,
using the procedure outlined in Ref.~\cite{Hoche:2018ouj}.
For these points we then evaluate the relative difference between
the real-emission correction and the infrared counterterms in the
CS and \Alaric subtraction scheme. This allows us to judge the quality
of the approximations in the logarithmically enhanced regions.
We present the results as box-and-whisker plots, with the boxes signifying
the second and third quartile, and the median indicated as a line inside the box.
The whiskers indicate the 5th and 95th percentile.

\begin{figure}[tp]
    \centering
    \includegraphics[width=0.45\textwidth]{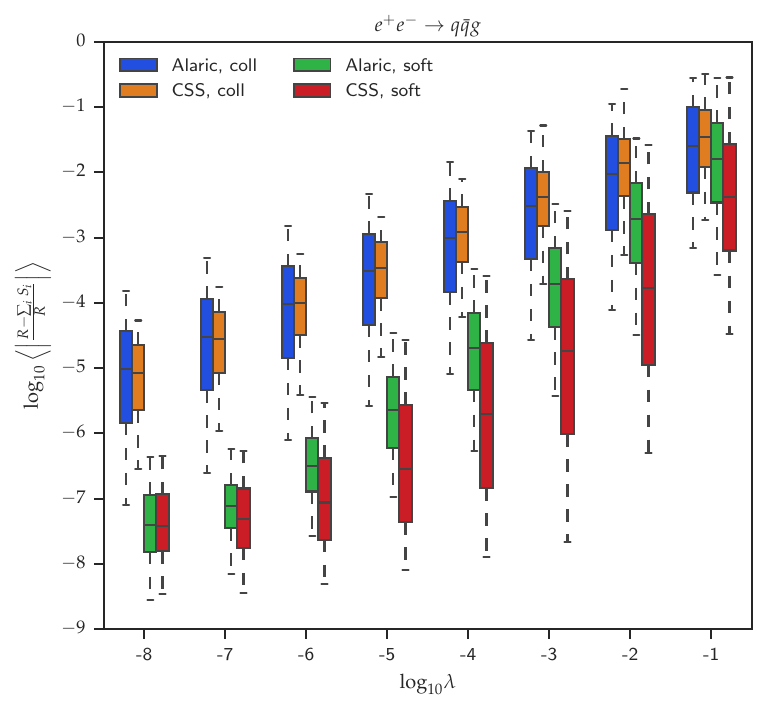}
    \includegraphics[width=0.45\textwidth]{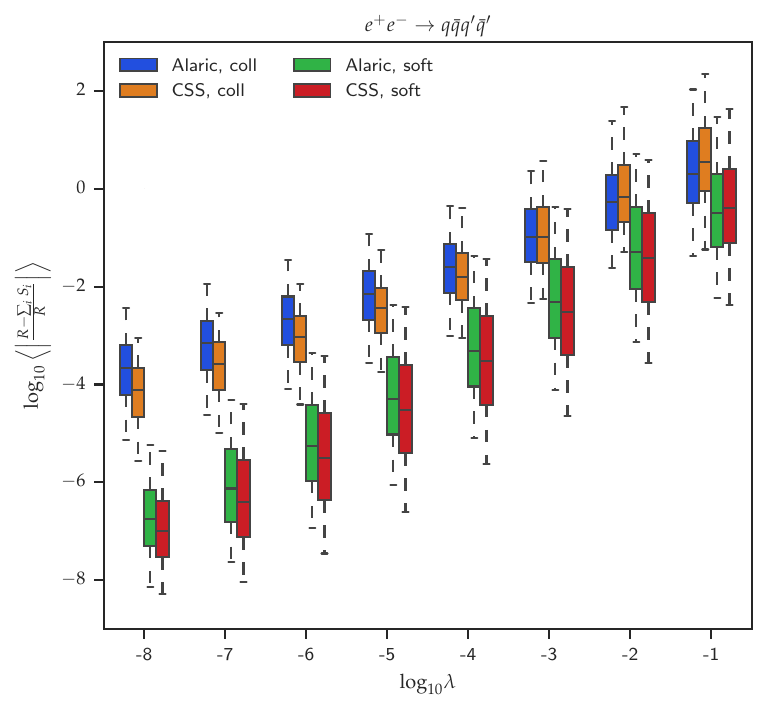}
    \caption{Left: Scaling of CS and Alaric subtraction for $ee \to q\bar{q}g$ with $i=g$, $j=d$ and $k=\bar{d}$.
    Right: Scaling of CS and Alaric subtraction for $ee \to q\bar{q}q^\prime\bar{q}^\prime$ with $i=d$, $j=\bar{d}$ and $k=u$.}
    \label{fig:eeqq-scaling}
\end{figure}
Starting with the simplest cases, we show in
Fig.~\ref{fig:eeqq-scaling} the scaling for $e^+e^- \to q \bar{q} g$
(left) and $e^+e^- \to q \bar{q} q^\prime \bar{q}^\prime$ (right),
testing the gluon emission and the gluon-to-quark splittings,
respectively.  Comparing CS and \Alaric subtraction we observe similar
performances in the deep infrared region, with both CS and \Alaric
subtraction tending to zero as expected.  In the collinear limit,
the \Alaric subtraction exhibits a slightly larger spread in the
distribution than the CS subtraction, whereas the opposite is true
in the soft limit.  For all practical purposes, these differences
can be disregarded.

\begin{figure}[tp]
    \centering
    \includegraphics[width=0.45\textwidth]{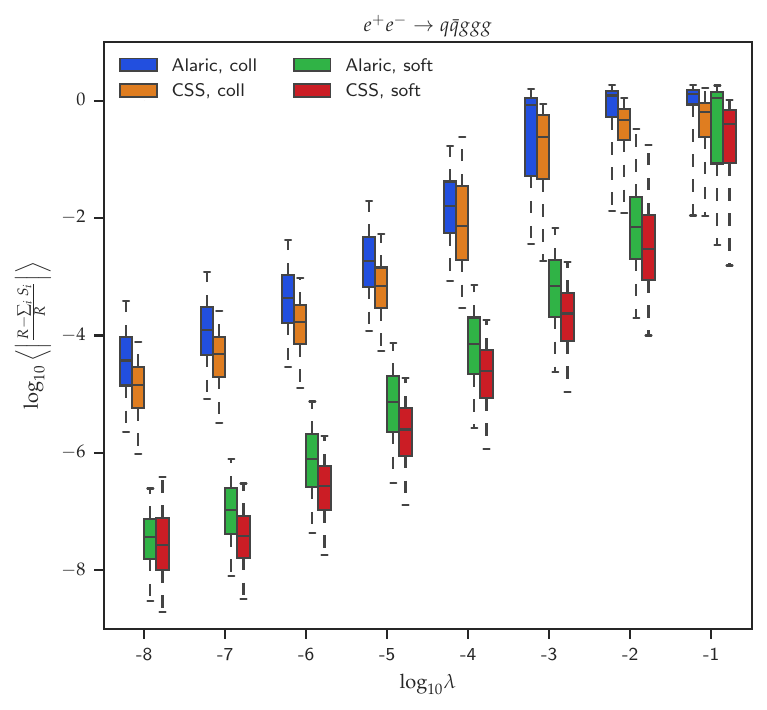}
    \includegraphics[width=0.45\textwidth]{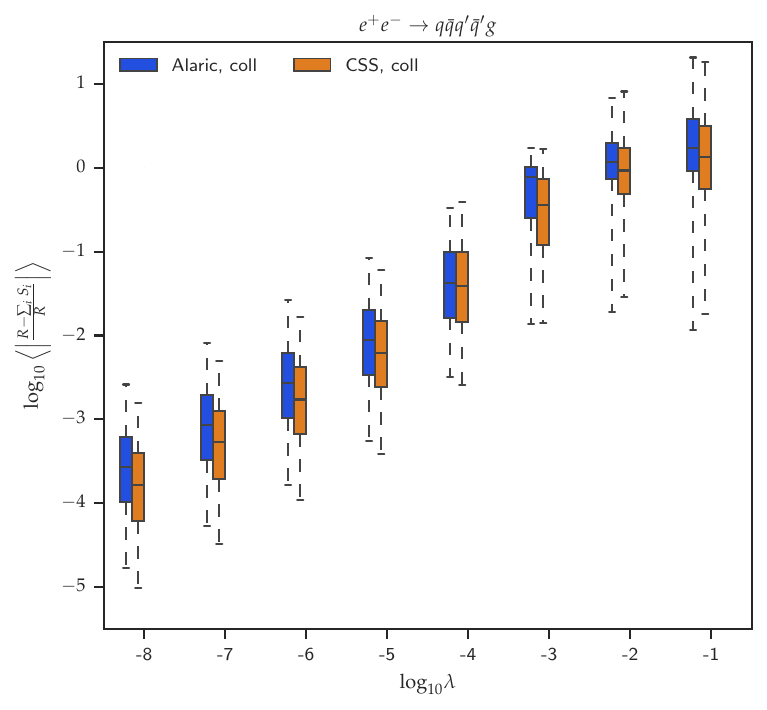}
    \caption{Left: Scaling of CS and Alaric subtraction for $ee \to q\bar{q}ggg$ with $i=g$, $j=g$ and $k=g$.
      Right: Scaling of CS and Alaric subtraction for $ee \to q\bar{q}q^\prime\bar{q}^\prime g$
      with $i=u$, $j=\bar{u}$ and $k=\bar{d}$.}\label{fig:eeqqgg-scaling}
\end{figure}
Analogously, we examine the gluon-to-gluon and the quark-to-quark
splittings using the $e^+e^- \to q \bar{q} g g g$ (left) and
$e^+e^- \to q \bar{q} q^\prime \bar{q}^\prime g$ (right) processes,
respectively. The results are shown in Fig.~\ref{fig:eeqqgg-scaling}.
Again we find similar scaling behavior for both subtraction schemes.

\FloatBarrier

\setlength{\tabcolsep}{5pt}
\renewcommand{\arraystretch}{1.25}
We validate the implementation of the integrated
counter-terms by comparing inclusive cross sections.  For clarity, we
only present numbers relevant for an informed comparison, i.e., the
integrated counter-terms and the real-subtracted cross-section for
each subtraction scheme.  In the trivial case of $e^+ e^- \to
2\ \mathrm{jets}$ at $\sqrt{s} = 91.2$~GeV, shown in Tab.~\ref{tab:2jet-Zjj},
we find excellent agreement, with about $0.8\sigma$ difference in total
cross section between the CS and \Alaric schemes.  In this case the integrated
counter term for the soft emission actually reduces to the exact same
expression in both cases, cf.\ the discussion in Sec.~\ref{sec:soft_integral}.
In order to verify the subtraction scheme for a non-trivial recoil dependence,
we investigate the cross section for $e^+ e^- \to Z H$ with a stable Z boson
and the Higgs boson decaying into gluons at a hypothetical FCC-ee
at $\sqrt{s} = 240$~GeV. Table~\ref{tab:2jet-Zjj} shows excellent agreement
between the two schemes, with a remaining difference of about $0.7\sigma$.

Turning to more complicated processes, we compute total cross sections
for $e^+ e^- \to 3\ \mathrm{jets}$, $e^+ e^- \to 4\ \mathrm{jets}$ and
$e^+ e^- \to 5\ \mathrm{jets}$, all at $\sqrt{s} = 91.2$ GeV. The results
are presented in Tab.~\ref{tab:njets}.  The errors of the individual cross
sections stay below the permille level, and we find agreement between the
schemes at the level of about $1\sigma$ for 3-jet production.  In the
4-jet case, the two results agree at the level of $1.85\sigma$.
For the production of 5 jets, we observe agreement at around
$0.6\sigma$.  These results confirm the robustness and consistency
of the \Alaric subtraction scheme up to high jet multiplicities.

\begin{table}[tp]
\begin{tabular}{p{4cm}|>{\raggedleft}p{15mm}|>{\raggedleft}p{15mm}|>{\raggedleft}p{15mm}|
>{\raggedleft}p{15mm}|>{\raggedleft}p{15mm}|>{\raggedleft}p{15mm}|>{\raggedleft\arraybackslash}p{15mm}}
  $e^+e^-\to 2$ jets &
  I [pb] & $\Delta$I [pb] & RS [pb] & $\Delta$RS [pb] & IRS [pb] & $\Delta$IRS [pb] & $\Delta$IRS/NLO \\
  $\sqrt{s}=91.2$ GeV, $\mu_R=\sqrt{s}$ & & & & & & & \\\hline
  CS & 131.9 & $<$0.01 & -505.79 & 0.19 & -373.89 & 0.19 & 0.004\permil\\
  \Alaric & 131.9 & $<$0.01 & -505.61 & 0.12 & -373.71 & 0.12 & 0.003\permil\\\hline
  CS-\Alaric & \multicolumn{4}{c|}{} & \multicolumn{3}{c}{-0.18 $\pm$ 0.22}
\end{tabular}\vskip 5mm
\begin{tabular}{p{4cm}|>{\raggedleft}p{15mm}|>{\raggedleft}p{15mm}|>{\raggedleft}p{15mm}|
>{\raggedleft}p{15mm}|>{\raggedleft}p{15mm}|>{\raggedleft}p{15mm}|>{\raggedleft\arraybackslash}p{15mm}}
  $e^+e^-\to ZH[\to gg]$ &
  I [pb] & $\Delta$I [pb] & RS [pb] & $\Delta$RS [pb] & IRS [pb] & $\Delta$IRS [pb] & $\Delta$IRS/NLO \\
  $\sqrt{s}=240$ GeV, $\mu_R=M_H$ & & & & & & & \\\hline
  CS & $-8.4\cdot10^{-4}$ & $0.03\cdot10^{-5}$ & $6.31\cdot10^{-5}$ & $0.07\cdot10^{-5}$ & $-7.8\cdot 10^{-4}$ & $0.08\cdot10^{-5}$ & 0.1\permil\\
  \Alaric & $-8.4\cdot10^{-4}$ & $0.03\cdot10^{-5}$ & $6.36\cdot10^{-5}$ & $0.07\cdot10^{-5}$ & $-7.8\cdot 10^{-4}$ & $0.08\cdot10^{-5}$ & 0.1\permil\\\hline
  CS-\Alaric & \multicolumn{4}{c|}{} & \multicolumn{3}{c}{$(0.005 \pm 0.011)\cdot 10^{-4}$}
\end{tabular}
\caption{Subtraction-scheme dependent cross-section contributions for $e^+e^-\to 2$ jets
and for $e^+e^-\to ZH$ with the Higgs boson decaying into gluons. In both cases the complete
hadronic final state defines the recoil system.\label{tab:2jet-Zjj}}
\end{table}

\begin{table}[tp]
\begin{tabular}{p{1cm}|p{2cm}|>{\raggedleft}p{15mm}|>{\raggedleft}p{15mm}|>{\raggedleft}p{15mm}|
>{\raggedleft}p{15mm}|>{\raggedleft}p{15mm}|>{\raggedleft}p{15mm}|>{\raggedleft\arraybackslash}p{15mm}}
  \multicolumn{2}{l|}{$e^+e^-\to n$ jets @ 91.2 GeV } &
  I [pb] & $\Delta$I [pb] & RS [pb] & $\Delta$RS [pb] & IRS [pb] & $\Delta$IRS [pb] & $\Delta$IRS/NLO \\
  \multicolumn{2}{l|}{$\mu_R=\sqrt{s}$, $y_{n,n-1}\!>\!(10\,{\rm GeV})^2/s$} & & & & & & & \\\hline
  & CS & 5118.85 &	0.37 & -1034.4 & 1.4	& 4084.45 &	1.4 & 0.1\permil\\
  $n$=3 & \Alaric & 5174.47 & 0.38 & -1087.6 & 1.8 & 4086.84 &	1.8 & 0.1\permil\\\cline{2-9}
  & CS-\Alaric & \multicolumn{4}{c|}{} & \multicolumn{3}{c}{-2.39 $\pm$ 2.32}\\\hline
  & CS & 938.97 & 0.12 & -94.45 & 0.73 & 844.52 & 0.74 & 0.5\permil\\
  $n$=4 & \Alaric & 922.36 & 0.12 & -79.75 &0.70 & 842.61 & 0.71 & 0.5\permil\\\cline{2-9}
  & CS-\Alaric & \multicolumn{4}{c|}{} & \multicolumn{3}{c}{1.91 $\pm$ 1.03}\\\hline
  & CS & 46.1544 & 0.015 & -2.341 & 0.058 & 43.813 & 0.060 & 1\permil\\
  $n$=5 & \Alaric & 44.6125 & 0.014 & -0.857 & 0.067 & 43.756 & 0.068 & 1\permil\\\cline{2-9}
  & CS-\Alaric & \multicolumn{4}{c|}{} & \multicolumn{3}{c}{0.057 $\pm$ 0.091}
\end{tabular}
\caption{Subtraction-scheme dependent cross-section     contributions for $e^+e^-\to nj$.
  The center-of-mass energy is set to $\sqrt{s}=91.2$~GeV. 
  Jets are defined according to the Durham algorithm~\cite{Catani:1991hj}
  and are required to satisfy a cut on the jet resolution from $n$ to $n-1$ jets,
  $y_{n,n-1}$.\label{tab:njets}}
\end{table}

\begin{figure}
  \begin{center}
    \includegraphics[width=0.42\textwidth]{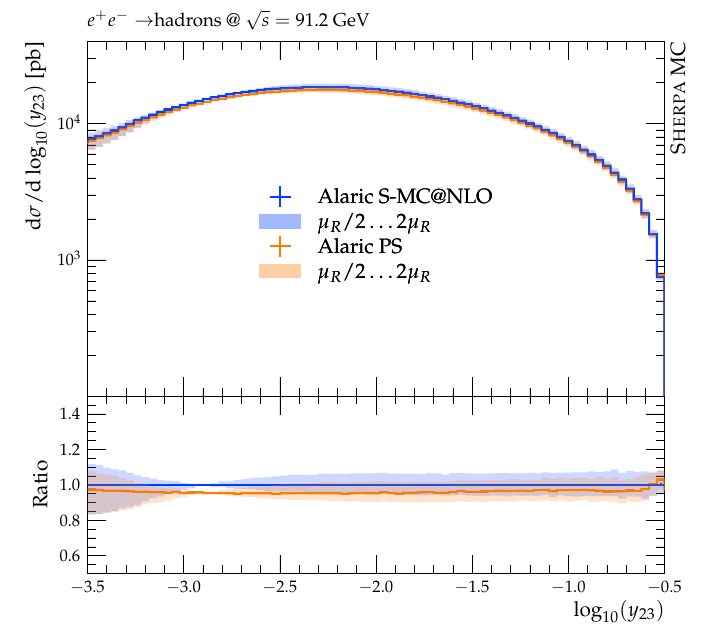}\nolinebreak
    \includegraphics[width=0.42\textwidth]{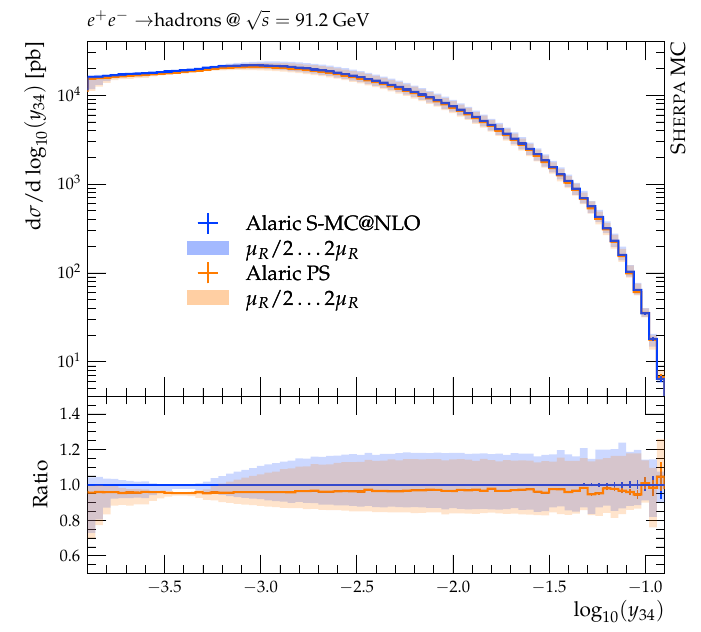}
  \end{center}
  \caption{Comparison of \SMCatNLO and pure parton-shower predictions for the differential
    $2\to3$- and $3\to4$-jet rates in the Durham algorithm at parton level.
    See the main text for details.}
  \label{fig:mcnlo}
\end{figure}
In order to validate our implementation of the \MCatNLO technology, we compare the differential
jet rates in the Durham algorithm between parton-shower predictions and matched results.
Figure~\ref{fig:mcnlo} displays the $2\to3$-jet rate (left) and the $3\to4$-jet rate (right)
as predicted by a simulation from the pure \Alaric parton shower, and from the next-to-leading
order matched \Alaric simulation. The increase in the total rate of the process by the
expected $K$-factor of $\alpha_s/\pi$ is clearly visible. In addition, we observe that the
scale uncertainty in this simulation is not reduced, because the real radiative corrections
that determine the jet rates are predicted (mostly) by the parton shower in both cases.
They are therefore endowed with variations determined by Eq.~\eqref{eq:scalar_nlo_contribution},
cf.\ the discussion in Sec.~\ref{sec:results}.

\begin{figure}
  \begin{center}
    \includegraphics[width=0.42\textwidth]{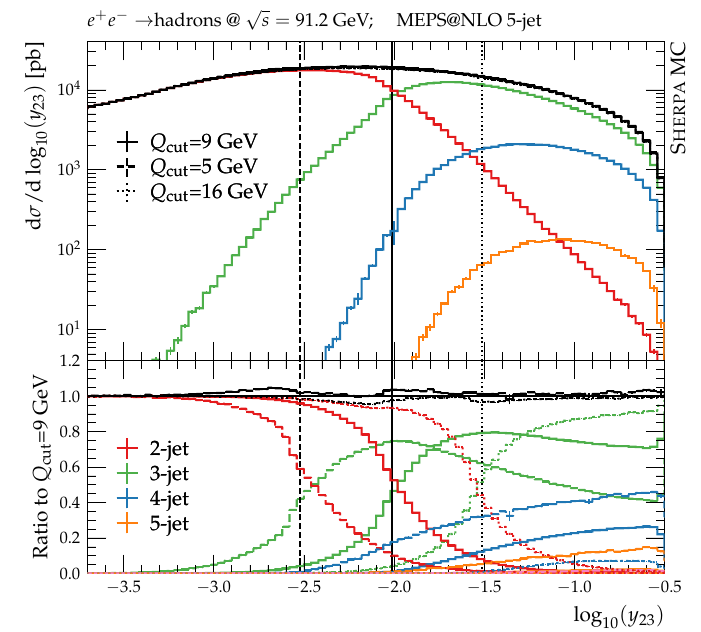}\nolinebreak
    \includegraphics[width=0.42\textwidth]{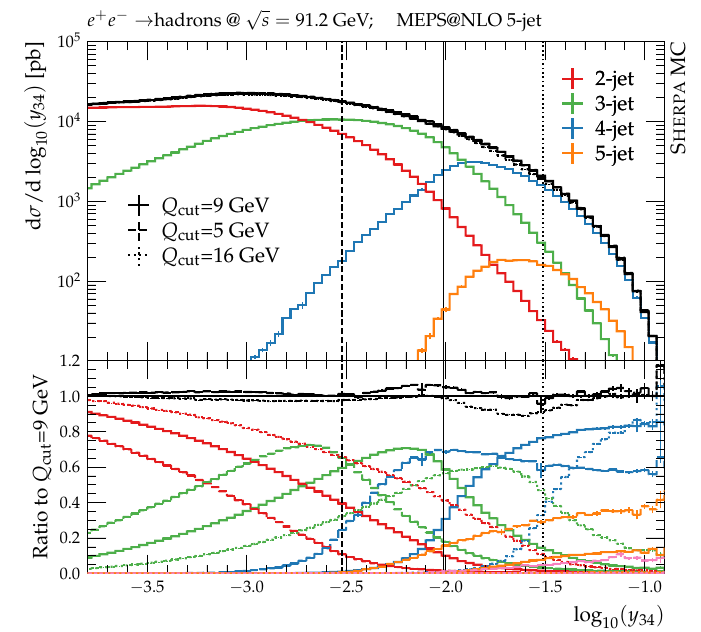}
  \end{center}
  \caption{$Q_{\rm cut}$ dependence of the differential $2\to3$- and $3\to4$-jet rates
    in the Durham algorithm at parton level. See the main text for details.}
  \label{fig:qcut}
\end{figure}
Next we investigate the stability of the multi-jet merging.
In Fig.~\ref{fig:qcut} we show the $2\to3$ and $3\to4$ jet rate in the
Durham $k_T$-algorithm~\cite{Catani:1991hj} for different values of the merging cut,
$Q_{\rm cut}$. The  $Q_{\rm cut}$ values are indicated in the figure by vertical lines and
are varied over a relatively large range. The colored histograms indicate the contributions to the
multi-jet merged result from sub-samples with $2$-jet through $5$-jet fixed-order calculations.
For the $2\to3$ jet rate, we expect a transition from the $2$-jet sub-sample to $3$-jet and
higher multiplicity sub-samples at the merging cut. For the $3\to4$ jet rate, we expect
a transition from the $2$-jet and $3$-jet sub-samples below the merging cut, to $4$-jet and
higher multiplicity matrix elements above the merging cut. We find that the variation of the
overall distribution (in black) around the merging cuts is of less than 10\%, and in most cases
even smaller. This indicates a good approximation of the matrix element by the parton shower,
even at relatively high resolution scale.

\begin{figure}
  \begin{center}
    \includegraphics[width=0.42\textwidth]{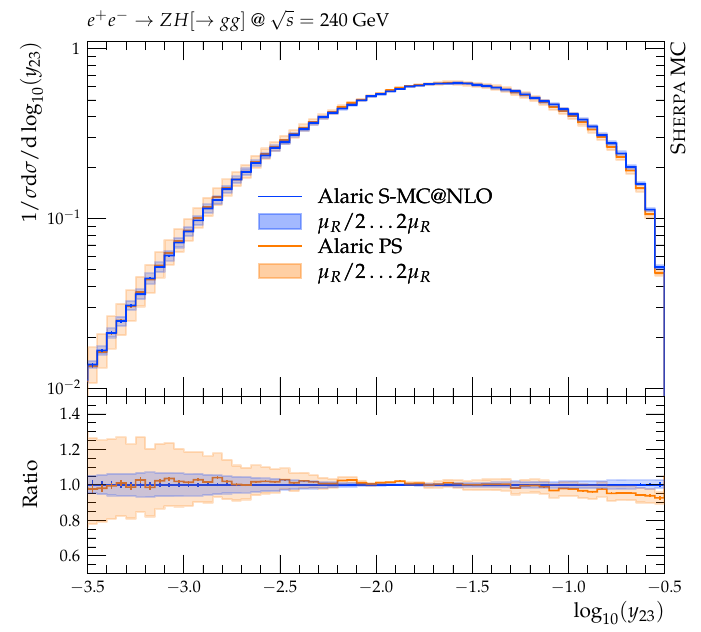}\nolinebreak
    \includegraphics[width=0.42\textwidth]{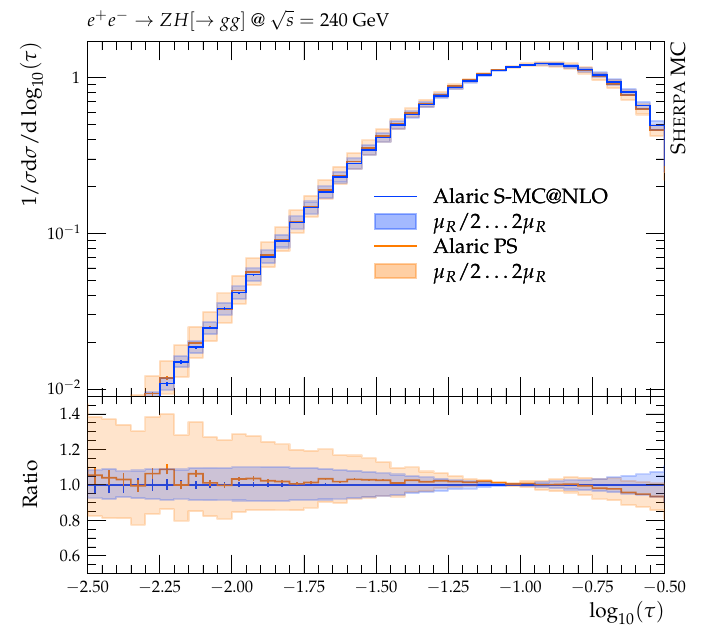}    
  \end{center}
  \caption{Comparison of \SMCatNLO and pure parton-shower predictions for the differential
    $2\to3$ jet rate in the Durham algorithm and the thrust observable in $e^+e^-\to ZH[\to gg]$.}\label{fig:zh}
\end{figure}

Finally, we illustrate the flexibility in choosing the recoil system by calculating matched
predictions for Higgs boson decays to gluons in the Higgstrahlung production mode $e^+e^-\to ZH$,
where we choose the Higgs as the recoil system.
Event shapes in this final state have in the past been computed with quark-gluon discrimination
at a potential future lepton collider like the FCC in mind \cite{Coloretti:2022jcl,Zhu:2023oka,
  Gehrmann-DeRidder:2023uld,Knobbe:2023njd,CampilloAveleira:2024fll,Gehrmann-DeRidder:2024avt,Fox:2025cuz}. 
We correspondingly choose a center of mass energy of $\sqrt{s} = 240$~GeV \cite{FCC:2025jtd}.
As renormalisation scale we use the Higgs mass $M_H$.
In Fig.~\ref{fig:zh} we show the normalized distributions for the Durham jet rate $y_{23}$ (left)
and thrust $\tau$ (right), calculated in the center of mass frame of the Higgs decay products, at \MCatNLO accuracy. 
We observe mild effects, at most of order 10\%,
in the region of large observable values where fixed order corrections are expected to be important. 
In the soft regions we observe no effects on the central values, but a significant reduction in scale
uncertainty, which is a consequence of the changes in the hard region by means of parton-shower unitarity.

\FloatBarrier
\section{Phenomenology}
\label{sec:results}
This section contains the first numerical results obtained with the matching and merging
techniques for the \Alaric parton shower.  We rely on the event generation framework
\Sherpa~\cite{Gleisberg:2003xi,Gleisberg:2008ta,Sherpa:2019gpd,Sherpa:2024mfk}
for the computation of the hard cross section~\cite{Krauss:2001iv,Gleisberg:2008fv},
the implementation of the generic components of the dipole subtraction~\cite{Gleisberg:2007md},
and the interface with the one-loop providers, in this case MCFM~\cite{Campbell:1999ah,
  Campbell:2011bn,Campbell:2015qma,Campbell:2019dru,Campbell:2021vlt}
and OpenLoops~\cite{Cascioli:2011va,Buccioni:2017yxi,Buccioni:2019sur}.
We use the color-correct \SMCatNLO matching method, which accounts for non-trivial
spin correlations across the hard process and the first splitting~\cite{Hoeche:2011fd,Hoeche:2012fm}.
We set $C_F=(N_c^2-1)/(2N_c)=4/3$ and $C_A=3$, and all quarks are considered to be massless. 
However, we implement heavy-flavor thresholds at $m_c=1.42$~GeV and $m_b=4.92$~GeV.  
The running coupling is evaluated at two-loop accuracy with $\alpha_s(m_z)=0.118$.  
Following standard practice to improve the logarithmic accuracy of the parton shower,
we include the effect of the two-loop cusp anomalous dimension through a rescaling
of the scalar part of the splitting function~\cite{Catani:1990rr}.
All analyses are performed with \Rivet~\cite{Buckley:2010ar,Bierlich:2019rhm,Bierlich:2024vqo}.

We adopt a scale variation scheme that reflects the NLL-correct
momentum mapping in the Alaric parton shower.  We first note that most
parton showers use a renormalization scale based on the transverse
momentum of the emission, which captures the effects of those
higher-order corrections that are due to phase-space restrictions of
real corrections in subsequent radiation~\cite{Amati:1980ch}.  While
the precise definition of the transverse momentum is unambiguous in
Ref.~\cite{Amati:1980ch}, there is considerable freedom to vary its functional
form if no more than NLL precision is to be achieved.  This freedom is
explored in existing parton shower models.  It is also known that the
finite part of higher-order corrections to soft gluon emission, which
stem from the collinear decay of the gluon, contribute a $K$-factor
proportional to the two-loop cusp anomalous
dimension~\cite{Kodaira:1981nh,Davies:1984hs,Davies:1984sp,Catani:1988vd}
that can be absorbed into the renormalization
scale~\cite{Catani:1990rr}.  This factor is of the form
\begin{equation}\label{eq:def_cmw_fac}
  K(t)=\left(\frac{67}{18}-\frac{\pi^2}{6}\right)C_A-\frac{10}{9}T_R\,n_f(t)\;,
\end{equation}
where $n_f(t)$ is the number of active flavors at scale $t$ and $t$
itself is of transverse momentum type. To be consistent with all leading
higher-order effects at NLL precision, the scalar splitting functions
in \Alaric must thus take the form
\begin{equation}\label{eq:scalar_nlo_contribution}
  V_{ij,k}^{(s)}\bigg[\,1+
  \frac{\alpha_s(t)}{2\pi}
  \bigg(\beta_0(t)\log\frac{k_T^2}{t}+K(t)\bigg)\bigg]\;,
\end{equation}
where $\beta_0(t)=11/3\,C_A-4/3\,T_R\,n_f(t)$.
When performing a leading-order multi-jet merging with \Alaric, the
renormalization scales to be used in the hard matrix elements are
determined by this constraint.  This is a consequence of the required
consistency between the parton-shower evolution and the fixed-order
calculation: to guarantee correct merging, the latter needs to only
contribute finite corrections, but must not include logarithmic
enhancements.  When performing a next-to-leading order merging, the
$\mathcal{O}(\alpha_s)$ corrections are fully incorporated by the
fixed-order calculation. To prevent double counting, they must thus
not be included in the renormalization scale setting.  When performing
scale variations in the following, we account for the above constraints
and begin with a variation of the renormalization scale in the
parton shower. The renormalization scale in the fixed-order calculation
is then chosen accordingly.

\begin{figure}[tp]
  \begin{center}
    \includegraphics[width=0.4\textwidth]{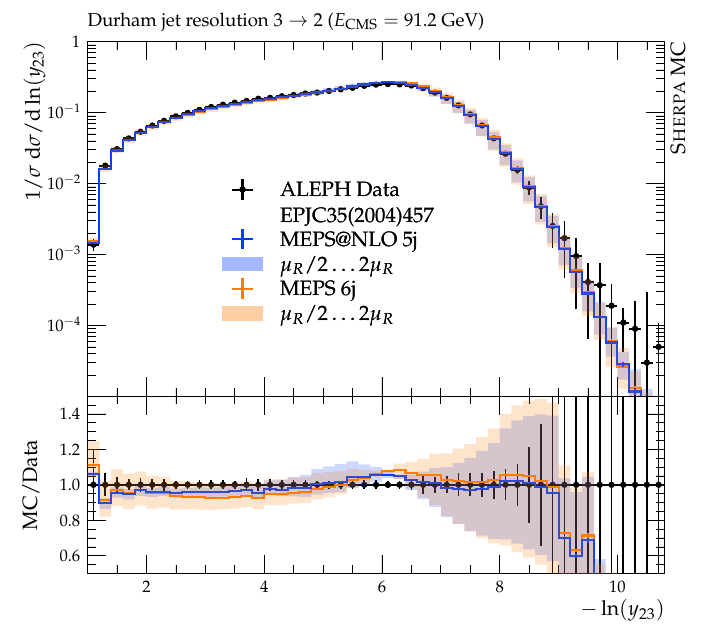}\nolinebreak
    \includegraphics[width=0.4\textwidth]{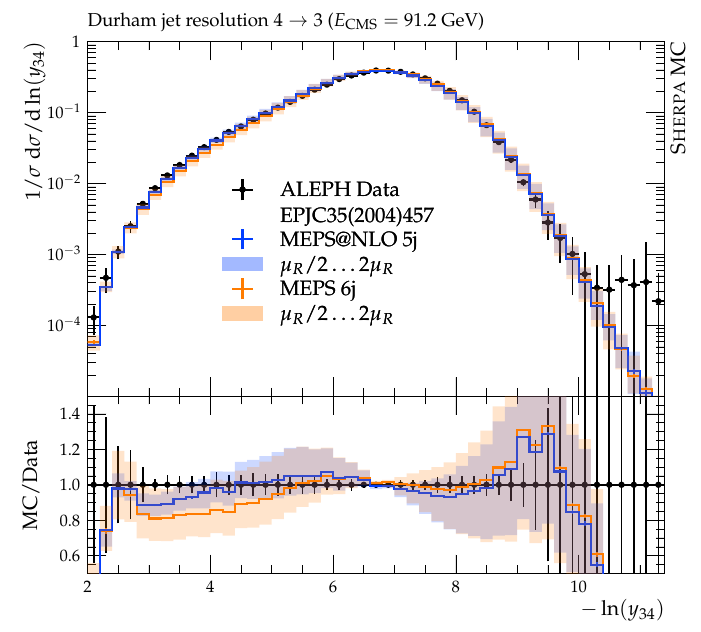}\\
    \includegraphics[width=0.4\textwidth]{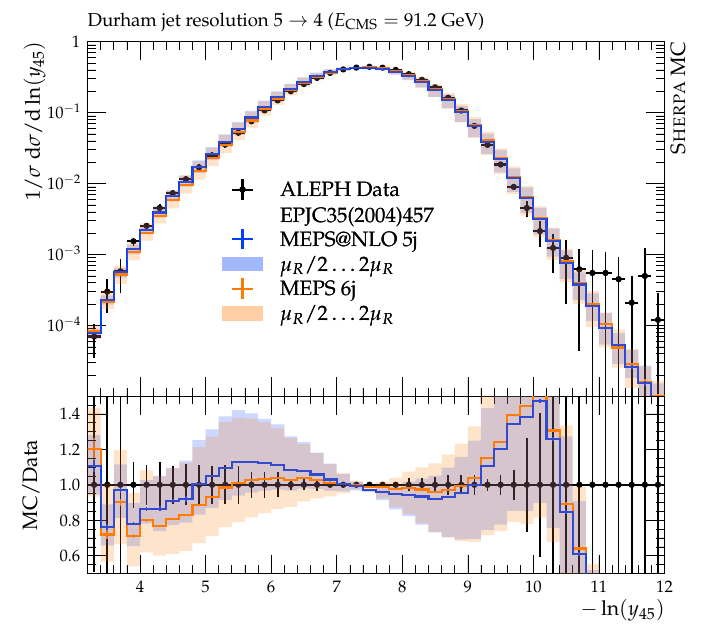}\nolinebreak
    \includegraphics[width=0.4\textwidth]{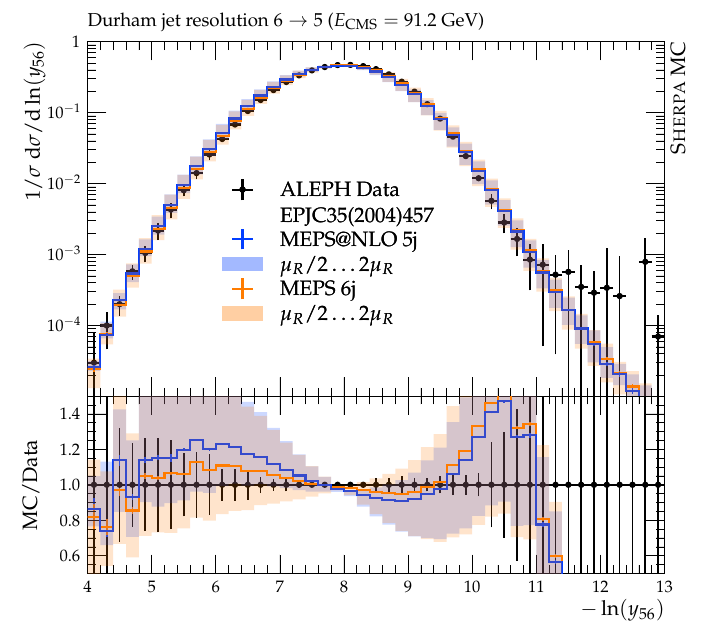}
  \end{center}
  \caption{Perturbative uncertainties in \MEPS and   \MEPSatNLO
           predictions of differential jet rates compared to data from
           \ALEPH~\cite{Heister:2003aj}.}
  \label{fig:jetrates_aleph}
\end{figure}
\begin{figure}[tp]
  \begin{center}
    \includegraphics[width=0.4\textwidth]{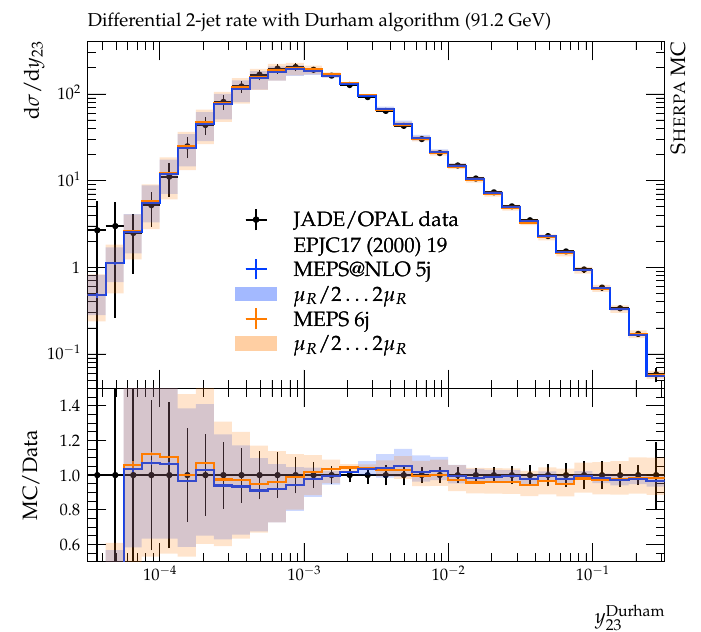}\nolinebreak
    \includegraphics[width=0.4\textwidth]{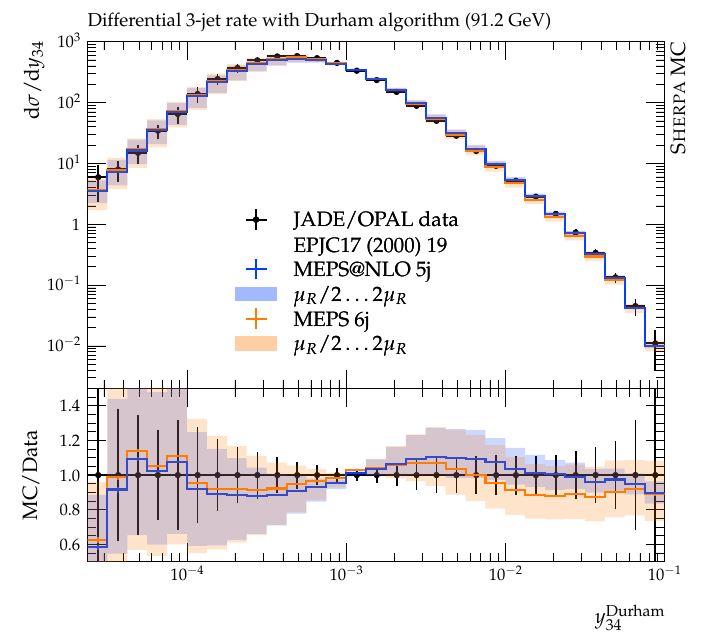}\\
    \includegraphics[width=0.4\textwidth]{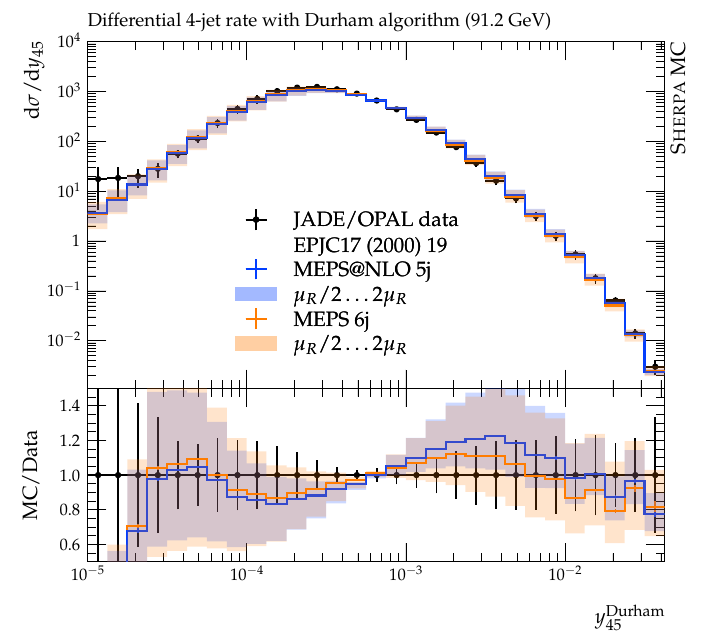}\nolinebreak
    \includegraphics[width=0.4\textwidth]{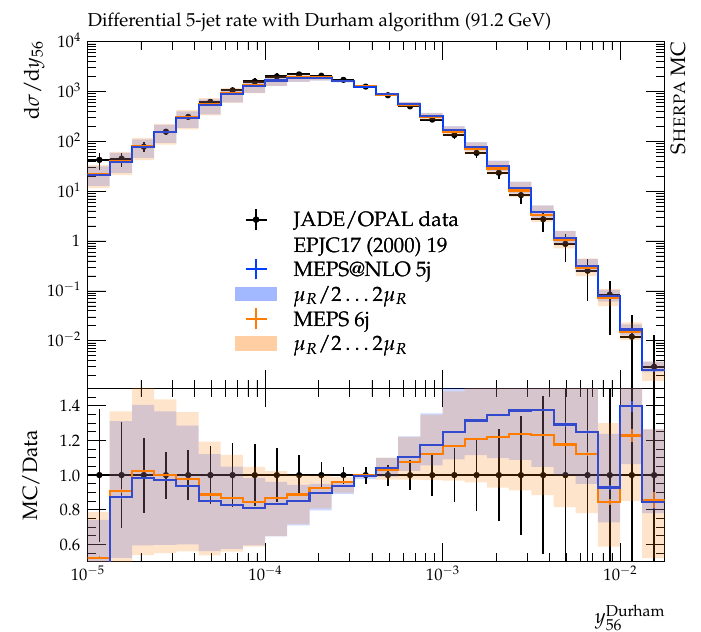}
  \end{center}
  \caption{Perturbative uncertainties in \MEPS and \MEPSatNLO
           predictions of differential jet rates compared to data from
           \JADE \& \OPAL~\cite{JADE:1999zar}.}
  \label{fig:jetrates_jadeopal}
\end{figure}
We first discuss differential jet rates in the Durham jet
algorithm~\cite{Catani:1991hj}.  Measurements of these quantities have
been performed by the \ALEPH collaboration~\cite{Heister:2003aj} and
by the \JADE and \OPAL collaborations~\cite{JADE:1999zar}.  The
\Alaric simulations include up to six jets from hard matrix
elements at the leading order, and up to five jets from hard matrix
elements at the next-to-leading order, with the merging cut set to
7~GeV.  The corresponding predictions from \Alaric in comparison to
the experimental measurements are shown in
Figs.~\ref{fig:jetrates_aleph} and~\ref{fig:jetrates_jadeopal}.  We
observe excellent agreement of the multi-jet merged results with both
experimental measurements, both at leading, and at next-to-leading
order.  It is worth noting that the systematic uncertainties from
scale variations in the hard, perturbative region of $y_{23}$ are much
smaller in the case of NLO merging than they are in the case of
leading-order merging.  This is due to the fact that the hard region
of this observable is described mostly at NLO precision by the 3-, 4-
and 5-jet fixed-order contributions to the merging, while yet higher
multiplicity contributions are very small due to restricted phase
space.  For $y_{34}$ there is still a considerable improvement, but it
is not as pronounced as in the case of $y_{23}$, because there is
significant phase space for 6-parton fixed-order contributions which
are described at leading order precision in our setup.  For $y_{45}$
and $y_{56}$ the improvement from the NLO merging is diminishing,
consistent with the fact that the 6-parton matrix elements are
pure leading order, and thus inherit the relatively large
renormalization scale dependence from the parton shower.

\begin{figure}[tp]
  \begin{center}
    \includegraphics[width=0.4\textwidth]{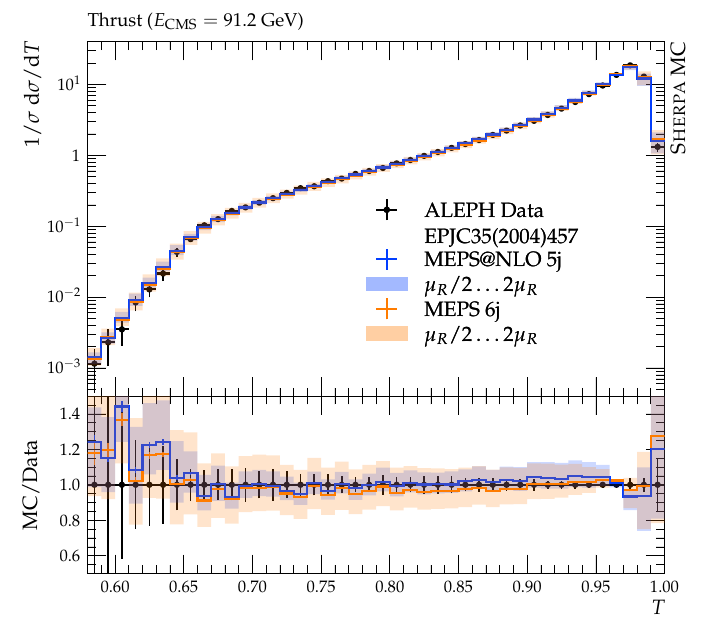}\nolinebreak
    \includegraphics[width=0.4\textwidth]{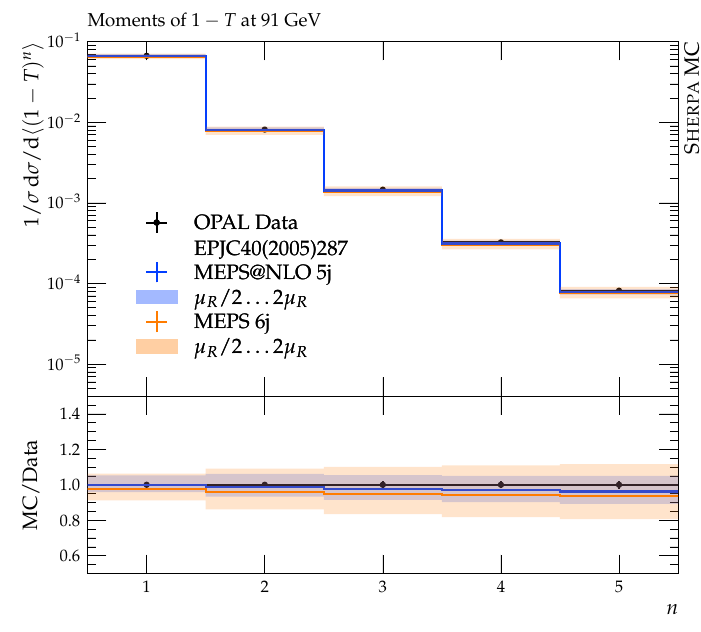}
  \end{center}
  \caption{Perturbative uncertainties in \MEPS and \MEPSatNLO
           predictions of thrust. Compared are the measurements for the event
           shape from \ALEPH~\cite{Heister:2003aj} and its moments from
           OPAL~\cite{Abbiendi:2004qz}.}
  \label{fig:thrust}
\end{figure}
Figure~\ref{fig:thrust} shows the thrust~\cite{Farhi:1977sg}
distribution as measured by the \ALEPH
collaboration~\cite{Heister:2003aj} (left), and the moments of thrust
as measured by the \OPAL collaboration~\cite{Abbiendi:2004qz} (right).
The thrust observable is known to very high accuracy in (resummed)
perturbation theory. Due to the intricate structure of
non-perturbative corrections~\cite{Nason:2023asn}, a coherent description
at the hadron level has, however, only been achieved recently.  Our prediction
consistently includes the differences in power corrections that occur
in the different regions, and it naturally extends beyond the Sudakov
shoulder~\cite{Catani:1997xc} due to the complete phase-space coverage
in the multi-jet calculations.  It is therefore not unexpected that
the predictions give excellent agreement with the experimental data.
We observe a reduction of the perturbative uncertainty by about a
factor two in the bulk of the phase space, and a smaller reduction in
the region beyond the Sudakov shoulder. This is in line with the
effects discussed for the Durham jet rates.

\begin{figure}[tp]
  \begin{center}
    \includegraphics[width=0.4\textwidth]{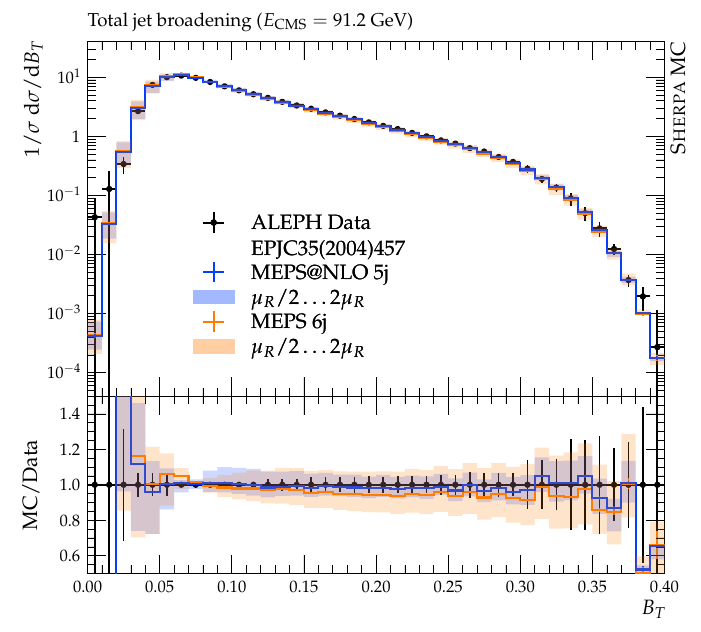}\nolinebreak
    \includegraphics[width=0.4\textwidth]{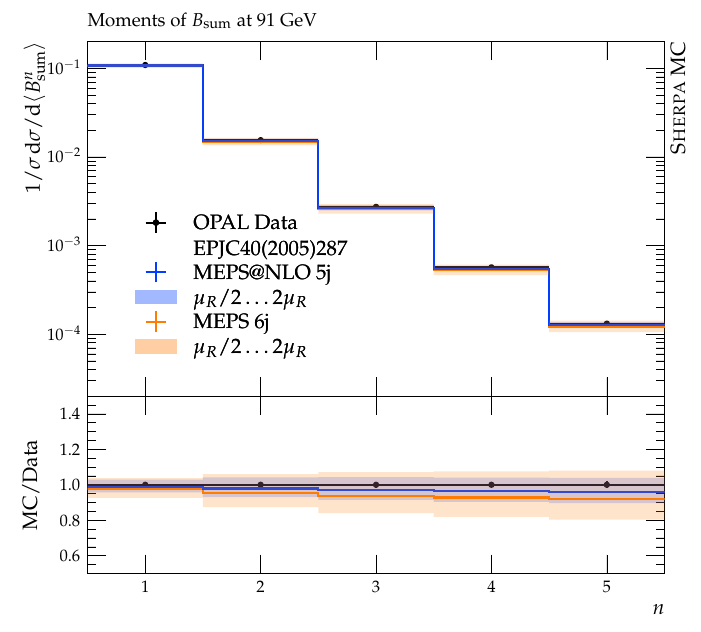}\\
    \includegraphics[width=0.4\textwidth]{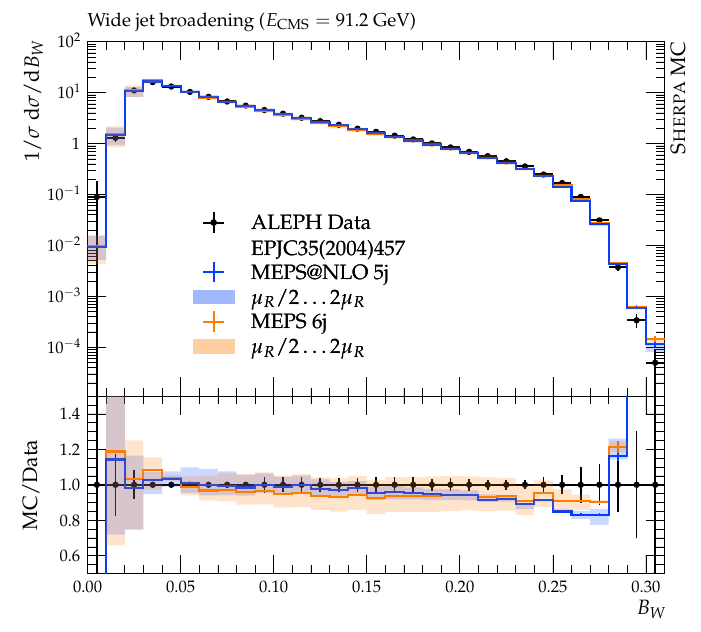}\nolinebreak
    \includegraphics[width=0.4\textwidth]{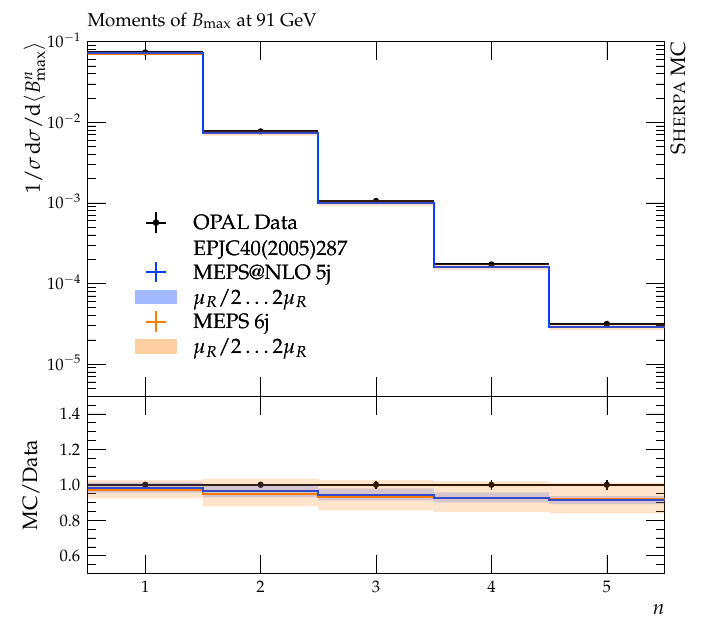}
  \end{center}
  \caption{Perturbative uncertainties in \MEPS and \MEPSatNLO
           predictions of total jet/hemisphere broadening (top) and
           wide jet/hemisphere broadening (bottom). Compared are the
           measurements from \ALEPH~\cite{Heister:2003aj} and \OPAL~\cite{Abbiendi:2004qz}.}
  \label{fig:broadening}
\end{figure}
The top panels of Fig.~\ref{fig:broadening} show the total jet
broadening distribution as measured by the \ALEPH
collaboration~\cite{Heister:2003aj} (left), and the moments of total
broadening as measured by the \OPAL
collaboration~\cite{Abbiendi:2004qz} (right).  Again, we observe a
significantly improved description of the experimental data with NLO
merged predictions, and a reduction of perturbative uncertainties by
about a factor of two.  Similar improvements are found for wide jet
broadening and its moments, shown in Fig.~\ref{fig:broadening}
(bottom).

\begin{figure}[tp]
  \begin{center}
    \includegraphics[width=0.4\textwidth]{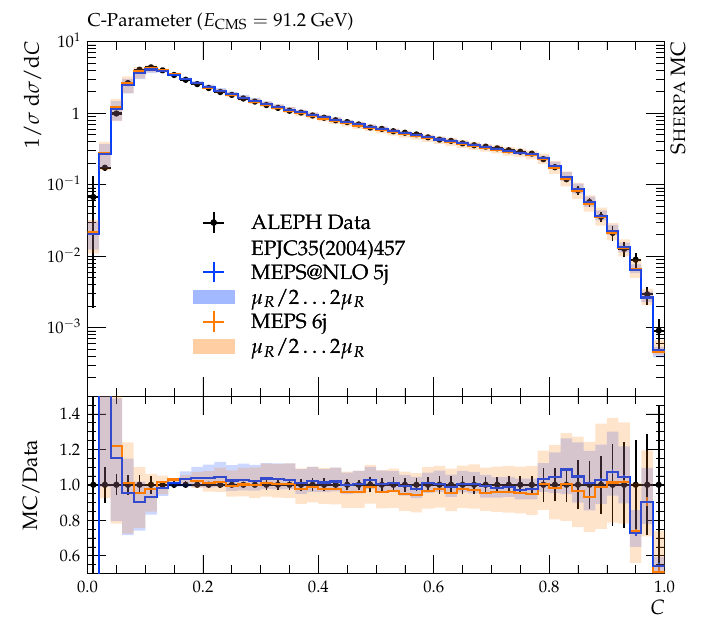}\nolinebreak
    \includegraphics[width=0.4\textwidth]{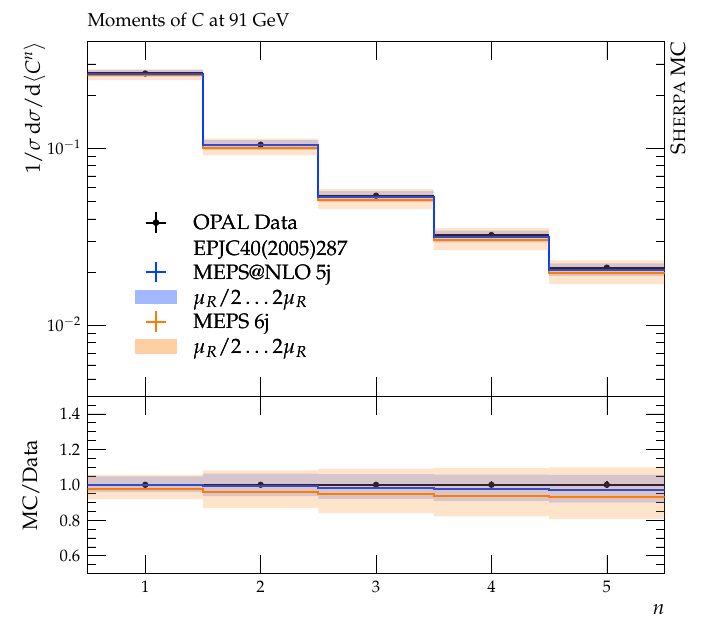}
  \end{center}
  \caption{Perturbative uncertainties in \MEPS and \MEPS@NLO
           predictions of the $C$-parameter. Compared are the
           measurements from \ALEPH~\cite{Heister:2003aj} and \OPAL~\cite{Abbiendi:2004qz}.}
  \label{fig:cparameter}
\end{figure}
Figure~\ref{fig:cparameter} shows the distribution of the
$C$-parameter as measured by the \ALEPH
collaboration~\cite{Heister:2003aj} (left), and the moments of the
$C$-parameter as measured by the \OPAL
collaboration~\cite{Abbiendi:2004qz} (right).  The description of the
distribution is achieved with similar quality in the case of leading-
and next-to-leading order merging; however, the moments show a clear
preference for the next-to-leading order merged result.  Again, the
renormalization scale uncertainty is reduced by about a factor of two
in both cases.

\begin{figure}[tp]
  \begin{center}
    \includegraphics[width=0.42\textwidth]{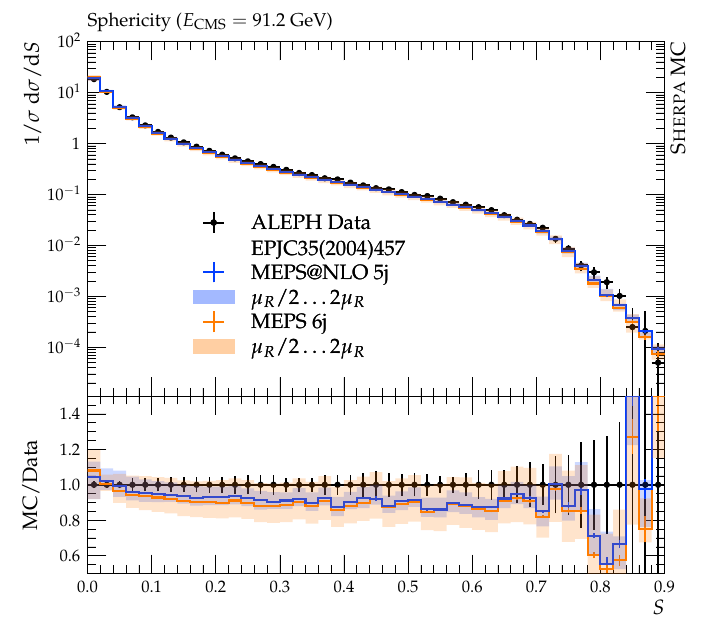}\nolinebreak
    \includegraphics[width=0.42\textwidth]{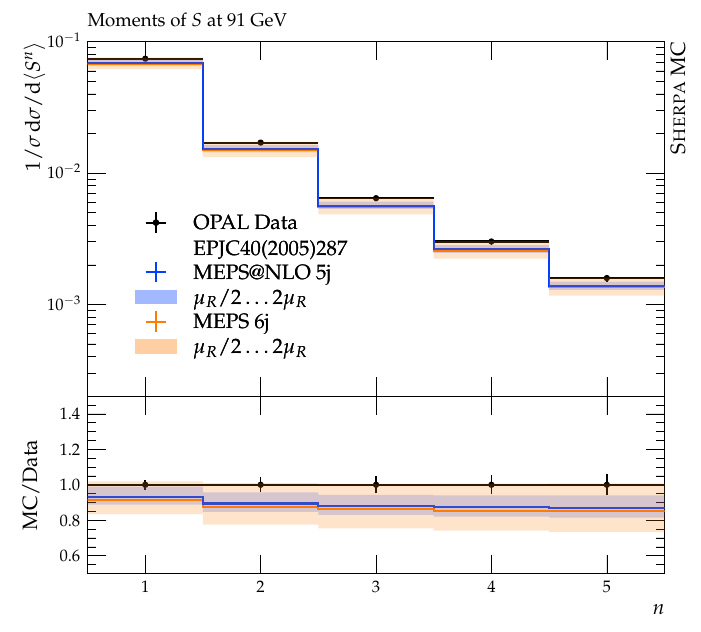}
  \end{center}
  \caption{Perturbative uncertainties in \MEPS and \MEPSatNLO
           predictions of sphericity. Compared are the
           measurements from \ALEPH~\cite{Heister:2003aj} and \OPAL~\cite{Abbiendi:2004qz}.}
  \label{fig:sphericity}
\end{figure}
As the last event shape variable, we investigate sphericity.
Figure~\ref{fig:sphericity} shows its distribution as measured by the
\ALEPH collaboration~\cite{Heister:2003aj} (left), and its moments as
measured by the \OPAL collaboration~\cite{Abbiendi:2004qz} (right).
We observe a significant improvement in the description of both the
distribution and its moments in the next-to-leading order merged
predictions.  Again, the perturbative uncertainties are reduced to
about half of their size at leading order.

\begin{figure}[tp]
  \begin{center}
    \includegraphics[width=0.42\textwidth]{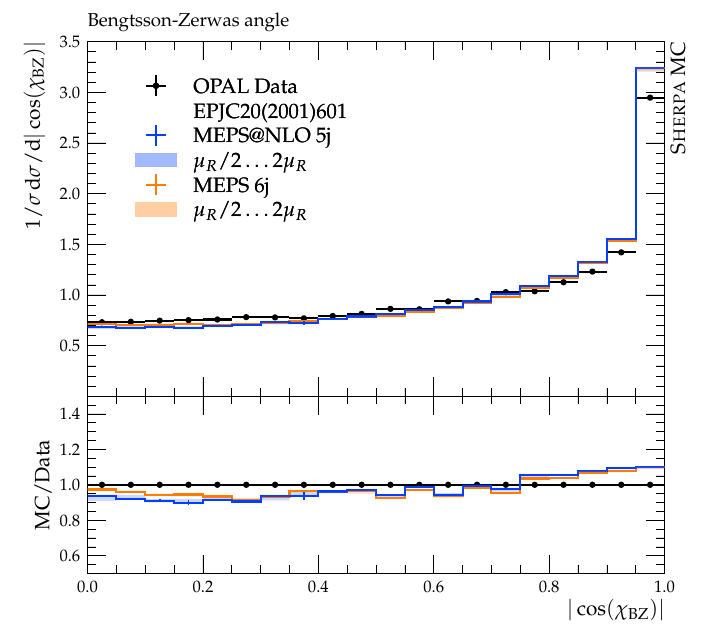}\nolinebreak
    \includegraphics[width=0.42\textwidth]{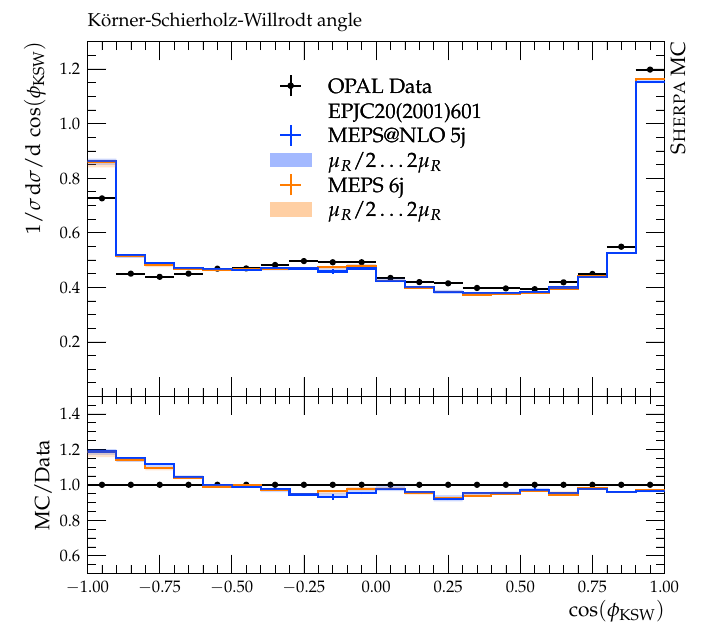}\\
    \includegraphics[width=0.42\textwidth]{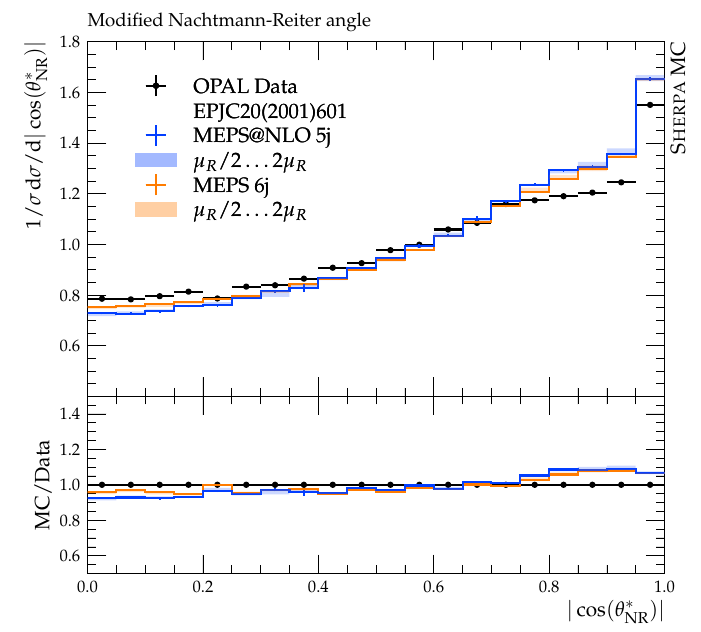}\nolinebreak
    \includegraphics[width=0.42\textwidth]{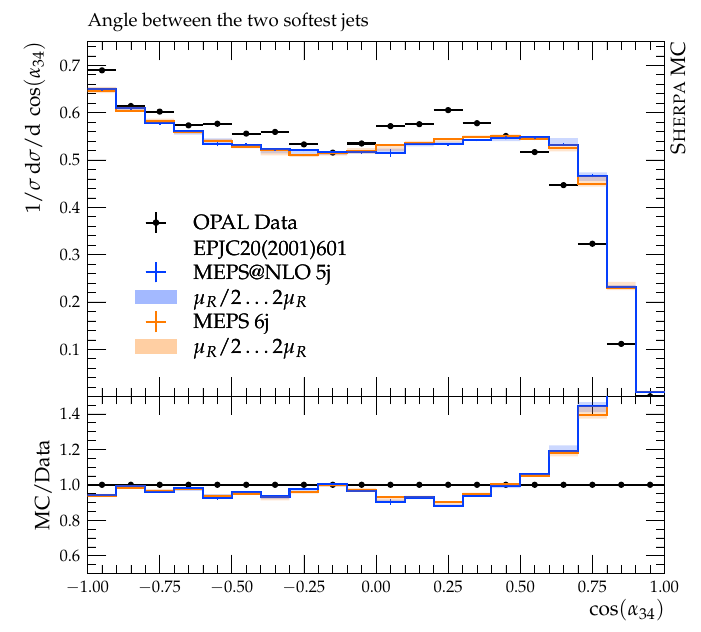}
  \end{center}
  \caption{Four-jet angles using the Durham algorithm compared to data from
           \OPAL~\cite{Abbiendi:2001qn}.}
  \label{fig:fourjetangles}
\end{figure}
Finally, we compare leading- and next-to-leading order merged
predictions for the four-jet angles measured by the \OPAL
collaboration~\cite{Abbiendi:2001qn}.  Figure~\ref{fig:fourjetangles}
shows that no improvement is obtained in the simulation using
next-to-leading order merging methods when compared to leading-order
merging.  It is difficult to judge the quality of the Monte-Carlo
predictions due to the missing experimental uncertainties.

\section{Outlook}
\label{sec:outlook}
In this manuscript, we have derived various analytic expressions needed
for a process-independent implementation of an infrared subtraction
scheme based on the \Alaric parton shower model.  We have validated
the subtraction in the processes of $e^+e^-\to n$~jets, with $n$ ranging
from two to five, which is the current state of the art for high-multiplicity
fixed-order multi-jet calculations at a lepton collider~\cite{Frederix:2010ne}.
Making use of the fixed-order subtraction, we have developed the first
\SMCatNLO matching method for the \Alaric parton shower, and applied it
to the phenomenologically important reactions of $e^+e^-\to$hadrons and
gluonic decays of the Higgs boson.  Utilizing the generic
implementation of next-to-leading order multi-jet merging methods in
the \Sherpa event generator, we combined the next-to-leading order
matched predictions for 2-, 3-, 4- and 5-jets and presented the first
multi-jet merged simulation for color singlet decays at the $Z$ pole
with a modern parton shower. The comparison to experimental data clearly
demonstrates the improvements obtained with next-to-leading order
multi-jet merging, and shows an encouraging reduction of the
perturbative uncertainties that will be crucial in view of a potential
future lepton collider such as the \FCC-ee.

In the near future we will extend the NLO matching of the \Alaric
parton-shower model to processes with colored particles in the
initial state, and include the matching for
final states with heavy quarks.  In addition, we expect to include
higher-order corrections to the evolution in the form of triple
collinear and double soft contributions to the splitting functions,
and more generally an extension towards higher logarithmic precision
of the parton shower model. The current state of the art simulations
will be made available in a next minor public version of the \Sherpa
event generator.

\section*{Acknowledgments}
This manuscript has been authored by Fermi Forward Discovery Group, LLC
under Contract No. 89243024CSC000002 with the U.S.\ Department of Energy,
Office of Science, Office of High Energy Physics.
The work of S.H. was supported by the U.S. Department of Energy,
Office of Science, Office of Advanced Scientific Computing Research,
Scientific Discovery through Advanced Computing (SciDAC-5) program,
grant “NeuCol”. F.K. gratefully acknowledges support by STFC under
grant agreement ST/P006744/1. P.M. is supported by the Swiss National
Science Foundation (SNF) under contract 200020-204200. D.R.\ is supported
by the European Union under the HORIZON program in
Marie Sk{\l}odowska-Curie project No. 101153541. During parts of this work, D.R. 
was further supported by STFC under grant agreement ST/P006744/1 as well as 
by the Durham University Physics Department Developing Talents Award and by 
the German Academic Exchange Service (DAAD).

\FloatBarrier

\bibliography{main}

\begin{thebibliography}{127}%
\makeatletter
\providecommand \@ifxundefined [1]{%
 \@ifx{#1\undefined}
}%
\providecommand \@ifnum [1]{%
 \ifnum #1\expandafter \@firstoftwo
 \else \expandafter \@secondoftwo
 \fi
}%
\providecommand \@ifx [1]{%
 \ifx #1\expandafter \@firstoftwo
 \else \expandafter \@secondoftwo
 \fi
}%
\providecommand \natexlab [1]{#1}%
\providecommand \enquote  [1]{``#1''}%
\providecommand \bibnamefont  [1]{#1}%
\providecommand \bibfnamefont [1]{#1}%
\providecommand \citenamefont [1]{#1}%
\providecommand \href@noop [0]{\@secondoftwo}%
\providecommand \href [0]{\begingroup \@sanitize@url \@href}%
\providecommand \@href[1]{\@@startlink{#1}\@@href}%
\providecommand \@@href[1]{\endgroup#1\@@endlink}%
\providecommand \@sanitize@url [0]{\catcode `\\12\catcode `\$12\catcode
  `\&12\catcode `\#12\catcode `\^12\catcode `\_12\catcode `\%12\relax}%
\providecommand \@@startlink[1]{}%
\providecommand \@@endlink[0]{}%
\providecommand \url  [0]{\begingroup\@sanitize@url \@url }%
\providecommand \@url [1]{\endgroup\@href {#1}{\urlprefix }}%
\providecommand \urlprefix  [0]{URL }%
\providecommand \Eprint [0]{\href }%
\providecommand \doibase [0]{http://dx.doi.org/}%
\providecommand \selectlanguage [0]{\@gobble}%
\providecommand \bibinfo  [0]{\@secondoftwo}%
\providecommand \bibfield  [0]{\@secondoftwo}%
\providecommand \translation [1]{[#1]}%
\providecommand \BibitemOpen [0]{}%
\providecommand \bibitemStop [0]{}%
\providecommand \bibitemNoStop [0]{.\EOS\space}%
\providecommand \EOS [0]{\spacefactor3000\relax}%
\providecommand \BibitemShut  [1]{\csname bibitem#1\endcsname}%
\let\auto@bib@innerbib\@empty
%</preamble>
\bibitem [{\citenamefont {Buckley}\ \emph {et~al.}(2011)\citenamefont {Buckley}
  \emph {et~al.}}]{Buckley:2011ms}%
  \BibitemOpen
  \bibfield  {author} {\bibinfo {author} {\bibfnamefont {A.}~\bibnamefont
  {Buckley}} \emph {et~al.},\ }\href {\doibase 10.1016/j.physrep.2011.03.005}
  {\bibfield  {journal} {\bibinfo  {journal} {Phys. Rept.}\ }\textbf {\bibinfo
  {volume} {504}},\ \bibinfo {pages} {145} (\bibinfo {year} {2011})},\ \Eprint
  {http://arxiv.org/abs/1101.2599} {arXiv:1101.2599 [hep-ph]} \BibitemShut
  {NoStop}%
\bibitem [{\citenamefont {Campbell}\ \emph {et~al.}(2024)\citenamefont
  {Campbell} \emph {et~al.}}]{Campbell:2022qmc}%
  \BibitemOpen
  \bibfield  {author} {\bibinfo {author} {\bibfnamefont {J.~M.}\ \bibnamefont
  {Campbell}} \emph {et~al.},\ }\href {\doibase 10.21468/SciPostPhys.16.5.130}
  {\bibfield  {journal} {\bibinfo  {journal} {SciPost Phys.}\ }\textbf
  {\bibinfo {volume} {16}},\ \bibinfo {pages} {130} (\bibinfo {year} {2024})},\
  \Eprint {http://arxiv.org/abs/2203.11110} {arXiv:2203.11110 [hep-ph]}
  \BibitemShut {NoStop}%
\bibitem [{\citenamefont {Aad}\ \emph {et~al.}(2008)\citenamefont {Aad} \emph
  {et~al.}}]{ATLAS:2008xda}%
  \BibitemOpen
  \bibfield  {author} {\bibinfo {author} {\bibfnamefont {G.}~\bibnamefont
  {Aad}} \emph {et~al.} (\bibinfo {collaboration} {ATLAS}),\ }\href {\doibase
  10.1088/1748-0221/3/08/S08003} {\bibfield  {journal} {\bibinfo  {journal}
  {JINST}\ }\textbf {\bibinfo {volume} {3}},\ \bibinfo {pages} {S08003}
  (\bibinfo {year} {2008})}\BibitemShut {NoStop}%
\bibitem [{\citenamefont {Chatrchyan}\ \emph {et~al.}(2008)\citenamefont
  {Chatrchyan} \emph {et~al.}}]{CMS:2008xjf}%
  \BibitemOpen
  \bibfield  {author} {\bibinfo {author} {\bibfnamefont {S.}~\bibnamefont
  {Chatrchyan}} \emph {et~al.} (\bibinfo {collaboration} {CMS}),\ }\href
  {\doibase 10.1088/1748-0221/3/08/S08004} {\bibfield  {journal} {\bibinfo
  {journal} {JINST}\ }\textbf {\bibinfo {volume} {3}},\ \bibinfo {pages}
  {S08004} (\bibinfo {year} {2008})}\BibitemShut {NoStop}%
\bibitem [{\citenamefont {Alves}\ \emph {et~al.}(2008)\citenamefont {Alves}
  \emph {et~al.}}]{LHCb:2008vvz}%
  \BibitemOpen
  \bibfield  {author} {\bibinfo {author} {\bibfnamefont {A.~A.}\ \bibnamefont
  {Alves}, \bibfnamefont {Jr.}} \emph {et~al.} (\bibinfo {collaboration}
  {LHCb}),\ }\href {\doibase 10.1088/1748-0221/3/08/S08005} {\bibfield
  {journal} {\bibinfo  {journal} {JINST}\ }\textbf {\bibinfo {volume} {3}},\
  \bibinfo {pages} {S08005} (\bibinfo {year} {2008})}\BibitemShut {NoStop}%
\bibitem [{\citenamefont {Aamodt}\ \emph {et~al.}(2008)\citenamefont {Aamodt}
  \emph {et~al.}}]{ALICE:2008ngc}%
  \BibitemOpen
  \bibfield  {author} {\bibinfo {author} {\bibfnamefont {K.}~\bibnamefont
  {Aamodt}} \emph {et~al.} (\bibinfo {collaboration} {ALICE}),\ }\href
  {\doibase 10.1088/1748-0221/3/08/S08002} {\bibfield  {journal} {\bibinfo
  {journal} {JINST}\ }\textbf {\bibinfo {volume} {3}},\ \bibinfo {pages}
  {S08002} (\bibinfo {year} {2008})}\BibitemShut {NoStop}%
\bibitem [{\citenamefont {Benedikt}\ \emph
  {et~al.}(2025{\natexlab{a}})\citenamefont {Benedikt} \emph
  {et~al.}}]{FCC:2025lpp}%
  \BibitemOpen
  \bibfield  {author} {\bibinfo {author} {\bibfnamefont {M.}~\bibnamefont
  {Benedikt}} \emph {et~al.} (\bibinfo {collaboration} {FCC}),\ }\href
  {\doibase 10.17181/CERN.9DKX.TDH9} {\  (\bibinfo {year}
  {2025}{\natexlab{a}}),\ 10.17181/CERN.9DKX.TDH9},\ \Eprint
  {http://arxiv.org/abs/2505.00272} {arXiv:2505.00272 [hep-ex]} \BibitemShut
  {NoStop}%
\bibitem [{\citenamefont {Benedikt}\ \emph
  {et~al.}(2025{\natexlab{b}})\citenamefont {Benedikt} \emph
  {et~al.}}]{FCC:2025uan}%
  \BibitemOpen
  \bibfield  {author} {\bibinfo {author} {\bibfnamefont {M.}~\bibnamefont
  {Benedikt}} \emph {et~al.} (\bibinfo {collaboration} {FCC}),\ }\href
  {\doibase 10.17181/CERN.EBAY.7W4X} {\  (\bibinfo {year}
  {2025}{\natexlab{b}}),\ 10.17181/CERN.EBAY.7W4X},\ \Eprint
  {http://arxiv.org/abs/2505.00274} {arXiv:2505.00274 [physics.acc-ph]}
  \BibitemShut {NoStop}%
\bibitem [{\citenamefont {Benedikt}\ \emph
  {et~al.}(2025{\natexlab{c}})\citenamefont {Benedikt} \emph
  {et~al.}}]{FCC:2025jtd}%
  \BibitemOpen
  \bibfield  {author} {\bibinfo {author} {\bibfnamefont {M.}~\bibnamefont
  {Benedikt}} \emph {et~al.} (\bibinfo {collaboration} {FCC}),\ }\href
  {\doibase 10.17181/CERN.I26X.V4VF} {\  (\bibinfo {year}
  {2025}{\natexlab{c}}),\ 10.17181/CERN.I26X.V4VF},\ \Eprint
  {http://arxiv.org/abs/2505.00273} {arXiv:2505.00273 [physics.acc-ph]}
  \BibitemShut {NoStop}%
\bibitem [{\citenamefont {Webber}(1984)}]{Webber:1983if}%
  \BibitemOpen
  \bibfield  {author} {\bibinfo {author} {\bibfnamefont {B.~R.}\ \bibnamefont
  {Webber}},\ }\href {\doibase 10.1016/0550-3213(84)90333-X} {\bibfield
  {journal} {\bibinfo  {journal} {Nucl. Phys. B}\ }\textbf {\bibinfo {volume}
  {238}},\ \bibinfo {pages} {492} (\bibinfo {year} {1984})}\BibitemShut
  {NoStop}%
\bibitem [{\citenamefont {Bengtsson}\ \emph {et~al.}(1986)\citenamefont
  {Bengtsson}, \citenamefont {Sj{\"o}strand},\ and\ \citenamefont {van
  Zijl}}]{Bengtsson:1986gz}%
  \BibitemOpen
  \bibfield  {author} {\bibinfo {author} {\bibfnamefont {M.}~\bibnamefont
  {Bengtsson}}, \bibinfo {author} {\bibfnamefont {T.}~\bibnamefont
  {Sj{\"o}strand}}, \ and\ \bibinfo {author} {\bibfnamefont {M.}~\bibnamefont
  {van Zijl}},\ }\href {\doibase 10.1007/BF01441353} {\bibfield  {journal}
  {\bibinfo  {journal} {Z. Phys. C}\ }\textbf {\bibinfo {volume} {32}},\
  \bibinfo {pages} {67} (\bibinfo {year} {1986})}\BibitemShut {NoStop}%
\bibitem [{\citenamefont {Bengtsson}\ and\ \citenamefont
  {Sj{\"o}strand}(1987{\natexlab{a}})}]{Bengtsson:1986et}%
  \BibitemOpen
  \bibfield  {author} {\bibinfo {author} {\bibfnamefont {M.}~\bibnamefont
  {Bengtsson}}\ and\ \bibinfo {author} {\bibfnamefont {T.}~\bibnamefont
  {Sj{\"o}strand}},\ }\href {\doibase 10.1016/0550-3213(87)90407-X} {\bibfield
  {journal} {\bibinfo  {journal} {Nucl. Phys. B}\ }\textbf {\bibinfo {volume}
  {289}},\ \bibinfo {pages} {810} (\bibinfo {year}
  {1987}{\natexlab{a}})}\BibitemShut {NoStop}%
\bibitem [{\citenamefont {Marchesini}\ and\ \citenamefont
  {Webber}(1988)}]{Marchesini:1987cf}%
  \BibitemOpen
  \bibfield  {author} {\bibinfo {author} {\bibfnamefont {G.}~\bibnamefont
  {Marchesini}}\ and\ \bibinfo {author} {\bibfnamefont {B.~R.}\ \bibnamefont
  {Webber}},\ }\href {\doibase 10.1016/0550-3213(88)90089-2} {\bibfield
  {journal} {\bibinfo  {journal} {Nucl. Phys. B}\ }\textbf {\bibinfo {volume}
  {310}},\ \bibinfo {pages} {461} (\bibinfo {year} {1988})}\BibitemShut
  {NoStop}%
\bibitem [{\citenamefont {Andersson}\ \emph {et~al.}(1990)\citenamefont
  {Andersson}, \citenamefont {Gustafson},\ and\ \citenamefont
  {L{\"o}nnblad}}]{Andersson:1989ki}%
  \BibitemOpen
  \bibfield  {author} {\bibinfo {author} {\bibfnamefont {B.}~\bibnamefont
  {Andersson}}, \bibinfo {author} {\bibfnamefont {G.}~\bibnamefont
  {Gustafson}}, \ and\ \bibinfo {author} {\bibfnamefont {L.}~\bibnamefont
  {L{\"o}nnblad}},\ }\href {\doibase 10.1016/0550-3213(90)90355-H} {\bibfield
  {journal} {\bibinfo  {journal} {Nucl. Phys. B}\ }\textbf {\bibinfo {volume}
  {339}},\ \bibinfo {pages} {393} (\bibinfo {year} {1990})}\BibitemShut
  {NoStop}%
\bibitem [{\citenamefont {Webber}(1986)}]{Webber:1986mc}%
  \BibitemOpen
  \bibfield  {author} {\bibinfo {author} {\bibfnamefont {B.~R.}\ \bibnamefont
  {Webber}},\ }\href {\doibase 10.1146/annurev.ns.36.120186.001345} {\bibfield
  {journal} {\bibinfo  {journal} {Ann. Rev. Nucl. Part. Sci.}\ }\textbf
  {\bibinfo {volume} {36}},\ \bibinfo {pages} {253} (\bibinfo {year}
  {1986})}\BibitemShut {NoStop}%
\bibitem [{\citenamefont {Bengtsson}\ and\ \citenamefont
  {Sj{\"o}strand}(1987{\natexlab{b}})}]{Bengtsson:1986hr}%
  \BibitemOpen
  \bibfield  {author} {\bibinfo {author} {\bibfnamefont {M.}~\bibnamefont
  {Bengtsson}}\ and\ \bibinfo {author} {\bibfnamefont {T.}~\bibnamefont
  {Sj{\"o}strand}},\ }\href {\doibase 10.1016/0370-2693(87)91031-8} {\bibfield
  {journal} {\bibinfo  {journal} {Phys. Lett. B}\ }\textbf {\bibinfo {volume}
  {185}},\ \bibinfo {pages} {435} (\bibinfo {year}
  {1987}{\natexlab{b}})}\BibitemShut {NoStop}%
\bibitem [{\citenamefont {Collins}(1988)}]{Collins:1987cp}%
  \BibitemOpen
  \bibfield  {author} {\bibinfo {author} {\bibfnamefont {J.~C.}\ \bibnamefont
  {Collins}},\ }\href {\doibase 10.1016/0550-3213(88)90654-2} {\bibfield
  {journal} {\bibinfo  {journal} {Nucl. Phys. B}\ }\textbf {\bibinfo {volume}
  {304}},\ \bibinfo {pages} {794} (\bibinfo {year} {1988})}\BibitemShut
  {NoStop}%
\bibitem [{\citenamefont {Knowles}(1988{\natexlab{a}})}]{Knowles:1987cu}%
  \BibitemOpen
  \bibfield  {author} {\bibinfo {author} {\bibfnamefont {I.~G.}\ \bibnamefont
  {Knowles}},\ }\href {\doibase 10.1016/0550-3213(88)90653-0} {\bibfield
  {journal} {\bibinfo  {journal} {Nucl. Phys. B}\ }\textbf {\bibinfo {volume}
  {304}},\ \bibinfo {pages} {767} (\bibinfo {year}
  {1988}{\natexlab{a}})}\BibitemShut {NoStop}%
\bibitem [{\citenamefont {Knowles}(1988{\natexlab{b}})}]{Knowles:1988vs}%
  \BibitemOpen
  \bibfield  {author} {\bibinfo {author} {\bibfnamefont {I.~G.}\ \bibnamefont
  {Knowles}},\ }\href {\doibase 10.1016/0550-3213(88)90092-2} {\bibfield
  {journal} {\bibinfo  {journal} {Nucl. Phys. B}\ }\textbf {\bibinfo {volume}
  {310}},\ \bibinfo {pages} {571} (\bibinfo {year}
  {1988}{\natexlab{b}})}\BibitemShut {NoStop}%
\bibitem [{\citenamefont {Knowles}(1990)}]{Knowles:1988hu}%
  \BibitemOpen
  \bibfield  {author} {\bibinfo {author} {\bibfnamefont {I.~G.}\ \bibnamefont
  {Knowles}},\ }\href {\doibase 10.1016/0010-4655(90)90063-7} {\bibfield
  {journal} {\bibinfo  {journal} {Comput. Phys. Commun.}\ }\textbf {\bibinfo
  {volume} {58}},\ \bibinfo {pages} {271} (\bibinfo {year} {1990})}\BibitemShut
  {NoStop}%
\bibitem [{\citenamefont {van Beekveld}\ \emph {et~al.}(2022)\citenamefont {van
  Beekveld}, \citenamefont {Ferrario~Ravasio}, \citenamefont {Hamilton},
  \citenamefont {Salam}, \citenamefont {Soto-Ontoso}, \citenamefont {Soyez},\
  and\ \citenamefont {Verheyen}}]{vanBeekveld:2022ukn}%
  \BibitemOpen
  \bibfield  {author} {\bibinfo {author} {\bibfnamefont {M.}~\bibnamefont {van
  Beekveld}}, \bibinfo {author} {\bibfnamefont {S.}~\bibnamefont
  {Ferrario~Ravasio}}, \bibinfo {author} {\bibfnamefont {K.}~\bibnamefont
  {Hamilton}}, \bibinfo {author} {\bibfnamefont {G.~P.}\ \bibnamefont {Salam}},
  \bibinfo {author} {\bibfnamefont {A.}~\bibnamefont {Soto-Ontoso}}, \bibinfo
  {author} {\bibfnamefont {G.}~\bibnamefont {Soyez}}, \ and\ \bibinfo {author}
  {\bibfnamefont {R.}~\bibnamefont {Verheyen}},\ }\href {\doibase
  10.1007/JHEP11(2022)020} {\bibfield  {journal} {\bibinfo  {journal} {JHEP}\
  }\textbf {\bibinfo {volume} {11}},\ \bibinfo {pages} {020} (\bibinfo {year}
  {2022})},\ \Eprint {http://arxiv.org/abs/2207.09467} {arXiv:2207.09467
  [hep-ph]} \BibitemShut {NoStop}%
\bibitem [{\citenamefont {Nagy}\ and\ \citenamefont
  {Soper}(2005)}]{Nagy:2005aa}%
  \BibitemOpen
  \bibfield  {author} {\bibinfo {author} {\bibfnamefont {Z.}~\bibnamefont
  {Nagy}}\ and\ \bibinfo {author} {\bibfnamefont {D.~E.}\ \bibnamefont
  {Soper}},\ }\href {\doibase 10.1088/1126-6708/2005/10/024} {\bibfield
  {journal} {\bibinfo  {journal} {JHEP}\ }\textbf {\bibinfo {volume} {10}},\
  \bibinfo {pages} {024} (\bibinfo {year} {2005})},\ \Eprint
  {http://arxiv.org/abs/hep-ph/0503053} {arXiv:hep-ph/0503053} \BibitemShut
  {NoStop}%
\bibitem [{\citenamefont {Nagy}\ and\ \citenamefont
  {Soper}(2006)}]{Nagy:2006kb}%
  \BibitemOpen
  \bibfield  {author} {\bibinfo {author} {\bibfnamefont {Z.}~\bibnamefont
  {Nagy}}\ and\ \bibinfo {author} {\bibfnamefont {D.~E.}\ \bibnamefont
  {Soper}},\ }in\ \href {\doibase 10.1142/9789812773524_0010} {\emph {\bibinfo
  {booktitle} {{Ringberg Workshop on New Trends in HERA Physics 2005}}}}\
  (\bibinfo {year} {2006})\ pp.\ \bibinfo {pages} {101--123},\ \Eprint
  {http://arxiv.org/abs/hep-ph/0601021} {arXiv:hep-ph/0601021} \BibitemShut
  {NoStop}%
\bibitem [{\citenamefont {Schumann}\ and\ \citenamefont
  {Krauss}(2008)}]{Schumann:2007mg}%
  \BibitemOpen
  \bibfield  {author} {\bibinfo {author} {\bibfnamefont {S.}~\bibnamefont
  {Schumann}}\ and\ \bibinfo {author} {\bibfnamefont {F.}~\bibnamefont
  {Krauss}},\ }\href {\doibase 10.1088/1126-6708/2008/03/038} {\bibfield
  {journal} {\bibinfo  {journal} {JHEP}\ }\textbf {\bibinfo {volume} {03}},\
  \bibinfo {pages} {038} (\bibinfo {year} {2008})},\ \Eprint
  {http://arxiv.org/abs/0709.1027} {arXiv:0709.1027 [hep-ph]} \BibitemShut
  {NoStop}%
\bibitem [{\citenamefont {Giele}\ \emph {et~al.}(2008)\citenamefont {Giele},
  \citenamefont {Kosower},\ and\ \citenamefont {Skands}}]{Giele:2007di}%
  \BibitemOpen
  \bibfield  {author} {\bibinfo {author} {\bibfnamefont {W.~T.}\ \bibnamefont
  {Giele}}, \bibinfo {author} {\bibfnamefont {D.~A.}\ \bibnamefont {Kosower}},
  \ and\ \bibinfo {author} {\bibfnamefont {P.~Z.}\ \bibnamefont {Skands}},\
  }\href {\doibase 10.1103/PhysRevD.78.014026} {\bibfield  {journal} {\bibinfo
  {journal} {Phys. Rev. D}\ }\textbf {\bibinfo {volume} {78}},\ \bibinfo
  {pages} {014026} (\bibinfo {year} {2008})},\ \Eprint
  {http://arxiv.org/abs/0707.3652} {arXiv:0707.3652 [hep-ph]} \BibitemShut
  {NoStop}%
\bibitem [{\citenamefont {Pl{\"a}tzer}\ and\ \citenamefont
  {Gieseke}(2011)}]{Platzer:2009jq}%
  \BibitemOpen
  \bibfield  {author} {\bibinfo {author} {\bibfnamefont {S.}~\bibnamefont
  {Pl{\"a}tzer}}\ and\ \bibinfo {author} {\bibfnamefont {S.}~\bibnamefont
  {Gieseke}},\ }\href {\doibase 10.1007/JHEP01(2011)024} {\bibfield  {journal}
  {\bibinfo  {journal} {JHEP}\ }\textbf {\bibinfo {volume} {01}},\ \bibinfo
  {pages} {024} (\bibinfo {year} {2011})},\ \Eprint
  {http://arxiv.org/abs/0909.5593} {arXiv:0909.5593 [hep-ph]} \BibitemShut
  {NoStop}%
\bibitem [{\citenamefont {H{\"o}che}\ and\ \citenamefont
  {Prestel}(2015)}]{Hoche:2015sya}%
  \BibitemOpen
  \bibfield  {author} {\bibinfo {author} {\bibfnamefont {S.}~\bibnamefont
  {H{\"o}che}}\ and\ \bibinfo {author} {\bibfnamefont {S.}~\bibnamefont
  {Prestel}},\ }\href {\doibase 10.1140/epjc/s10052-015-3684-2} {\bibfield
  {journal} {\bibinfo  {journal} {Eur. Phys. J. C}\ }\textbf {\bibinfo {volume}
  {75}},\ \bibinfo {pages} {461} (\bibinfo {year} {2015})},\ \Eprint
  {http://arxiv.org/abs/1506.05057} {arXiv:1506.05057 [hep-ph]} \BibitemShut
  {NoStop}%
\bibitem [{\citenamefont {Fischer}\ \emph {et~al.}(2016)\citenamefont
  {Fischer}, \citenamefont {Prestel}, \citenamefont {Ritzmann},\ and\
  \citenamefont {Skands}}]{Fischer:2016vfv}%
  \BibitemOpen
  \bibfield  {author} {\bibinfo {author} {\bibfnamefont {N.}~\bibnamefont
  {Fischer}}, \bibinfo {author} {\bibfnamefont {S.}~\bibnamefont {Prestel}},
  \bibinfo {author} {\bibfnamefont {M.}~\bibnamefont {Ritzmann}}, \ and\
  \bibinfo {author} {\bibfnamefont {P.}~\bibnamefont {Skands}},\ }\href
  {\doibase 10.1140/epjc/s10052-016-4429-6} {\bibfield  {journal} {\bibinfo
  {journal} {Eur. Phys. J. C}\ }\textbf {\bibinfo {volume} {76}},\ \bibinfo
  {pages} {589} (\bibinfo {year} {2016})},\ \Eprint
  {http://arxiv.org/abs/1605.06142} {arXiv:1605.06142 [hep-ph]} \BibitemShut
  {NoStop}%
\bibitem [{\citenamefont {Cabouat}\ and\ \citenamefont
  {Sj{\"o}strand}(2018)}]{Cabouat:2017rzi}%
  \BibitemOpen
  \bibfield  {author} {\bibinfo {author} {\bibfnamefont {B.}~\bibnamefont
  {Cabouat}}\ and\ \bibinfo {author} {\bibfnamefont {T.}~\bibnamefont
  {Sj{\"o}strand}},\ }\href {\doibase 10.1140/epjc/s10052-018-5645-z}
  {\bibfield  {journal} {\bibinfo  {journal} {Eur. Phys. J. C}\ }\textbf
  {\bibinfo {volume} {78}},\ \bibinfo {pages} {226} (\bibinfo {year} {2018})},\
  \Eprint {http://arxiv.org/abs/1710.00391} {arXiv:1710.00391 [hep-ph]}
  \BibitemShut {NoStop}%
\bibitem [{\citenamefont {Nagy}\ and\ \citenamefont
  {Soper}(2012)}]{Nagy:2012bt}%
  \BibitemOpen
  \bibfield  {author} {\bibinfo {author} {\bibfnamefont {Z.}~\bibnamefont
  {Nagy}}\ and\ \bibinfo {author} {\bibfnamefont {D.~E.}\ \bibnamefont
  {Soper}},\ }\href {\doibase 10.1007/JHEP06(2012)044} {\bibfield  {journal}
  {\bibinfo  {journal} {JHEP}\ }\textbf {\bibinfo {volume} {06}},\ \bibinfo
  {pages} {044} (\bibinfo {year} {2012})},\ \Eprint
  {http://arxiv.org/abs/1202.4496} {arXiv:1202.4496 [hep-ph]} \BibitemShut
  {NoStop}%
\bibitem [{\citenamefont {Pl{\"a}tzer}\ and\ \citenamefont
  {Sj{\"o}dahl}(2012)}]{Platzer:2012np}%
  \BibitemOpen
  \bibfield  {author} {\bibinfo {author} {\bibfnamefont {S.}~\bibnamefont
  {Pl{\"a}tzer}}\ and\ \bibinfo {author} {\bibfnamefont {M.}~\bibnamefont
  {Sj{\"o}dahl}},\ }\href {\doibase 10.1007/JHEP07(2012)042} {\bibfield
  {journal} {\bibinfo  {journal} {JHEP}\ }\textbf {\bibinfo {volume} {07}},\
  \bibinfo {pages} {042} (\bibinfo {year} {2012})},\ \Eprint
  {http://arxiv.org/abs/1201.0260} {arXiv:1201.0260 [hep-ph]} \BibitemShut
  {NoStop}%
\bibitem [{\citenamefont {Nagy}\ and\ \citenamefont
  {Soper}(2014)}]{Nagy:2014mqa}%
  \BibitemOpen
  \bibfield  {author} {\bibinfo {author} {\bibfnamefont {Z.}~\bibnamefont
  {Nagy}}\ and\ \bibinfo {author} {\bibfnamefont {D.~E.}\ \bibnamefont
  {Soper}},\ }\href {\doibase 10.1007/JHEP06(2014)097} {\bibfield  {journal}
  {\bibinfo  {journal} {JHEP}\ }\textbf {\bibinfo {volume} {06}},\ \bibinfo
  {pages} {097} (\bibinfo {year} {2014})},\ \Eprint
  {http://arxiv.org/abs/1401.6364} {arXiv:1401.6364 [hep-ph]} \BibitemShut
  {NoStop}%
\bibitem [{\citenamefont {Nagy}\ and\ \citenamefont
  {Soper}(2015)}]{Nagy:2015hwa}%
  \BibitemOpen
  \bibfield  {author} {\bibinfo {author} {\bibfnamefont {Z.}~\bibnamefont
  {Nagy}}\ and\ \bibinfo {author} {\bibfnamefont {D.~E.}\ \bibnamefont
  {Soper}},\ }\href {\doibase 10.1007/JHEP07(2015)119} {\bibfield  {journal}
  {\bibinfo  {journal} {JHEP}\ }\textbf {\bibinfo {volume} {07}},\ \bibinfo
  {pages} {119} (\bibinfo {year} {2015})},\ \Eprint
  {http://arxiv.org/abs/1501.00778} {arXiv:1501.00778 [hep-ph]} \BibitemShut
  {NoStop}%
\bibitem [{\citenamefont {Pl{\"a}tzer}\ \emph {et~al.}(2018)\citenamefont
  {Pl{\"a}tzer}, \citenamefont {Sj{\"o}dahl},\ and\ \citenamefont
  {Thor{\'e}n}}]{Platzer:2018pmd}%
  \BibitemOpen
  \bibfield  {author} {\bibinfo {author} {\bibfnamefont {S.}~\bibnamefont
  {Pl{\"a}tzer}}, \bibinfo {author} {\bibfnamefont {M.}~\bibnamefont
  {Sj{\"o}dahl}}, \ and\ \bibinfo {author} {\bibfnamefont {J.}~\bibnamefont
  {Thor{\'e}n}},\ }\href {\doibase 10.1007/JHEP11(2018)009} {\bibfield
  {journal} {\bibinfo  {journal} {JHEP}\ }\textbf {\bibinfo {volume} {11}},\
  \bibinfo {pages} {009} (\bibinfo {year} {2018})},\ \Eprint
  {http://arxiv.org/abs/1808.00332} {arXiv:1808.00332 [hep-ph]} \BibitemShut
  {NoStop}%
\bibitem [{\citenamefont {Isaacson}\ and\ \citenamefont
  {Prestel}(2019)}]{Isaacson:2018zdi}%
  \BibitemOpen
  \bibfield  {author} {\bibinfo {author} {\bibfnamefont {J.}~\bibnamefont
  {Isaacson}}\ and\ \bibinfo {author} {\bibfnamefont {S.}~\bibnamefont
  {Prestel}},\ }\href {\doibase 10.1103/PhysRevD.99.014021} {\bibfield
  {journal} {\bibinfo  {journal} {Phys. Rev. D}\ }\textbf {\bibinfo {volume}
  {99}},\ \bibinfo {pages} {014021} (\bibinfo {year} {2019})},\ \Eprint
  {http://arxiv.org/abs/1806.10102} {arXiv:1806.10102 [hep-ph]} \BibitemShut
  {NoStop}%
\bibitem [{\citenamefont {Nagy}\ and\ \citenamefont
  {Soper}(2019{\natexlab{a}})}]{Nagy:2019rwb}%
  \BibitemOpen
  \bibfield  {author} {\bibinfo {author} {\bibfnamefont {Z.}~\bibnamefont
  {Nagy}}\ and\ \bibinfo {author} {\bibfnamefont {D.~E.}\ \bibnamefont
  {Soper}},\ }\href {\doibase 10.1103/PhysRevD.100.074005} {\bibfield
  {journal} {\bibinfo  {journal} {Phys. Rev. D}\ }\textbf {\bibinfo {volume}
  {100}},\ \bibinfo {pages} {074005} (\bibinfo {year} {2019}{\natexlab{a}})},\
  \Eprint {http://arxiv.org/abs/1908.11420} {arXiv:1908.11420 [hep-ph]}
  \BibitemShut {NoStop}%
\bibitem [{\citenamefont {Nagy}\ and\ \citenamefont
  {Soper}(2019{\natexlab{b}})}]{Nagy:2019pjp}%
  \BibitemOpen
  \bibfield  {author} {\bibinfo {author} {\bibfnamefont {Z.}~\bibnamefont
  {Nagy}}\ and\ \bibinfo {author} {\bibfnamefont {D.~E.}\ \bibnamefont
  {Soper}},\ }\href {\doibase 10.1103/PhysRevD.99.054009} {\bibfield  {journal}
  {\bibinfo  {journal} {Phys. Rev. D}\ }\textbf {\bibinfo {volume} {99}},\
  \bibinfo {pages} {054009} (\bibinfo {year} {2019}{\natexlab{b}})},\ \Eprint
  {http://arxiv.org/abs/1902.02105} {arXiv:1902.02105 [hep-ph]} \BibitemShut
  {NoStop}%
\bibitem [{\citenamefont {Forshaw}\ \emph {et~al.}(2019)\citenamefont
  {Forshaw}, \citenamefont {Holguin},\ and\ \citenamefont
  {Pl{\"a}tzer}}]{Forshaw:2019ver}%
  \BibitemOpen
  \bibfield  {author} {\bibinfo {author} {\bibfnamefont {J.~R.}\ \bibnamefont
  {Forshaw}}, \bibinfo {author} {\bibfnamefont {J.}~\bibnamefont {Holguin}}, \
  and\ \bibinfo {author} {\bibfnamefont {S.}~\bibnamefont {Pl{\"a}tzer}},\
  }\href {\doibase 10.1007/JHEP08(2019)145} {\bibfield  {journal} {\bibinfo
  {journal} {JHEP}\ }\textbf {\bibinfo {volume} {08}},\ \bibinfo {pages} {145}
  (\bibinfo {year} {2019})},\ \Eprint {http://arxiv.org/abs/1905.08686}
  {arXiv:1905.08686 [hep-ph]} \BibitemShut {NoStop}%
\bibitem [{\citenamefont {H{\"o}che}\ and\ \citenamefont
  {Reichelt}(2021)}]{Hoche:2020pxj}%
  \BibitemOpen
  \bibfield  {author} {\bibinfo {author} {\bibfnamefont {S.}~\bibnamefont
  {H{\"o}che}}\ and\ \bibinfo {author} {\bibfnamefont {D.}~\bibnamefont
  {Reichelt}},\ }\href {\doibase 10.1103/PhysRevD.104.034006} {\bibfield
  {journal} {\bibinfo  {journal} {Phys. Rev. D}\ }\textbf {\bibinfo {volume}
  {104}},\ \bibinfo {pages} {034006} (\bibinfo {year} {2021})},\ \Eprint
  {http://arxiv.org/abs/2001.11492} {arXiv:2001.11492 [hep-ph]} \BibitemShut
  {NoStop}%
\bibitem [{\citenamefont {De~Angelis}\ \emph {et~al.}(2021)\citenamefont
  {De~Angelis}, \citenamefont {Forshaw},\ and\ \citenamefont
  {Pl{\"a}tzer}}]{DeAngelis:2020rvq}%
  \BibitemOpen
  \bibfield  {author} {\bibinfo {author} {\bibfnamefont {M.}~\bibnamefont
  {De~Angelis}}, \bibinfo {author} {\bibfnamefont {J.~R.}\ \bibnamefont
  {Forshaw}}, \ and\ \bibinfo {author} {\bibfnamefont {S.}~\bibnamefont
  {Pl{\"a}tzer}},\ }\href {\doibase 10.1103/PhysRevLett.126.112001} {\bibfield
  {journal} {\bibinfo  {journal} {Phys. Rev. Lett.}\ }\textbf {\bibinfo
  {volume} {126}},\ \bibinfo {pages} {112001} (\bibinfo {year} {2021})},\
  \Eprint {http://arxiv.org/abs/2007.09648} {arXiv:2007.09648 [hep-ph]}
  \BibitemShut {NoStop}%
\bibitem [{\citenamefont {Holguin}\ \emph {et~al.}(2021)\citenamefont
  {Holguin}, \citenamefont {Forshaw},\ and\ \citenamefont
  {Pl{\"a}tzer}}]{Holguin:2020joq}%
  \BibitemOpen
  \bibfield  {author} {\bibinfo {author} {\bibfnamefont {J.}~\bibnamefont
  {Holguin}}, \bibinfo {author} {\bibfnamefont {J.~R.}\ \bibnamefont
  {Forshaw}}, \ and\ \bibinfo {author} {\bibfnamefont {S.}~\bibnamefont
  {Pl{\"a}tzer}},\ }\href {\doibase 10.1140/epjc/s10052-021-09145-1} {\bibfield
   {journal} {\bibinfo  {journal} {Eur. Phys. J. C}\ }\textbf {\bibinfo
  {volume} {81}},\ \bibinfo {pages} {364} (\bibinfo {year} {2021})},\ \Eprint
  {http://arxiv.org/abs/2011.15087} {arXiv:2011.15087 [hep-ph]} \BibitemShut
  {NoStop}%
\bibitem [{\citenamefont {Gustafson}(1993)}]{Gustafson:1992uh}%
  \BibitemOpen
  \bibfield  {author} {\bibinfo {author} {\bibfnamefont {G.}~\bibnamefont
  {Gustafson}},\ }\href {\doibase 10.1016/0550-3213(93)90203-2} {\bibfield
  {journal} {\bibinfo  {journal} {Nucl. Phys. B}\ }\textbf {\bibinfo {volume}
  {392}},\ \bibinfo {pages} {251} (\bibinfo {year} {1993})}\BibitemShut
  {NoStop}%
\bibitem [{\citenamefont {Hamilton}\ \emph {et~al.}(2021)\citenamefont
  {Hamilton}, \citenamefont {Medves}, \citenamefont {Salam}, \citenamefont
  {Scyboz},\ and\ \citenamefont {Soyez}}]{Hamilton:2020rcu}%
  \BibitemOpen
  \bibfield  {author} {\bibinfo {author} {\bibfnamefont {K.}~\bibnamefont
  {Hamilton}}, \bibinfo {author} {\bibfnamefont {R.}~\bibnamefont {Medves}},
  \bibinfo {author} {\bibfnamefont {G.~P.}\ \bibnamefont {Salam}}, \bibinfo
  {author} {\bibfnamefont {L.}~\bibnamefont {Scyboz}}, \ and\ \bibinfo {author}
  {\bibfnamefont {G.}~\bibnamefont {Soyez}},\ }\href {\doibase
  10.1007/JHEP03(2021)041} {\bibfield  {journal} {\bibinfo  {journal} {JHEP}\
  }\textbf {\bibinfo {volume} {03}},\ \bibinfo {pages} {041} (\bibinfo {year}
  {2021})},\ \Eprint {http://arxiv.org/abs/2011.10054} {arXiv:2011.10054
  [hep-ph]} \BibitemShut {NoStop}%
\bibitem [{\citenamefont {Dasgupta}\ \emph {et~al.}(2018)\citenamefont
  {Dasgupta}, \citenamefont {Dreyer}, \citenamefont {Hamilton}, \citenamefont
  {Monni},\ and\ \citenamefont {Salam}}]{Dasgupta:2018nvj}%
  \BibitemOpen
  \bibfield  {author} {\bibinfo {author} {\bibfnamefont {M.}~\bibnamefont
  {Dasgupta}}, \bibinfo {author} {\bibfnamefont {F.~A.}\ \bibnamefont
  {Dreyer}}, \bibinfo {author} {\bibfnamefont {K.}~\bibnamefont {Hamilton}},
  \bibinfo {author} {\bibfnamefont {P.~F.}\ \bibnamefont {Monni}}, \ and\
  \bibinfo {author} {\bibfnamefont {G.~P.}\ \bibnamefont {Salam}},\ }\href
  {\doibase 10.1007/JHEP09(2018)033} {\bibfield  {journal} {\bibinfo  {journal}
  {JHEP}\ }\textbf {\bibinfo {volume} {09}},\ \bibinfo {pages} {033} (\bibinfo
  {year} {2018})},\ \bibinfo {note} {[Erratum: JHEP 03, 083 (2020)]},\ \Eprint
  {http://arxiv.org/abs/1805.09327} {arXiv:1805.09327 [hep-ph]} \BibitemShut
  {NoStop}%
\bibitem [{\citenamefont {Dasgupta}\ \emph {et~al.}(2020)\citenamefont
  {Dasgupta}, \citenamefont {Dreyer}, \citenamefont {Hamilton}, \citenamefont
  {Monni}, \citenamefont {Salam},\ and\ \citenamefont
  {Soyez}}]{Dasgupta:2020fwr}%
  \BibitemOpen
  \bibfield  {author} {\bibinfo {author} {\bibfnamefont {M.}~\bibnamefont
  {Dasgupta}}, \bibinfo {author} {\bibfnamefont {F.~A.}\ \bibnamefont
  {Dreyer}}, \bibinfo {author} {\bibfnamefont {K.}~\bibnamefont {Hamilton}},
  \bibinfo {author} {\bibfnamefont {P.~F.}\ \bibnamefont {Monni}}, \bibinfo
  {author} {\bibfnamefont {G.~P.}\ \bibnamefont {Salam}}, \ and\ \bibinfo
  {author} {\bibfnamefont {G.}~\bibnamefont {Soyez}},\ }\href {\doibase
  10.1103/PhysRevLett.125.052002} {\bibfield  {journal} {\bibinfo  {journal}
  {Phys. Rev. Lett.}\ }\textbf {\bibinfo {volume} {125}},\ \bibinfo {pages}
  {052002} (\bibinfo {year} {2020})},\ \Eprint
  {http://arxiv.org/abs/2002.11114} {arXiv:2002.11114 [hep-ph]} \BibitemShut
  {NoStop}%
\bibitem [{\citenamefont {Hamilton}\ \emph {et~al.}(2022)\citenamefont
  {Hamilton}, \citenamefont {Karlberg}, \citenamefont {Salam}, \citenamefont
  {Scyboz},\ and\ \citenamefont {Verheyen}}]{Hamilton:2021dyz}%
  \BibitemOpen
  \bibfield  {author} {\bibinfo {author} {\bibfnamefont {K.}~\bibnamefont
  {Hamilton}}, \bibinfo {author} {\bibfnamefont {A.}~\bibnamefont {Karlberg}},
  \bibinfo {author} {\bibfnamefont {G.~P.}\ \bibnamefont {Salam}}, \bibinfo
  {author} {\bibfnamefont {L.}~\bibnamefont {Scyboz}}, \ and\ \bibinfo {author}
  {\bibfnamefont {R.}~\bibnamefont {Verheyen}},\ }\href {\doibase
  10.1007/JHEP03(2022)193} {\bibfield  {journal} {\bibinfo  {journal} {JHEP}\
  }\textbf {\bibinfo {volume} {03}},\ \bibinfo {pages} {193} (\bibinfo {year}
  {2022})},\ \Eprint {http://arxiv.org/abs/2111.01161} {arXiv:2111.01161
  [hep-ph]} \BibitemShut {NoStop}%
\bibitem [{\citenamefont {Herren}\ \emph {et~al.}(2023)\citenamefont {Herren},
  \citenamefont {H{\"o}che}, \citenamefont {Krauss}, \citenamefont {Reichelt},\
  and\ \citenamefont {Sch{\"o}nherr}}]{Herren:2022jej}%
  \BibitemOpen
  \bibfield  {author} {\bibinfo {author} {\bibfnamefont {F.}~\bibnamefont
  {Herren}}, \bibinfo {author} {\bibfnamefont {S.}~\bibnamefont {H{\"o}che}},
  \bibinfo {author} {\bibfnamefont {F.}~\bibnamefont {Krauss}}, \bibinfo
  {author} {\bibfnamefont {D.}~\bibnamefont {Reichelt}}, \ and\ \bibinfo
  {author} {\bibfnamefont {M.}~\bibnamefont {Sch{\"o}nherr}},\ }\href {\doibase
  10.1007/JHEP10(2023)091} {\bibfield  {journal} {\bibinfo  {journal} {JHEP}\
  }\textbf {\bibinfo {volume} {10}},\ \bibinfo {pages} {091} (\bibinfo {year}
  {2023})},\ \Eprint {http://arxiv.org/abs/2208.06057} {arXiv:2208.06057
  [hep-ph]} \BibitemShut {NoStop}%
\bibitem [{\citenamefont {Assi}\ and\ \citenamefont
  {H{\"o}che}(2024)}]{Assi:2023rbu}%
  \BibitemOpen
  \bibfield  {author} {\bibinfo {author} {\bibfnamefont {B.}~\bibnamefont
  {Assi}}\ and\ \bibinfo {author} {\bibfnamefont {S.}~\bibnamefont
  {H{\"o}che}},\ }\href {\doibase 10.1103/PhysRevD.109.114008} {\bibfield
  {journal} {\bibinfo  {journal} {Phys. Rev. D}\ }\textbf {\bibinfo {volume}
  {109}},\ \bibinfo {pages} {114008} (\bibinfo {year} {2024})},\ \Eprint
  {http://arxiv.org/abs/2307.00728} {arXiv:2307.00728 [hep-ph]} \BibitemShut
  {NoStop}%
\bibitem [{\citenamefont {H{\"o}che}\ \emph {et~al.}(2025)\citenamefont
  {H{\"o}che}, \citenamefont {Krauss},\ and\ \citenamefont
  {Reichelt}}]{Hoche:2024dee}%
  \BibitemOpen
  \bibfield  {author} {\bibinfo {author} {\bibfnamefont {S.}~\bibnamefont
  {H{\"o}che}}, \bibinfo {author} {\bibfnamefont {F.}~\bibnamefont {Krauss}}, \
  and\ \bibinfo {author} {\bibfnamefont {D.}~\bibnamefont {Reichelt}},\ }\href
  {\doibase 10.1103/PhysRevD.111.094032} {\bibfield  {journal} {\bibinfo
  {journal} {Phys. Rev. D}\ }\textbf {\bibinfo {volume} {111}},\ \bibinfo
  {pages} {094032} (\bibinfo {year} {2025})},\ \Eprint
  {http://arxiv.org/abs/2404.14360} {arXiv:2404.14360 [hep-ph]} \BibitemShut
  {NoStop}%
\bibitem [{\citenamefont {Preuss}(2024)}]{Preuss:2024vyu}%
  \BibitemOpen
  \bibfield  {author} {\bibinfo {author} {\bibfnamefont {C.~T.}\ \bibnamefont
  {Preuss}},\ }\href {\doibase 10.1007/JHEP07(2024)161} {\bibfield  {journal}
  {\bibinfo  {journal} {JHEP}\ }\textbf {\bibinfo {volume} {07}},\ \bibinfo
  {pages} {161} (\bibinfo {year} {2024})},\ \Eprint
  {http://arxiv.org/abs/2403.19452} {arXiv:2403.19452 [hep-ph]} \BibitemShut
  {NoStop}%
\bibitem [{\citenamefont {H{\"o}che}\ and\ \citenamefont
  {Prestel}(2017)}]{Hoche:2017iem}%
  \BibitemOpen
  \bibfield  {author} {\bibinfo {author} {\bibfnamefont {S.}~\bibnamefont
  {H{\"o}che}}\ and\ \bibinfo {author} {\bibfnamefont {S.}~\bibnamefont
  {Prestel}},\ }\href {\doibase 10.1103/PhysRevD.96.074017} {\bibfield
  {journal} {\bibinfo  {journal} {Phys. Rev. D}\ }\textbf {\bibinfo {volume}
  {96}},\ \bibinfo {pages} {074017} (\bibinfo {year} {2017})},\ \Eprint
  {http://arxiv.org/abs/1705.00742} {arXiv:1705.00742 [hep-ph]} \BibitemShut
  {NoStop}%
\bibitem [{\citenamefont {Dulat}\ \emph {et~al.}(2018)\citenamefont {Dulat},
  \citenamefont {H{\"o}che},\ and\ \citenamefont {Prestel}}]{Dulat:2018vuy}%
  \BibitemOpen
  \bibfield  {author} {\bibinfo {author} {\bibfnamefont {F.}~\bibnamefont
  {Dulat}}, \bibinfo {author} {\bibfnamefont {S.}~\bibnamefont {H{\"o}che}}, \
  and\ \bibinfo {author} {\bibfnamefont {S.}~\bibnamefont {Prestel}},\ }\href
  {\doibase 10.1103/PhysRevD.98.074013} {\bibfield  {journal} {\bibinfo
  {journal} {Phys. Rev. D}\ }\textbf {\bibinfo {volume} {98}},\ \bibinfo
  {pages} {074013} (\bibinfo {year} {2018})},\ \Eprint
  {http://arxiv.org/abs/1805.03757} {arXiv:1805.03757 [hep-ph]} \BibitemShut
  {NoStop}%
\bibitem [{\citenamefont {Gellersen}\ \emph {et~al.}(2022)\citenamefont
  {Gellersen}, \citenamefont {H{\"o}che},\ and\ \citenamefont
  {Prestel}}]{Gellersen:2021eci}%
  \BibitemOpen
  \bibfield  {author} {\bibinfo {author} {\bibfnamefont {L.}~\bibnamefont
  {Gellersen}}, \bibinfo {author} {\bibfnamefont {S.}~\bibnamefont
  {H{\"o}che}}, \ and\ \bibinfo {author} {\bibfnamefont {S.}~\bibnamefont
  {Prestel}},\ }\href {\doibase 10.1103/PhysRevD.105.114012} {\bibfield
  {journal} {\bibinfo  {journal} {Phys. Rev. D}\ }\textbf {\bibinfo {volume}
  {105}},\ \bibinfo {pages} {114012} (\bibinfo {year} {2022})},\ \Eprint
  {http://arxiv.org/abs/2110.05964} {arXiv:2110.05964 [hep-ph]} \BibitemShut
  {NoStop}%
\bibitem [{\citenamefont {Ferrario~Ravasio}\ \emph {et~al.}(2023)\citenamefont
  {Ferrario~Ravasio}, \citenamefont {Hamilton}, \citenamefont {Karlberg},
  \citenamefont {Salam}, \citenamefont {Scyboz},\ and\ \citenamefont
  {Soyez}}]{FerrarioRavasio:2023kyg}%
  \BibitemOpen
  \bibfield  {author} {\bibinfo {author} {\bibfnamefont {S.}~\bibnamefont
  {Ferrario~Ravasio}}, \bibinfo {author} {\bibfnamefont {K.}~\bibnamefont
  {Hamilton}}, \bibinfo {author} {\bibfnamefont {A.}~\bibnamefont {Karlberg}},
  \bibinfo {author} {\bibfnamefont {G.~P.}\ \bibnamefont {Salam}}, \bibinfo
  {author} {\bibfnamefont {L.}~\bibnamefont {Scyboz}}, \ and\ \bibinfo {author}
  {\bibfnamefont {G.}~\bibnamefont {Soyez}},\ }\href {\doibase
  10.1103/PhysRevLett.131.161906} {\bibfield  {journal} {\bibinfo  {journal}
  {Phys. Rev. Lett.}\ }\textbf {\bibinfo {volume} {131}},\ \bibinfo {pages}
  {161906} (\bibinfo {year} {2023})},\ \Eprint
  {http://arxiv.org/abs/2307.11142} {arXiv:2307.11142 [hep-ph]} \BibitemShut
  {NoStop}%
\bibitem [{\citenamefont {van Beekveld}\ \emph {et~al.}(2025)\citenamefont {van
  Beekveld}, \citenamefont {Dasgupta}, \citenamefont {El-Menoufi},
  \citenamefont {Helliwell}, \citenamefont {Monni},\ and\ \citenamefont
  {Salam}}]{vanBeekveld:2024qxs}%
  \BibitemOpen
  \bibfield  {author} {\bibinfo {author} {\bibfnamefont {M.}~\bibnamefont {van
  Beekveld}}, \bibinfo {author} {\bibfnamefont {M.}~\bibnamefont {Dasgupta}},
  \bibinfo {author} {\bibfnamefont {B.~K.}\ \bibnamefont {El-Menoufi}},
  \bibinfo {author} {\bibfnamefont {J.}~\bibnamefont {Helliwell}}, \bibinfo
  {author} {\bibfnamefont {P.~F.}\ \bibnamefont {Monni}}, \ and\ \bibinfo
  {author} {\bibfnamefont {G.~P.}\ \bibnamefont {Salam}},\ }\href {\doibase
  10.1007/JHEP03(2025)209} {\bibfield  {journal} {\bibinfo  {journal} {JHEP}\
  }\textbf {\bibinfo {volume} {03}},\ \bibinfo {pages} {209} (\bibinfo {year}
  {2025})},\ \Eprint {http://arxiv.org/abs/2409.08316} {arXiv:2409.08316
  [hep-ph]} \BibitemShut {NoStop}%
\bibitem [{\citenamefont {Frixione}\ and\ \citenamefont
  {Webber}(2002)}]{Frixione:2002ik}%
  \BibitemOpen
  \bibfield  {author} {\bibinfo {author} {\bibfnamefont {S.}~\bibnamefont
  {Frixione}}\ and\ \bibinfo {author} {\bibfnamefont {B.~R.}\ \bibnamefont
  {Webber}},\ }\href {\doibase 10.1088/1126-6708/2002/06/029} {\bibfield
  {journal} {\bibinfo  {journal} {JHEP}\ }\textbf {\bibinfo {volume} {06}},\
  \bibinfo {pages} {029} (\bibinfo {year} {2002})},\ \Eprint
  {http://arxiv.org/abs/hep-ph/0204244} {arXiv:hep-ph/0204244} \BibitemShut
  {NoStop}%
\bibitem [{\citenamefont {Nason}(2004)}]{Nason:2004rx}%
  \BibitemOpen
  \bibfield  {author} {\bibinfo {author} {\bibfnamefont {P.}~\bibnamefont
  {Nason}},\ }\href {\doibase 10.1088/1126-6708/2004/11/040} {\bibfield
  {journal} {\bibinfo  {journal} {JHEP}\ }\textbf {\bibinfo {volume} {11}},\
  \bibinfo {pages} {040} (\bibinfo {year} {2004})},\ \Eprint
  {http://arxiv.org/abs/hep-ph/0409146} {arXiv:hep-ph/0409146} \BibitemShut
  {NoStop}%
\bibitem [{\citenamefont {Frixione}\ \emph {et~al.}(2007)\citenamefont
  {Frixione}, \citenamefont {Nason},\ and\ \citenamefont
  {Oleari}}]{Frixione:2007vw}%
  \BibitemOpen
  \bibfield  {author} {\bibinfo {author} {\bibfnamefont {S.}~\bibnamefont
  {Frixione}}, \bibinfo {author} {\bibfnamefont {P.}~\bibnamefont {Nason}}, \
  and\ \bibinfo {author} {\bibfnamefont {C.}~\bibnamefont {Oleari}},\ }\href
  {\doibase 10.1088/1126-6708/2007/11/070} {\bibfield  {journal} {\bibinfo
  {journal} {JHEP}\ }\textbf {\bibinfo {volume} {11}},\ \bibinfo {pages} {070}
  (\bibinfo {year} {2007})},\ \Eprint {http://arxiv.org/abs/0709.2092}
  {arXiv:0709.2092 [hep-ph]} \BibitemShut {NoStop}%
\bibitem [{\citenamefont {Alioli}\ \emph {et~al.}(2010)\citenamefont {Alioli},
  \citenamefont {Nason}, \citenamefont {Oleari},\ and\ \citenamefont
  {Re}}]{Alioli:2010xd}%
  \BibitemOpen
  \bibfield  {author} {\bibinfo {author} {\bibfnamefont {S.}~\bibnamefont
  {Alioli}}, \bibinfo {author} {\bibfnamefont {P.}~\bibnamefont {Nason}},
  \bibinfo {author} {\bibfnamefont {C.}~\bibnamefont {Oleari}}, \ and\ \bibinfo
  {author} {\bibfnamefont {E.}~\bibnamefont {Re}},\ }\href {\doibase
  10.1007/JHEP06(2010)043} {\bibfield  {journal} {\bibinfo  {journal} {JHEP}\
  }\textbf {\bibinfo {volume} {06}},\ \bibinfo {pages} {043} (\bibinfo {year}
  {2010})},\ \Eprint {http://arxiv.org/abs/1002.2581} {arXiv:1002.2581
  [hep-ph]} \BibitemShut {NoStop}%
\bibitem [{\citenamefont {H{\"o}che}\ \emph {et~al.}(2011)\citenamefont
  {H{\"o}che}, \citenamefont {Krauss}, \citenamefont {Sch{\"o}nherr},\ and\
  \citenamefont {Siegert}}]{Hoche:2010pf}%
  \BibitemOpen
  \bibfield  {author} {\bibinfo {author} {\bibfnamefont {S.}~\bibnamefont
  {H{\"o}che}}, \bibinfo {author} {\bibfnamefont {F.}~\bibnamefont {Krauss}},
  \bibinfo {author} {\bibfnamefont {M.}~\bibnamefont {Sch{\"o}nherr}}, \ and\
  \bibinfo {author} {\bibfnamefont {F.}~\bibnamefont {Siegert}},\ }\href
  {\doibase 10.1007/JHEP04(2011)024} {\bibfield  {journal} {\bibinfo  {journal}
  {JHEP}\ }\textbf {\bibinfo {volume} {04}},\ \bibinfo {pages} {024} (\bibinfo
  {year} {2011})},\ \Eprint {http://arxiv.org/abs/1008.5399} {arXiv:1008.5399
  [hep-ph]} \BibitemShut {NoStop}%
\bibitem [{\citenamefont {H{\"o}che}\ \emph {et~al.}(2012)\citenamefont
  {H{\"o}che}, \citenamefont {Krauss}, \citenamefont {Sch{\"o}nherr},\ and\
  \citenamefont {Siegert}}]{Hoeche:2011fd}%
  \BibitemOpen
  \bibfield  {author} {\bibinfo {author} {\bibfnamefont {S.}~\bibnamefont
  {H{\"o}che}}, \bibinfo {author} {\bibfnamefont {F.}~\bibnamefont {Krauss}},
  \bibinfo {author} {\bibfnamefont {M.}~\bibnamefont {Sch{\"o}nherr}}, \ and\
  \bibinfo {author} {\bibfnamefont {F.}~\bibnamefont {Siegert}},\ }\href
  {\doibase 10.1007/JHEP09(2012)049} {\bibfield  {journal} {\bibinfo  {journal}
  {JHEP}\ }\textbf {\bibinfo {volume} {09}},\ \bibinfo {pages} {049} (\bibinfo
  {year} {2012})},\ \Eprint {http://arxiv.org/abs/1111.1220} {arXiv:1111.1220
  [hep-ph]} \BibitemShut {NoStop}%
\bibitem [{\citenamefont {Alwall}\ \emph {et~al.}(2014)\citenamefont {Alwall},
  \citenamefont {Frederix}, \citenamefont {Frixione}, \citenamefont {Hirschi},
  \citenamefont {Maltoni}, \citenamefont {Mattelaer}, \citenamefont {Shao},
  \citenamefont {Stelzer}, \citenamefont {Torrielli},\ and\ \citenamefont
  {Zaro}}]{Alwall:2014hca}%
  \BibitemOpen
  \bibfield  {author} {\bibinfo {author} {\bibfnamefont {J.}~\bibnamefont
  {Alwall}}, \bibinfo {author} {\bibfnamefont {R.}~\bibnamefont {Frederix}},
  \bibinfo {author} {\bibfnamefont {S.}~\bibnamefont {Frixione}}, \bibinfo
  {author} {\bibfnamefont {V.}~\bibnamefont {Hirschi}}, \bibinfo {author}
  {\bibfnamefont {F.}~\bibnamefont {Maltoni}}, \bibinfo {author} {\bibfnamefont
  {O.}~\bibnamefont {Mattelaer}}, \bibinfo {author} {\bibfnamefont {H.~S.}\
  \bibnamefont {Shao}}, \bibinfo {author} {\bibfnamefont {T.}~\bibnamefont
  {Stelzer}}, \bibinfo {author} {\bibfnamefont {P.}~\bibnamefont {Torrielli}},
  \ and\ \bibinfo {author} {\bibfnamefont {M.}~\bibnamefont {Zaro}},\ }\href
  {\doibase 10.1007/JHEP07(2014)079} {\bibfield  {journal} {\bibinfo  {journal}
  {JHEP}\ }\textbf {\bibinfo {volume} {07}},\ \bibinfo {pages} {079} (\bibinfo
  {year} {2014})},\ \Eprint {http://arxiv.org/abs/1405.0301} {arXiv:1405.0301
  [hep-ph]} \BibitemShut {NoStop}%
\bibitem [{\citenamefont {Andre}\ and\ \citenamefont
  {Sj{\"o}strand}(1998)}]{Andre:1997vh}%
  \BibitemOpen
  \bibfield  {author} {\bibinfo {author} {\bibfnamefont {J.}~\bibnamefont
  {Andre}}\ and\ \bibinfo {author} {\bibfnamefont {T.}~\bibnamefont
  {Sj{\"o}strand}},\ }\href {\doibase 10.1103/PhysRevD.57.5767} {\bibfield
  {journal} {\bibinfo  {journal} {Phys. Rev. D}\ }\textbf {\bibinfo {volume}
  {57}},\ \bibinfo {pages} {5767} (\bibinfo {year} {1998})},\ \Eprint
  {http://arxiv.org/abs/hep-ph/9708390} {arXiv:hep-ph/9708390} \BibitemShut
  {NoStop}%
\bibitem [{\citenamefont {Catani}\ \emph {et~al.}(2001)\citenamefont {Catani},
  \citenamefont {Krauss}, \citenamefont {Kuhn},\ and\ \citenamefont
  {Webber}}]{Catani:2001cc}%
  \BibitemOpen
  \bibfield  {author} {\bibinfo {author} {\bibfnamefont {S.}~\bibnamefont
  {Catani}}, \bibinfo {author} {\bibfnamefont {F.}~\bibnamefont {Krauss}},
  \bibinfo {author} {\bibfnamefont {R.}~\bibnamefont {Kuhn}}, \ and\ \bibinfo
  {author} {\bibfnamefont {B.~R.}\ \bibnamefont {Webber}},\ }\href {\doibase
  10.1088/1126-6708/2001/11/063} {\bibfield  {journal} {\bibinfo  {journal}
  {JHEP}\ }\textbf {\bibinfo {volume} {11}},\ \bibinfo {pages} {063} (\bibinfo
  {year} {2001})},\ \Eprint {http://arxiv.org/abs/hep-ph/0109231}
  {arXiv:hep-ph/0109231} \BibitemShut {NoStop}%
\bibitem [{\citenamefont {L{\"o}nnblad}(2002)}]{Lonnblad:2001iq}%
  \BibitemOpen
  \bibfield  {author} {\bibinfo {author} {\bibfnamefont {L.}~\bibnamefont
  {L{\"o}nnblad}},\ }\href {\doibase 10.1088/1126-6708/2002/05/046} {\bibfield
  {journal} {\bibinfo  {journal} {JHEP}\ }\textbf {\bibinfo {volume} {05}},\
  \bibinfo {pages} {046} (\bibinfo {year} {2002})},\ \Eprint
  {http://arxiv.org/abs/hep-ph/0112284} {arXiv:hep-ph/0112284} \BibitemShut
  {NoStop}%
\bibitem [{\citenamefont {Mangano}\ \emph {et~al.}(2002)\citenamefont
  {Mangano}, \citenamefont {Moretti},\ and\ \citenamefont
  {Pittau}}]{Mangano:2001xp}%
  \BibitemOpen
  \bibfield  {author} {\bibinfo {author} {\bibfnamefont {M.~L.}\ \bibnamefont
  {Mangano}}, \bibinfo {author} {\bibfnamefont {M.}~\bibnamefont {Moretti}}, \
  and\ \bibinfo {author} {\bibfnamefont {R.}~\bibnamefont {Pittau}},\ }\href
  {\doibase 10.1016/S0550-3213(02)00249-3} {\bibfield  {journal} {\bibinfo
  {journal} {Nucl. Phys. B}\ }\textbf {\bibinfo {volume} {632}},\ \bibinfo
  {pages} {343} (\bibinfo {year} {2002})},\ \Eprint
  {http://arxiv.org/abs/hep-ph/0108069} {arXiv:hep-ph/0108069} \BibitemShut
  {NoStop}%
\bibitem [{\citenamefont {Krauss}(2002)}]{Krauss:2002up}%
  \BibitemOpen
  \bibfield  {author} {\bibinfo {author} {\bibfnamefont {F.}~\bibnamefont
  {Krauss}},\ }\href {\doibase 10.1088/1126-6708/2002/08/015} {\bibfield
  {journal} {\bibinfo  {journal} {JHEP}\ }\textbf {\bibinfo {volume} {08}},\
  \bibinfo {pages} {015} (\bibinfo {year} {2002})},\ \Eprint
  {http://arxiv.org/abs/hep-ph/0205283} {arXiv:hep-ph/0205283} \BibitemShut
  {NoStop}%
\bibitem [{\citenamefont {Lavesson}\ and\ \citenamefont
  {L{\"o}nnblad}(2008{\natexlab{a}})}]{Lavesson:2007uu}%
  \BibitemOpen
  \bibfield  {author} {\bibinfo {author} {\bibfnamefont {N.}~\bibnamefont
  {Lavesson}}\ and\ \bibinfo {author} {\bibfnamefont {L.}~\bibnamefont
  {L{\"o}nnblad}},\ }\href {\doibase 10.1088/1126-6708/2008/04/085} {\bibfield
  {journal} {\bibinfo  {journal} {JHEP}\ }\textbf {\bibinfo {volume} {04}},\
  \bibinfo {pages} {085} (\bibinfo {year} {2008}{\natexlab{a}})},\ \Eprint
  {http://arxiv.org/abs/0712.2966} {arXiv:0712.2966 [hep-ph]} \BibitemShut
  {NoStop}%
\bibitem [{\citenamefont {H{\"o}che}\ \emph {et~al.}(2009)\citenamefont
  {H{\"o}che}, \citenamefont {Krauss}, \citenamefont {Schumann},\ and\
  \citenamefont {Siegert}}]{Hoeche:2009rj}%
  \BibitemOpen
  \bibfield  {author} {\bibinfo {author} {\bibfnamefont {S.}~\bibnamefont
  {H{\"o}che}}, \bibinfo {author} {\bibfnamefont {F.}~\bibnamefont {Krauss}},
  \bibinfo {author} {\bibfnamefont {S.}~\bibnamefont {Schumann}}, \ and\
  \bibinfo {author} {\bibfnamefont {F.}~\bibnamefont {Siegert}},\ }\href
  {\doibase 10.1088/1126-6708/2009/05/053} {\bibfield  {journal} {\bibinfo
  {journal} {JHEP}\ }\textbf {\bibinfo {volume} {05}},\ \bibinfo {pages} {053}
  (\bibinfo {year} {2009})},\ \Eprint {http://arxiv.org/abs/0903.1219}
  {arXiv:0903.1219 [hep-ph]} \BibitemShut {NoStop}%
\bibitem [{\citenamefont {Hamilton}\ \emph {et~al.}(2009)\citenamefont
  {Hamilton}, \citenamefont {Richardson},\ and\ \citenamefont
  {Tully}}]{Hamilton:2009ne}%
  \BibitemOpen
  \bibfield  {author} {\bibinfo {author} {\bibfnamefont {K.}~\bibnamefont
  {Hamilton}}, \bibinfo {author} {\bibfnamefont {P.}~\bibnamefont
  {Richardson}}, \ and\ \bibinfo {author} {\bibfnamefont {J.}~\bibnamefont
  {Tully}},\ }\href {\doibase 10.1088/1126-6708/2009/11/038} {\bibfield
  {journal} {\bibinfo  {journal} {JHEP}\ }\textbf {\bibinfo {volume} {11}},\
  \bibinfo {pages} {038} (\bibinfo {year} {2009})},\ \Eprint
  {http://arxiv.org/abs/0905.3072} {arXiv:0905.3072 [hep-ph]} \BibitemShut
  {NoStop}%
\bibitem [{\citenamefont {L{\"o}nnblad}\ and\ \citenamefont
  {Prestel}(2012)}]{Lonnblad:2011xx}%
  \BibitemOpen
  \bibfield  {author} {\bibinfo {author} {\bibfnamefont {L.}~\bibnamefont
  {L{\"o}nnblad}}\ and\ \bibinfo {author} {\bibfnamefont {S.}~\bibnamefont
  {Prestel}},\ }\href {\doibase 10.1007/JHEP03(2012)019} {\bibfield  {journal}
  {\bibinfo  {journal} {JHEP}\ }\textbf {\bibinfo {volume} {03}},\ \bibinfo
  {pages} {019} (\bibinfo {year} {2012})},\ \Eprint
  {http://arxiv.org/abs/1109.4829} {arXiv:1109.4829 [hep-ph]} \BibitemShut
  {NoStop}%
\bibitem [{\citenamefont {L{\"o}nnblad}\ and\ \citenamefont
  {Prestel}(2013{\natexlab{a}})}]{Lonnblad:2012ng}%
  \BibitemOpen
  \bibfield  {author} {\bibinfo {author} {\bibfnamefont {L.}~\bibnamefont
  {L{\"o}nnblad}}\ and\ \bibinfo {author} {\bibfnamefont {S.}~\bibnamefont
  {Prestel}},\ }\href {\doibase 10.1007/JHEP02(2013)094} {\bibfield  {journal}
  {\bibinfo  {journal} {JHEP}\ }\textbf {\bibinfo {volume} {02}},\ \bibinfo
  {pages} {094} (\bibinfo {year} {2013}{\natexlab{a}})},\ \Eprint
  {http://arxiv.org/abs/1211.4827} {arXiv:1211.4827 [hep-ph]} \BibitemShut
  {NoStop}%
\bibitem [{\citenamefont {Pl{\"a}tzer}(2013)}]{Platzer:2012bs}%
  \BibitemOpen
  \bibfield  {author} {\bibinfo {author} {\bibfnamefont {S.}~\bibnamefont
  {Pl{\"a}tzer}},\ }\href {\doibase 10.1007/JHEP08(2013)114} {\bibfield
  {journal} {\bibinfo  {journal} {JHEP}\ }\textbf {\bibinfo {volume} {08}},\
  \bibinfo {pages} {114} (\bibinfo {year} {2013})},\ \Eprint
  {http://arxiv.org/abs/1211.5467} {arXiv:1211.5467 [hep-ph]} \BibitemShut
  {NoStop}%
\bibitem [{\citenamefont {H{\"o}che}\ \emph
  {et~al.}(2019{\natexlab{a}})\citenamefont {H{\"o}che}, \citenamefont
  {Krause},\ and\ \citenamefont {Siegert}}]{Hoche:2019ncc}%
  \BibitemOpen
  \bibfield  {author} {\bibinfo {author} {\bibfnamefont {S.}~\bibnamefont
  {H{\"o}che}}, \bibinfo {author} {\bibfnamefont {J.}~\bibnamefont {Krause}}, \
  and\ \bibinfo {author} {\bibfnamefont {F.}~\bibnamefont {Siegert}},\ }\href
  {\doibase 10.1103/PhysRevD.100.014011} {\bibfield  {journal} {\bibinfo
  {journal} {Phys. Rev. D}\ }\textbf {\bibinfo {volume} {100}},\ \bibinfo
  {pages} {014011} (\bibinfo {year} {2019}{\natexlab{a}})},\ \Eprint
  {http://arxiv.org/abs/1904.09382} {arXiv:1904.09382 [hep-ph]} \BibitemShut
  {NoStop}%
\bibitem [{\citenamefont {Lavesson}\ and\ \citenamefont
  {L{\"o}nnblad}(2008{\natexlab{b}})}]{Lavesson:2008ah}%
  \BibitemOpen
  \bibfield  {author} {\bibinfo {author} {\bibfnamefont {N.}~\bibnamefont
  {Lavesson}}\ and\ \bibinfo {author} {\bibfnamefont {L.}~\bibnamefont
  {L{\"o}nnblad}},\ }\href {\doibase 10.1088/1126-6708/2008/12/070} {\bibfield
  {journal} {\bibinfo  {journal} {JHEP}\ }\textbf {\bibinfo {volume} {12}},\
  \bibinfo {pages} {070} (\bibinfo {year} {2008}{\natexlab{b}})},\ \Eprint
  {http://arxiv.org/abs/0811.2912} {arXiv:0811.2912 [hep-ph]} \BibitemShut
  {NoStop}%
\bibitem [{\citenamefont {Gehrmann}\ \emph {et~al.}(2013)\citenamefont
  {Gehrmann}, \citenamefont {H{\"o}che}, \citenamefont {Krauss}, \citenamefont
  {Sch{\"o}nherr},\ and\ \citenamefont {Siegert}}]{Gehrmann:2012yg}%
  \BibitemOpen
  \bibfield  {author} {\bibinfo {author} {\bibfnamefont {T.}~\bibnamefont
  {Gehrmann}}, \bibinfo {author} {\bibfnamefont {S.}~\bibnamefont {H{\"o}che}},
  \bibinfo {author} {\bibfnamefont {F.}~\bibnamefont {Krauss}}, \bibinfo
  {author} {\bibfnamefont {M.}~\bibnamefont {Sch{\"o}nherr}}, \ and\ \bibinfo
  {author} {\bibfnamefont {F.}~\bibnamefont {Siegert}},\ }\href {\doibase
  10.1007/JHEP01(2013)144} {\bibfield  {journal} {\bibinfo  {journal} {JHEP}\
  }\textbf {\bibinfo {volume} {01}},\ \bibinfo {pages} {144} (\bibinfo {year}
  {2013})},\ \Eprint {http://arxiv.org/abs/1207.5031} {arXiv:1207.5031
  [hep-ph]} \BibitemShut {NoStop}%
\bibitem [{\citenamefont {H{\"o}che}\ \emph {et~al.}(2013)\citenamefont
  {H{\"o}che}, \citenamefont {Krauss}, \citenamefont {Sch{\"o}nherr},\ and\
  \citenamefont {Siegert}}]{Hoeche:2012yf}%
  \BibitemOpen
  \bibfield  {author} {\bibinfo {author} {\bibfnamefont {S.}~\bibnamefont
  {H{\"o}che}}, \bibinfo {author} {\bibfnamefont {F.}~\bibnamefont {Krauss}},
  \bibinfo {author} {\bibfnamefont {M.}~\bibnamefont {Sch{\"o}nherr}}, \ and\
  \bibinfo {author} {\bibfnamefont {F.}~\bibnamefont {Siegert}},\ }\href
  {\doibase 10.1007/JHEP04(2013)027} {\bibfield  {journal} {\bibinfo  {journal}
  {JHEP}\ }\textbf {\bibinfo {volume} {04}},\ \bibinfo {pages} {027} (\bibinfo
  {year} {2013})},\ \Eprint {http://arxiv.org/abs/1207.5030} {arXiv:1207.5030
  [hep-ph]} \BibitemShut {NoStop}%
\bibitem [{\citenamefont {Frederix}\ and\ \citenamefont
  {Frixione}(2012)}]{Frederix:2012ps}%
  \BibitemOpen
  \bibfield  {author} {\bibinfo {author} {\bibfnamefont {R.}~\bibnamefont
  {Frederix}}\ and\ \bibinfo {author} {\bibfnamefont {S.}~\bibnamefont
  {Frixione}},\ }\href {\doibase 10.1007/JHEP12(2012)061} {\bibfield  {journal}
  {\bibinfo  {journal} {JHEP}\ }\textbf {\bibinfo {volume} {12}},\ \bibinfo
  {pages} {061} (\bibinfo {year} {2012})},\ \Eprint
  {http://arxiv.org/abs/1209.6215} {arXiv:1209.6215 [hep-ph]} \BibitemShut
  {NoStop}%
\bibitem [{\citenamefont {L{\"o}nnblad}\ and\ \citenamefont
  {Prestel}(2013{\natexlab{b}})}]{Lonnblad:2012ix}%
  \BibitemOpen
  \bibfield  {author} {\bibinfo {author} {\bibfnamefont {L.}~\bibnamefont
  {L{\"o}nnblad}}\ and\ \bibinfo {author} {\bibfnamefont {S.}~\bibnamefont
  {Prestel}},\ }\href {\doibase 10.1007/JHEP03(2013)166} {\bibfield  {journal}
  {\bibinfo  {journal} {JHEP}\ }\textbf {\bibinfo {volume} {03}},\ \bibinfo
  {pages} {166} (\bibinfo {year} {2013}{\natexlab{b}})},\ \Eprint
  {http://arxiv.org/abs/1211.7278} {arXiv:1211.7278 [hep-ph]} \BibitemShut
  {NoStop}%
\bibitem [{\citenamefont {Catani}\ and\ \citenamefont
  {Seymour}(1997)}]{Catani:1996vz}%
  \BibitemOpen
  \bibfield  {author} {\bibinfo {author} {\bibfnamefont {S.}~\bibnamefont
  {Catani}}\ and\ \bibinfo {author} {\bibfnamefont {M.~H.}\ \bibnamefont
  {Seymour}},\ }\href {\doibase 10.1016/S0550-3213(96)00589-5} {\bibfield
  {journal} {\bibinfo  {journal} {Nucl. Phys. B}\ }\textbf {\bibinfo {volume}
  {485}},\ \bibinfo {pages} {291} (\bibinfo {year} {1997})},\ \bibinfo {note}
  {[Erratum: Nucl.Phys.B 510, 503--504 (1998)]},\ \Eprint
  {http://arxiv.org/abs/hep-ph/9605323} {arXiv:hep-ph/9605323} \BibitemShut
  {NoStop}%
\bibitem [{\citenamefont {Catani}\ \emph {et~al.}(2002)\citenamefont {Catani},
  \citenamefont {Dittmaier}, \citenamefont {Seymour},\ and\ \citenamefont
  {Trocsanyi}}]{Catani:2002hc}%
  \BibitemOpen
  \bibfield  {author} {\bibinfo {author} {\bibfnamefont {S.}~\bibnamefont
  {Catani}}, \bibinfo {author} {\bibfnamefont {S.}~\bibnamefont {Dittmaier}},
  \bibinfo {author} {\bibfnamefont {M.~H.}\ \bibnamefont {Seymour}}, \ and\
  \bibinfo {author} {\bibfnamefont {Z.}~\bibnamefont {Trocsanyi}},\ }\href
  {\doibase 10.1016/S0550-3213(02)00098-6} {\bibfield  {journal} {\bibinfo
  {journal} {Nucl. Phys. B}\ }\textbf {\bibinfo {volume} {627}},\ \bibinfo
  {pages} {189} (\bibinfo {year} {2002})},\ \Eprint
  {http://arxiv.org/abs/hep-ph/0201036} {arXiv:hep-ph/0201036} \BibitemShut
  {NoStop}%
\bibitem [{\citenamefont {Bassetto}\ \emph {et~al.}(1983)\citenamefont
  {Bassetto}, \citenamefont {Ciafaloni},\ and\ \citenamefont
  {Marchesini}}]{Bassetto:1983mvz}%
  \BibitemOpen
  \bibfield  {author} {\bibinfo {author} {\bibfnamefont {A.}~\bibnamefont
  {Bassetto}}, \bibinfo {author} {\bibfnamefont {M.}~\bibnamefont {Ciafaloni}},
  \ and\ \bibinfo {author} {\bibfnamefont {G.}~\bibnamefont {Marchesini}},\
  }\href {\doibase 10.1016/0370-1573(83)90083-2} {\bibfield  {journal}
  {\bibinfo  {journal} {Phys. Rept.}\ }\textbf {\bibinfo {volume} {100}},\
  \bibinfo {pages} {201} (\bibinfo {year} {1983})}\BibitemShut {NoStop}%
\bibitem [{\citenamefont {H{\"o}che}\ and\ \citenamefont
  {Sch{\"o}nherr}(2012)}]{Hoeche:2012fm}%
  \BibitemOpen
  \bibfield  {author} {\bibinfo {author} {\bibfnamefont {S.}~\bibnamefont
  {H{\"o}che}}\ and\ \bibinfo {author} {\bibfnamefont {M.}~\bibnamefont
  {Sch{\"o}nherr}},\ }\href {\doibase 10.1103/PhysRevD.86.094042} {\bibfield
  {journal} {\bibinfo  {journal} {Phys. Rev. D}\ }\textbf {\bibinfo {volume}
  {86}},\ \bibinfo {pages} {094042} (\bibinfo {year} {2012})},\ \Eprint
  {http://arxiv.org/abs/1208.2815} {arXiv:1208.2815 [hep-ph]} \BibitemShut
  {NoStop}%
\bibitem [{\citenamefont {Campbell}\ \emph {et~al.}(2025)\citenamefont
  {Campbell}, \citenamefont {H{\"o}che}, \citenamefont {Knobbe}, \citenamefont
  {Preuss},\ and\ \citenamefont {Reichelt}}]{Campbell:2025lrs}%
  \BibitemOpen
  \bibfield  {author} {\bibinfo {author} {\bibfnamefont {J.~M.}\ \bibnamefont
  {Campbell}}, \bibinfo {author} {\bibfnamefont {S.}~\bibnamefont {H{\"o}che}},
  \bibinfo {author} {\bibfnamefont {M.}~\bibnamefont {Knobbe}}, \bibinfo
  {author} {\bibfnamefont {C.~T.}\ \bibnamefont {Preuss}}, \ and\ \bibinfo
  {author} {\bibfnamefont {D.}~\bibnamefont {Reichelt}},\ }\href@noop {} {\
  (\bibinfo {year} {2025})},\ \Eprint {http://arxiv.org/abs/2505.10408}
  {arXiv:2505.10408 [hep-ph]} \BibitemShut {NoStop}%
\bibitem [{\citenamefont {Gell-Mann}\ and\ \citenamefont
  {Goldberger}(1954)}]{Gell-Mann:1954wra}%
  \BibitemOpen
  \bibfield  {author} {\bibinfo {author} {\bibfnamefont {M.}~\bibnamefont
  {Gell-Mann}}\ and\ \bibinfo {author} {\bibfnamefont {M.~L.}\ \bibnamefont
  {Goldberger}},\ }\href {\doibase 10.1103/PhysRev.96.1433} {\bibfield
  {journal} {\bibinfo  {journal} {Phys. Rev.}\ }\textbf {\bibinfo {volume}
  {96}},\ \bibinfo {pages} {1433} (\bibinfo {year} {1954})}\BibitemShut
  {NoStop}%
\bibitem [{\citenamefont {Brown}\ and\ \citenamefont
  {Goble}(1968)}]{Brown:1968dzy}%
  \BibitemOpen
  \bibfield  {author} {\bibinfo {author} {\bibfnamefont {L.~S.}\ \bibnamefont
  {Brown}}\ and\ \bibinfo {author} {\bibfnamefont {R.~L.}\ \bibnamefont
  {Goble}},\ }\href {\doibase 10.1103/PhysRev.173.1505} {\bibfield  {journal}
  {\bibinfo  {journal} {Phys. Rev.}\ }\textbf {\bibinfo {volume} {173}},\
  \bibinfo {pages} {1505} (\bibinfo {year} {1968})}\BibitemShut {NoStop}%
\bibitem [{\citenamefont {Gleisberg}\ and\ \citenamefont
  {Krauss}(2008)}]{Gleisberg:2007md}%
  \BibitemOpen
  \bibfield  {author} {\bibinfo {author} {\bibfnamefont {T.}~\bibnamefont
  {Gleisberg}}\ and\ \bibinfo {author} {\bibfnamefont {F.}~\bibnamefont
  {Krauss}},\ }\href {\doibase 10.1140/epjc/s10052-007-0495-0} {\bibfield
  {journal} {\bibinfo  {journal} {Eur. Phys. J. C}\ }\textbf {\bibinfo {volume}
  {53}},\ \bibinfo {pages} {501} (\bibinfo {year} {2008})},\ \Eprint
  {http://arxiv.org/abs/0709.2881} {arXiv:0709.2881 [hep-ph]} \BibitemShut
  {NoStop}%
\bibitem [{\citenamefont {H{\"o}che}\ \emph
  {et~al.}(2019{\natexlab{b}})\citenamefont {H{\"o}che}, \citenamefont
  {Liebschner},\ and\ \citenamefont {Siegert}}]{Hoche:2018ouj}%
  \BibitemOpen
  \bibfield  {author} {\bibinfo {author} {\bibfnamefont {S.}~\bibnamefont
  {H{\"o}che}}, \bibinfo {author} {\bibfnamefont {S.}~\bibnamefont
  {Liebschner}}, \ and\ \bibinfo {author} {\bibfnamefont {F.}~\bibnamefont
  {Siegert}},\ }\href {\doibase 10.1140/epjc/s10052-019-7212-7} {\bibfield
  {journal} {\bibinfo  {journal} {Eur. Phys. J. C}\ }\textbf {\bibinfo {volume}
  {79}},\ \bibinfo {pages} {728} (\bibinfo {year} {2019}{\natexlab{b}})},\
  \Eprint {http://arxiv.org/abs/1807.04348} {arXiv:1807.04348 [hep-ph]}
  \BibitemShut {NoStop}%
\bibitem [{\citenamefont {Catani}\ \emph
  {et~al.}(1991{\natexlab{a}})\citenamefont {Catani}, \citenamefont
  {Dokshitzer}, \citenamefont {Olsson}, \citenamefont {Turnock},\ and\
  \citenamefont {Webber}}]{Catani:1991hj}%
  \BibitemOpen
  \bibfield  {author} {\bibinfo {author} {\bibfnamefont {S.}~\bibnamefont
  {Catani}}, \bibinfo {author} {\bibfnamefont {Y.~L.}\ \bibnamefont
  {Dokshitzer}}, \bibinfo {author} {\bibfnamefont {M.}~\bibnamefont {Olsson}},
  \bibinfo {author} {\bibfnamefont {G.}~\bibnamefont {Turnock}}, \ and\
  \bibinfo {author} {\bibfnamefont {B.~R.}\ \bibnamefont {Webber}},\ }\href
  {\doibase 10.1016/0370-2693(91)90196-W} {\bibfield  {journal} {\bibinfo
  {journal} {Phys. Lett. B}\ }\textbf {\bibinfo {volume} {269}},\ \bibinfo
  {pages} {432} (\bibinfo {year} {1991}{\natexlab{a}})}\BibitemShut {NoStop}%
\bibitem [{\citenamefont {Coloretti}\ \emph {et~al.}(2022)\citenamefont
  {Coloretti}, \citenamefont {Gehrmann-De~Ridder},\ and\ \citenamefont
  {Preuss}}]{Coloretti:2022jcl}%
  \BibitemOpen
  \bibfield  {author} {\bibinfo {author} {\bibfnamefont {G.}~\bibnamefont
  {Coloretti}}, \bibinfo {author} {\bibfnamefont {A.}~\bibnamefont
  {Gehrmann-De~Ridder}}, \ and\ \bibinfo {author} {\bibfnamefont {C.~T.}\
  \bibnamefont {Preuss}},\ }\href {\doibase 10.1007/JHEP06(2022)009} {\bibfield
   {journal} {\bibinfo  {journal} {JHEP}\ }\textbf {\bibinfo {volume} {06}},\
  \bibinfo {pages} {009} (\bibinfo {year} {2022})},\ \Eprint
  {http://arxiv.org/abs/2202.07333} {arXiv:2202.07333 [hep-ph]} \BibitemShut
  {NoStop}%
\bibitem [{\citenamefont {Zhu}\ \emph {et~al.}(2025)\citenamefont {Zhu},
  \citenamefont {Song}, \citenamefont {Gao}, \citenamefont {Kang},\ and\
  \citenamefont {Maji}}]{Zhu:2023oka}%
  \BibitemOpen
  \bibfield  {author} {\bibinfo {author} {\bibfnamefont {J.}~\bibnamefont
  {Zhu}}, \bibinfo {author} {\bibfnamefont {Y.}~\bibnamefont {Song}}, \bibinfo
  {author} {\bibfnamefont {J.}~\bibnamefont {Gao}}, \bibinfo {author}
  {\bibfnamefont {D.}~\bibnamefont {Kang}}, \ and\ \bibinfo {author}
  {\bibfnamefont {T.}~\bibnamefont {Maji}},\ }\href {\doibase
  10.1088/1674-1137/ad94e0} {\bibfield  {journal} {\bibinfo  {journal} {Chin.
  Phys. C}\ }\textbf {\bibinfo {volume} {49}},\ \bibinfo {pages} {023106}
  (\bibinfo {year} {2025})},\ \Eprint {http://arxiv.org/abs/2311.07282}
  {arXiv:2311.07282 [hep-ph]} \BibitemShut {NoStop}%
\bibitem [{\citenamefont {Gehrmann-De~Ridder}\ \emph
  {et~al.}(2024{\natexlab{a}})\citenamefont {Gehrmann-De~Ridder}, \citenamefont
  {Preuss},\ and\ \citenamefont {Williams}}]{Gehrmann-DeRidder:2023uld}%
  \BibitemOpen
  \bibfield  {author} {\bibinfo {author} {\bibfnamefont {A.}~\bibnamefont
  {Gehrmann-De~Ridder}}, \bibinfo {author} {\bibfnamefont {C.~T.}\ \bibnamefont
  {Preuss}}, \ and\ \bibinfo {author} {\bibfnamefont {C.}~\bibnamefont
  {Williams}},\ }\href {\doibase 10.1007/JHEP03(2024)104} {\bibfield  {journal}
  {\bibinfo  {journal} {JHEP}\ }\textbf {\bibinfo {volume} {03}},\ \bibinfo
  {pages} {104} (\bibinfo {year} {2024}{\natexlab{a}})},\ \Eprint
  {http://arxiv.org/abs/2310.09354} {arXiv:2310.09354 [hep-ph]} \BibitemShut
  {NoStop}%
\bibitem [{\citenamefont {Knobbe}\ \emph {et~al.}(2024)\citenamefont {Knobbe},
  \citenamefont {Krauss}, \citenamefont {Reichelt},\ and\ \citenamefont
  {Schumann}}]{Knobbe:2023njd}%
  \BibitemOpen
  \bibfield  {author} {\bibinfo {author} {\bibfnamefont {M.}~\bibnamefont
  {Knobbe}}, \bibinfo {author} {\bibfnamefont {F.}~\bibnamefont {Krauss}},
  \bibinfo {author} {\bibfnamefont {D.}~\bibnamefont {Reichelt}}, \ and\
  \bibinfo {author} {\bibfnamefont {S.}~\bibnamefont {Schumann}},\ }\href
  {\doibase 10.1140/epjc/s10052-024-12430-4} {\bibfield  {journal} {\bibinfo
  {journal} {Eur. Phys. J. C}\ }\textbf {\bibinfo {volume} {84}},\ \bibinfo
  {pages} {83} (\bibinfo {year} {2024})},\ \Eprint
  {http://arxiv.org/abs/2306.03682} {arXiv:2306.03682 [hep-ph]} \BibitemShut
  {NoStop}%
\bibitem [{\citenamefont {Campillo~Aveleira}\ \emph {et~al.}(2024)\citenamefont
  {Campillo~Aveleira}, \citenamefont {Gehrmann-De~Ridder},\ and\ \citenamefont
  {Preuss}}]{CampilloAveleira:2024fll}%
  \BibitemOpen
  \bibfield  {author} {\bibinfo {author} {\bibfnamefont {B.}~\bibnamefont
  {Campillo~Aveleira}}, \bibinfo {author} {\bibfnamefont {A.}~\bibnamefont
  {Gehrmann-De~Ridder}}, \ and\ \bibinfo {author} {\bibfnamefont {C.~T.}\
  \bibnamefont {Preuss}},\ }\href {\doibase 10.1140/epjc/s10052-024-13127-4}
  {\bibfield  {journal} {\bibinfo  {journal} {Eur. Phys. J. C}\ }\textbf
  {\bibinfo {volume} {84}},\ \bibinfo {pages} {789} (\bibinfo {year} {2024})},\
  \Eprint {http://arxiv.org/abs/2402.17379} {arXiv:2402.17379 [hep-ph]}
  \BibitemShut {NoStop}%
\bibitem [{\citenamefont {Gehrmann-De~Ridder}\ \emph
  {et~al.}(2024{\natexlab{b}})\citenamefont {Gehrmann-De~Ridder}, \citenamefont
  {Preuss}, \citenamefont {Reichelt},\ and\ \citenamefont
  {Schumann}}]{Gehrmann-DeRidder:2024avt}%
  \BibitemOpen
  \bibfield  {author} {\bibinfo {author} {\bibfnamefont {A.}~\bibnamefont
  {Gehrmann-De~Ridder}}, \bibinfo {author} {\bibfnamefont {C.~T.}\ \bibnamefont
  {Preuss}}, \bibinfo {author} {\bibfnamefont {D.}~\bibnamefont {Reichelt}}, \
  and\ \bibinfo {author} {\bibfnamefont {S.}~\bibnamefont {Schumann}},\ }\href
  {\doibase 10.1007/JHEP07(2024)160} {\bibfield  {journal} {\bibinfo  {journal}
  {JHEP}\ }\textbf {\bibinfo {volume} {07}},\ \bibinfo {pages} {160} (\bibinfo
  {year} {2024}{\natexlab{b}})},\ \Eprint {http://arxiv.org/abs/2403.06929}
  {arXiv:2403.06929 [hep-ph]} \BibitemShut {NoStop}%
\bibitem [{\citenamefont {Fox}\ \emph {et~al.}(2025)\citenamefont {Fox},
  \citenamefont {Gehrmann-De~Ridder}, \citenamefont {Gehrmann}, \citenamefont
  {Glover}, \citenamefont {Marcoli},\ and\ \citenamefont
  {Preuss}}]{Fox:2025cuz}%
  \BibitemOpen
  \bibfield  {author} {\bibinfo {author} {\bibfnamefont {E.}~\bibnamefont
  {Fox}}, \bibinfo {author} {\bibfnamefont {A.}~\bibnamefont
  {Gehrmann-De~Ridder}}, \bibinfo {author} {\bibfnamefont {T.}~\bibnamefont
  {Gehrmann}}, \bibinfo {author} {\bibfnamefont {N.}~\bibnamefont {Glover}},
  \bibinfo {author} {\bibfnamefont {M.}~\bibnamefont {Marcoli}}, \ and\
  \bibinfo {author} {\bibfnamefont {C.~T.}\ \bibnamefont {Preuss}},\ }\href
  {\doibase 10.1103/1znh-sm96} {\bibfield  {journal} {\bibinfo  {journal}
  {Phys. Rev. Lett.}\ }\textbf {\bibinfo {volume} {134}},\ \bibinfo {pages}
  {251905} (\bibinfo {year} {2025})},\ \Eprint
  {http://arxiv.org/abs/2502.17333} {arXiv:2502.17333 [hep-ph]} \BibitemShut
  {NoStop}%
\bibitem [{\citenamefont {Gleisberg}\ \emph {et~al.}(2004)\citenamefont
  {Gleisberg}, \citenamefont {H{\"o}che}, \citenamefont {Krauss}, \citenamefont
  {Sch{\"a}licke}, \citenamefont {Schumann},\ and\ \citenamefont
  {Winter}}]{Gleisberg:2003xi}%
  \BibitemOpen
  \bibfield  {author} {\bibinfo {author} {\bibfnamefont {T.}~\bibnamefont
  {Gleisberg}}, \bibinfo {author} {\bibfnamefont {S.}~\bibnamefont
  {H{\"o}che}}, \bibinfo {author} {\bibfnamefont {F.}~\bibnamefont {Krauss}},
  \bibinfo {author} {\bibfnamefont {A.}~\bibnamefont {Sch{\"a}licke}}, \bibinfo
  {author} {\bibfnamefont {S.}~\bibnamefont {Schumann}}, \ and\ \bibinfo
  {author} {\bibfnamefont {J.-C.}\ \bibnamefont {Winter}},\ }\href {\doibase
  10.1088/1126-6708/2004/02/056} {\bibfield  {journal} {\bibinfo  {journal}
  {JHEP}\ }\textbf {\bibinfo {volume} {02}},\ \bibinfo {pages} {056} (\bibinfo
  {year} {2004})},\ \Eprint {http://arxiv.org/abs/hep-ph/0311263}
  {arXiv:hep-ph/0311263} \BibitemShut {NoStop}%
\bibitem [{\citenamefont {Gleisberg}\ \emph {et~al.}(2009)\citenamefont
  {Gleisberg}, \citenamefont {H{\"o}che}, \citenamefont {Krauss}, \citenamefont
  {Sch{\"o}nherr}, \citenamefont {Schumann}, \citenamefont {Siegert},\ and\
  \citenamefont {Winter}}]{Gleisberg:2008ta}%
  \BibitemOpen
  \bibfield  {author} {\bibinfo {author} {\bibfnamefont {T.}~\bibnamefont
  {Gleisberg}}, \bibinfo {author} {\bibfnamefont {S.}~\bibnamefont
  {H{\"o}che}}, \bibinfo {author} {\bibfnamefont {F.}~\bibnamefont {Krauss}},
  \bibinfo {author} {\bibfnamefont {M.}~\bibnamefont {Sch{\"o}nherr}}, \bibinfo
  {author} {\bibfnamefont {S.}~\bibnamefont {Schumann}}, \bibinfo {author}
  {\bibfnamefont {F.}~\bibnamefont {Siegert}}, \ and\ \bibinfo {author}
  {\bibfnamefont {J.}~\bibnamefont {Winter}},\ }\href {\doibase
  10.1088/1126-6708/2009/02/007} {\bibfield  {journal} {\bibinfo  {journal}
  {JHEP}\ }\textbf {\bibinfo {volume} {02}},\ \bibinfo {pages} {007} (\bibinfo
  {year} {2009})},\ \Eprint {http://arxiv.org/abs/0811.4622} {arXiv:0811.4622
  [hep-ph]} \BibitemShut {NoStop}%
\bibitem [{\citenamefont {Bothmann}\ \emph {et~al.}(2019)\citenamefont
  {Bothmann} \emph {et~al.}}]{Sherpa:2019gpd}%
  \BibitemOpen
  \bibfield  {author} {\bibinfo {author} {\bibfnamefont {E.}~\bibnamefont
  {Bothmann}} \emph {et~al.} (\bibinfo {collaboration} {Sherpa}),\ }\href
  {\doibase 10.21468/SciPostPhys.7.3.034} {\bibfield  {journal} {\bibinfo
  {journal} {SciPost Phys.}\ }\textbf {\bibinfo {volume} {7}},\ \bibinfo
  {pages} {034} (\bibinfo {year} {2019})},\ \Eprint
  {http://arxiv.org/abs/1905.09127} {arXiv:1905.09127 [hep-ph]} \BibitemShut
  {NoStop}%
\bibitem [{\citenamefont {Bothmann}\ \emph {et~al.}(2024)\citenamefont
  {Bothmann} \emph {et~al.}}]{Sherpa:2024mfk}%
  \BibitemOpen
  \bibfield  {author} {\bibinfo {author} {\bibfnamefont {E.}~\bibnamefont
  {Bothmann}} \emph {et~al.} (\bibinfo {collaboration} {Sherpa}),\ }\href
  {\doibase 10.1007/JHEP12(2024)156} {\bibfield  {journal} {\bibinfo  {journal}
  {JHEP}\ }\textbf {\bibinfo {volume} {12}},\ \bibinfo {pages} {156} (\bibinfo
  {year} {2024})},\ \Eprint {http://arxiv.org/abs/2410.22148} {arXiv:2410.22148
  [hep-ph]} \BibitemShut {NoStop}%
\bibitem [{\citenamefont {Krauss}\ \emph {et~al.}(2002)\citenamefont {Krauss},
  \citenamefont {Kuhn},\ and\ \citenamefont {Soff}}]{Krauss:2001iv}%
  \BibitemOpen
  \bibfield  {author} {\bibinfo {author} {\bibfnamefont {F.}~\bibnamefont
  {Krauss}}, \bibinfo {author} {\bibfnamefont {R.}~\bibnamefont {Kuhn}}, \ and\
  \bibinfo {author} {\bibfnamefont {G.}~\bibnamefont {Soff}},\ }\href {\doibase
  10.1088/1126-6708/2002/02/044} {\bibfield  {journal} {\bibinfo  {journal}
  {JHEP}\ }\textbf {\bibinfo {volume} {02}},\ \bibinfo {pages} {044} (\bibinfo
  {year} {2002})},\ \Eprint {http://arxiv.org/abs/hep-ph/0109036}
  {arXiv:hep-ph/0109036} \BibitemShut {NoStop}%
\bibitem [{\citenamefont {Gleisberg}\ and\ \citenamefont
  {H{\"o}che}(2008)}]{Gleisberg:2008fv}%
  \BibitemOpen
  \bibfield  {author} {\bibinfo {author} {\bibfnamefont {T.}~\bibnamefont
  {Gleisberg}}\ and\ \bibinfo {author} {\bibfnamefont {S.}~\bibnamefont
  {H{\"o}che}},\ }\href {\doibase 10.1088/1126-6708/2008/12/039} {\bibfield
  {journal} {\bibinfo  {journal} {JHEP}\ }\textbf {\bibinfo {volume} {12}},\
  \bibinfo {pages} {039} (\bibinfo {year} {2008})},\ \Eprint
  {http://arxiv.org/abs/0808.3674} {arXiv:0808.3674 [hep-ph]} \BibitemShut
  {NoStop}%
\bibitem [{\citenamefont {Campbell}\ and\ \citenamefont
  {Ellis}(1999)}]{Campbell:1999ah}%
  \BibitemOpen
  \bibfield  {author} {\bibinfo {author} {\bibfnamefont {J.~M.}\ \bibnamefont
  {Campbell}}\ and\ \bibinfo {author} {\bibfnamefont {R.~K.}\ \bibnamefont
  {Ellis}},\ }\href {\doibase 10.1103/PhysRevD.60.113006} {\bibfield  {journal}
  {\bibinfo  {journal} {Phys. Rev. D}\ }\textbf {\bibinfo {volume} {60}},\
  \bibinfo {pages} {113006} (\bibinfo {year} {1999})},\ \Eprint
  {http://arxiv.org/abs/hep-ph/9905386} {arXiv:hep-ph/9905386} \BibitemShut
  {NoStop}%
\bibitem [{\citenamefont {Campbell}\ \emph {et~al.}(2011)\citenamefont
  {Campbell}, \citenamefont {Ellis},\ and\ \citenamefont
  {Williams}}]{Campbell:2011bn}%
  \BibitemOpen
  \bibfield  {author} {\bibinfo {author} {\bibfnamefont {J.~M.}\ \bibnamefont
  {Campbell}}, \bibinfo {author} {\bibfnamefont {R.~K.}\ \bibnamefont {Ellis}},
  \ and\ \bibinfo {author} {\bibfnamefont {C.}~\bibnamefont {Williams}},\
  }\href {\doibase 10.1007/JHEP07(2011)018} {\bibfield  {journal} {\bibinfo
  {journal} {JHEP}\ }\textbf {\bibinfo {volume} {07}},\ \bibinfo {pages} {018}
  (\bibinfo {year} {2011})},\ \Eprint {http://arxiv.org/abs/1105.0020}
  {arXiv:1105.0020 [hep-ph]} \BibitemShut {NoStop}%
\bibitem [{\citenamefont {Campbell}\ \emph {et~al.}(2015)\citenamefont
  {Campbell}, \citenamefont {Ellis},\ and\ \citenamefont
  {Giele}}]{Campbell:2015qma}%
  \BibitemOpen
  \bibfield  {author} {\bibinfo {author} {\bibfnamefont {J.~M.}\ \bibnamefont
  {Campbell}}, \bibinfo {author} {\bibfnamefont {R.~K.}\ \bibnamefont {Ellis}},
  \ and\ \bibinfo {author} {\bibfnamefont {W.~T.}\ \bibnamefont {Giele}},\
  }\href {\doibase 10.1140/epjc/s10052-015-3461-2} {\bibfield  {journal}
  {\bibinfo  {journal} {Eur. Phys. J. C}\ }\textbf {\bibinfo {volume} {75}},\
  \bibinfo {pages} {246} (\bibinfo {year} {2015})},\ \Eprint
  {http://arxiv.org/abs/1503.06182} {arXiv:1503.06182 [physics.comp-ph]}
  \BibitemShut {NoStop}%
\bibitem [{\citenamefont {Campbell}\ and\ \citenamefont
  {Neumann}(2019)}]{Campbell:2019dru}%
  \BibitemOpen
  \bibfield  {author} {\bibinfo {author} {\bibfnamefont {J.}~\bibnamefont
  {Campbell}}\ and\ \bibinfo {author} {\bibfnamefont {T.}~\bibnamefont
  {Neumann}},\ }\href {\doibase 10.1007/JHEP12(2019)034} {\bibfield  {journal}
  {\bibinfo  {journal} {JHEP}\ }\textbf {\bibinfo {volume} {12}},\ \bibinfo
  {pages} {034} (\bibinfo {year} {2019})},\ \Eprint
  {http://arxiv.org/abs/1909.09117} {arXiv:1909.09117 [hep-ph]} \BibitemShut
  {NoStop}%
\bibitem [{\citenamefont {Campbell}\ \emph {et~al.}(2021)\citenamefont
  {Campbell}, \citenamefont {H{\"o}che},\ and\ \citenamefont
  {Preuss}}]{Campbell:2021vlt}%
  \BibitemOpen
  \bibfield  {author} {\bibinfo {author} {\bibfnamefont {J.~M.}\ \bibnamefont
  {Campbell}}, \bibinfo {author} {\bibfnamefont {S.}~\bibnamefont {H{\"o}che}},
  \ and\ \bibinfo {author} {\bibfnamefont {C.~T.}\ \bibnamefont {Preuss}},\
  }\href {\doibase 10.1140/epjc/s10052-021-09885-0} {\bibfield  {journal}
  {\bibinfo  {journal} {Eur. Phys. J. C}\ }\textbf {\bibinfo {volume} {81}},\
  \bibinfo {pages} {1117} (\bibinfo {year} {2021})},\ \Eprint
  {http://arxiv.org/abs/2107.04472} {arXiv:2107.04472 [hep-ph]} \BibitemShut
  {NoStop}%
\bibitem [{\citenamefont {Cascioli}\ \emph {et~al.}(2012)\citenamefont
  {Cascioli}, \citenamefont {Maierhofer},\ and\ \citenamefont
  {Pozzorini}}]{Cascioli:2011va}%
  \BibitemOpen
  \bibfield  {author} {\bibinfo {author} {\bibfnamefont {F.}~\bibnamefont
  {Cascioli}}, \bibinfo {author} {\bibfnamefont {P.}~\bibnamefont
  {Maierhofer}}, \ and\ \bibinfo {author} {\bibfnamefont {S.}~\bibnamefont
  {Pozzorini}},\ }\href {\doibase 10.1103/PhysRevLett.108.111601} {\bibfield
  {journal} {\bibinfo  {journal} {Phys. Rev. Lett.}\ }\textbf {\bibinfo
  {volume} {108}},\ \bibinfo {pages} {111601} (\bibinfo {year} {2012})},\
  \Eprint {http://arxiv.org/abs/1111.5206} {arXiv:1111.5206 [hep-ph]}
  \BibitemShut {NoStop}%
\bibitem [{\citenamefont {Buccioni}\ \emph {et~al.}(2018)\citenamefont
  {Buccioni}, \citenamefont {Pozzorini},\ and\ \citenamefont
  {Zoller}}]{Buccioni:2017yxi}%
  \BibitemOpen
  \bibfield  {author} {\bibinfo {author} {\bibfnamefont {F.}~\bibnamefont
  {Buccioni}}, \bibinfo {author} {\bibfnamefont {S.}~\bibnamefont {Pozzorini}},
  \ and\ \bibinfo {author} {\bibfnamefont {M.}~\bibnamefont {Zoller}},\ }\href
  {\doibase 10.1140/epjc/s10052-018-5562-1} {\bibfield  {journal} {\bibinfo
  {journal} {Eur. Phys. J. C}\ }\textbf {\bibinfo {volume} {78}},\ \bibinfo
  {pages} {70} (\bibinfo {year} {2018})},\ \Eprint
  {http://arxiv.org/abs/1710.11452} {arXiv:1710.11452 [hep-ph]} \BibitemShut
  {NoStop}%
\bibitem [{\citenamefont {Buccioni}\ \emph {et~al.}(2019)\citenamefont
  {Buccioni}, \citenamefont {Lang}, \citenamefont {Lindert}, \citenamefont
  {Maierh{\"o}fer}, \citenamefont {Pozzorini}, \citenamefont {Zhang},\ and\
  \citenamefont {Zoller}}]{Buccioni:2019sur}%
  \BibitemOpen
  \bibfield  {author} {\bibinfo {author} {\bibfnamefont {F.}~\bibnamefont
  {Buccioni}}, \bibinfo {author} {\bibfnamefont {J.-N.}\ \bibnamefont {Lang}},
  \bibinfo {author} {\bibfnamefont {J.~M.}\ \bibnamefont {Lindert}}, \bibinfo
  {author} {\bibfnamefont {P.}~\bibnamefont {Maierh{\"o}fer}}, \bibinfo
  {author} {\bibfnamefont {S.}~\bibnamefont {Pozzorini}}, \bibinfo {author}
  {\bibfnamefont {H.}~\bibnamefont {Zhang}}, \ and\ \bibinfo {author}
  {\bibfnamefont {M.~F.}\ \bibnamefont {Zoller}},\ }\href {\doibase
  10.1140/epjc/s10052-019-7306-2} {\bibfield  {journal} {\bibinfo  {journal}
  {Eur. Phys. J. C}\ }\textbf {\bibinfo {volume} {79}},\ \bibinfo {pages} {866}
  (\bibinfo {year} {2019})},\ \Eprint {http://arxiv.org/abs/1907.13071}
  {arXiv:1907.13071 [hep-ph]} \BibitemShut {NoStop}%
\bibitem [{\citenamefont {Catani}\ \emph
  {et~al.}(1991{\natexlab{b}})\citenamefont {Catani}, \citenamefont {Webber},\
  and\ \citenamefont {Marchesini}}]{Catani:1990rr}%
  \BibitemOpen
  \bibfield  {author} {\bibinfo {author} {\bibfnamefont {S.}~\bibnamefont
  {Catani}}, \bibinfo {author} {\bibfnamefont {B.~R.}\ \bibnamefont {Webber}},
  \ and\ \bibinfo {author} {\bibfnamefont {G.}~\bibnamefont {Marchesini}},\
  }\href {\doibase 10.1016/0550-3213(91)90390-J} {\bibfield  {journal}
  {\bibinfo  {journal} {Nucl. Phys. B}\ }\textbf {\bibinfo {volume} {349}},\
  \bibinfo {pages} {635} (\bibinfo {year} {1991}{\natexlab{b}})}\BibitemShut
  {NoStop}%
\bibitem [{\citenamefont {Buckley}\ \emph {et~al.}(2013)\citenamefont
  {Buckley}, \citenamefont {Butterworth}, \citenamefont {Grellscheid},
  \citenamefont {Hoeth}, \citenamefont {L{\"o}nnblad}, \citenamefont {Monk},
  \citenamefont {Schulz},\ and\ \citenamefont {Siegert}}]{Buckley:2010ar}%
  \BibitemOpen
  \bibfield  {author} {\bibinfo {author} {\bibfnamefont {A.}~\bibnamefont
  {Buckley}}, \bibinfo {author} {\bibfnamefont {J.}~\bibnamefont
  {Butterworth}}, \bibinfo {author} {\bibfnamefont {D.}~\bibnamefont
  {Grellscheid}}, \bibinfo {author} {\bibfnamefont {H.}~\bibnamefont {Hoeth}},
  \bibinfo {author} {\bibfnamefont {L.}~\bibnamefont {L{\"o}nnblad}}, \bibinfo
  {author} {\bibfnamefont {J.}~\bibnamefont {Monk}}, \bibinfo {author}
  {\bibfnamefont {H.}~\bibnamefont {Schulz}}, \ and\ \bibinfo {author}
  {\bibfnamefont {F.}~\bibnamefont {Siegert}},\ }\href {\doibase
  10.1016/j.cpc.2013.05.021} {\bibfield  {journal} {\bibinfo  {journal}
  {Comput. Phys. Commun.}\ }\textbf {\bibinfo {volume} {184}},\ \bibinfo
  {pages} {2803} (\bibinfo {year} {2013})},\ \Eprint
  {http://arxiv.org/abs/1003.0694} {arXiv:1003.0694 [hep-ph]} \BibitemShut
  {NoStop}%
\bibitem [{\citenamefont {Bierlich}\ \emph {et~al.}(2020)\citenamefont
  {Bierlich} \emph {et~al.}}]{Bierlich:2019rhm}%
  \BibitemOpen
  \bibfield  {author} {\bibinfo {author} {\bibfnamefont {C.}~\bibnamefont
  {Bierlich}} \emph {et~al.},\ }\href {\doibase 10.21468/SciPostPhys.8.2.026}
  {\bibfield  {journal} {\bibinfo  {journal} {SciPost Phys.}\ }\textbf
  {\bibinfo {volume} {8}},\ \bibinfo {pages} {026} (\bibinfo {year} {2020})},\
  \Eprint {http://arxiv.org/abs/1912.05451} {arXiv:1912.05451 [hep-ph]}
  \BibitemShut {NoStop}%
\bibitem [{\citenamefont {Bierlich}\ \emph {et~al.}(2024)\citenamefont
  {Bierlich}, \citenamefont {Buckley}, \citenamefont {Butterworth},
  \citenamefont {Gutschow}, \citenamefont {L{\"o}nnblad}, \citenamefont
  {Procter}, \citenamefont {Richardson},\ and\ \citenamefont
  {Yeh}}]{Bierlich:2024vqo}%
  \BibitemOpen
  \bibfield  {author} {\bibinfo {author} {\bibfnamefont {C.}~\bibnamefont
  {Bierlich}}, \bibinfo {author} {\bibfnamefont {A.}~\bibnamefont {Buckley}},
  \bibinfo {author} {\bibfnamefont {J.~M.}\ \bibnamefont {Butterworth}},
  \bibinfo {author} {\bibfnamefont {C.}~\bibnamefont {Gutschow}}, \bibinfo
  {author} {\bibfnamefont {L.}~\bibnamefont {L{\"o}nnblad}}, \bibinfo {author}
  {\bibfnamefont {T.}~\bibnamefont {Procter}}, \bibinfo {author} {\bibfnamefont
  {P.}~\bibnamefont {Richardson}}, \ and\ \bibinfo {author} {\bibfnamefont
  {Y.}~\bibnamefont {Yeh}},\ }\href {\doibase 10.21468/SciPostPhysCodeb.36}
  {\bibfield  {journal} {\bibinfo  {journal} {SciPost Phys. Codeb.}\ }\textbf
  {\bibinfo {volume} {36}},\ \bibinfo {pages} {1} (\bibinfo {year} {2024})},\
  \Eprint {http://arxiv.org/abs/2404.15984} {arXiv:2404.15984 [hep-ph]}
  \BibitemShut {NoStop}%
\bibitem [{\citenamefont {Amati}\ \emph {et~al.}(1980)\citenamefont {Amati},
  \citenamefont {Bassetto}, \citenamefont {Ciafaloni}, \citenamefont
  {Marchesini},\ and\ \citenamefont {Veneziano}}]{Amati:1980ch}%
  \BibitemOpen
  \bibfield  {author} {\bibinfo {author} {\bibfnamefont {D.}~\bibnamefont
  {Amati}}, \bibinfo {author} {\bibfnamefont {A.}~\bibnamefont {Bassetto}},
  \bibinfo {author} {\bibfnamefont {M.}~\bibnamefont {Ciafaloni}}, \bibinfo
  {author} {\bibfnamefont {G.}~\bibnamefont {Marchesini}}, \ and\ \bibinfo
  {author} {\bibfnamefont {G.}~\bibnamefont {Veneziano}},\ }\href {\doibase
  10.1016/0550-3213(80)90012-7} {\bibfield  {journal} {\bibinfo  {journal}
  {Nucl. Phys. B}\ }\textbf {\bibinfo {volume} {173}},\ \bibinfo {pages} {429}
  (\bibinfo {year} {1980})}\BibitemShut {NoStop}%
\bibitem [{\citenamefont {Kodaira}\ and\ \citenamefont
  {Trentadue}(1982)}]{Kodaira:1981nh}%
  \BibitemOpen
  \bibfield  {author} {\bibinfo {author} {\bibfnamefont {J.}~\bibnamefont
  {Kodaira}}\ and\ \bibinfo {author} {\bibfnamefont {L.}~\bibnamefont
  {Trentadue}},\ }\href {\doibase 10.1016/0370-2693(82)90907-8} {\bibfield
  {journal} {\bibinfo  {journal} {Phys. Lett. B}\ }\textbf {\bibinfo {volume}
  {112}},\ \bibinfo {pages} {66} (\bibinfo {year} {1982})}\BibitemShut
  {NoStop}%
\bibitem [{\citenamefont {Davies}\ and\ \citenamefont
  {Stirling}(1984)}]{Davies:1984hs}%
  \BibitemOpen
  \bibfield  {author} {\bibinfo {author} {\bibfnamefont {C.~T.~H.}\
  \bibnamefont {Davies}}\ and\ \bibinfo {author} {\bibfnamefont {W.~J.}\
  \bibnamefont {Stirling}},\ }\href {\doibase 10.1016/0550-3213(84)90316-X}
  {\bibfield  {journal} {\bibinfo  {journal} {Nucl. Phys. B}\ }\textbf
  {\bibinfo {volume} {244}},\ \bibinfo {pages} {337} (\bibinfo {year}
  {1984})}\BibitemShut {NoStop}%
\bibitem [{\citenamefont {Davies}\ \emph {et~al.}(1984)\citenamefont {Davies},
  \citenamefont {Webber},\ and\ \citenamefont {Stirling}}]{Davies:1984sp}%
  \BibitemOpen
  \bibfield  {author} {\bibinfo {author} {\bibfnamefont {C.~T.~H.}\
  \bibnamefont {Davies}}, \bibinfo {author} {\bibfnamefont {B.~R.}\
  \bibnamefont {Webber}}, \ and\ \bibinfo {author} {\bibfnamefont {W.~J.}\
  \bibnamefont {Stirling}},\ }\href {\doibase 10.1016/0550-3213(85)90402-X} {\
  \textbf {\bibinfo {volume} {1}},\ \bibinfo {pages} {I.95} (\bibinfo {year}
  {1984})}\BibitemShut {NoStop}%
\bibitem [{\citenamefont {Catani}\ \emph {et~al.}(1988)\citenamefont {Catani},
  \citenamefont {D'Emilio},\ and\ \citenamefont {Trentadue}}]{Catani:1988vd}%
  \BibitemOpen
  \bibfield  {author} {\bibinfo {author} {\bibfnamefont {S.}~\bibnamefont
  {Catani}}, \bibinfo {author} {\bibfnamefont {E.}~\bibnamefont {D'Emilio}}, \
  and\ \bibinfo {author} {\bibfnamefont {L.}~\bibnamefont {Trentadue}},\ }\href
  {\doibase 10.1016/0370-2693(88)90912-4} {\bibfield  {journal} {\bibinfo
  {journal} {Phys. Lett. B}\ }\textbf {\bibinfo {volume} {211}},\ \bibinfo
  {pages} {335} (\bibinfo {year} {1988})}\BibitemShut {NoStop}%
\bibitem [{\citenamefont {Heister}\ \emph {et~al.}(2004)\citenamefont {Heister}
  \emph {et~al.}}]{Heister:2003aj}%
  \BibitemOpen
  \bibfield  {author} {\bibinfo {author} {\bibfnamefont {A.}~\bibnamefont
  {Heister}} \emph {et~al.} (\bibinfo {collaboration} {ALEPH}),\ }\href
  {\doibase 10.1140/epjc/s2004-01891-4} {\bibfield  {journal} {\bibinfo
  {journal} {Eur. Phys. J. C}\ }\textbf {\bibinfo {volume} {35}},\ \bibinfo
  {pages} {457} (\bibinfo {year} {2004})}\BibitemShut {NoStop}%
\bibitem [{\citenamefont {Pfeifenschneider}\ \emph {et~al.}(2000)\citenamefont
  {Pfeifenschneider} \emph {et~al.}}]{JADE:1999zar}%
  \BibitemOpen
  \bibfield  {author} {\bibinfo {author} {\bibfnamefont {P.}~\bibnamefont
  {Pfeifenschneider}} \emph {et~al.} (\bibinfo {collaboration} {JADE, OPAL}),\
  }\href {\doibase 10.1007/s100520000432} {\bibfield  {journal} {\bibinfo
  {journal} {Eur. Phys. J. C}\ }\textbf {\bibinfo {volume} {17}},\ \bibinfo
  {pages} {19} (\bibinfo {year} {2000})},\ \Eprint
  {http://arxiv.org/abs/hep-ex/0001055} {arXiv:hep-ex/0001055} \BibitemShut
  {NoStop}%
\bibitem [{\citenamefont {Abbiendi}\ \emph {et~al.}(2005)\citenamefont
  {Abbiendi} \emph {et~al.}}]{Abbiendi:2004qz}%
  \BibitemOpen
  \bibfield  {author} {\bibinfo {author} {\bibfnamefont {G.}~\bibnamefont
  {Abbiendi}} \emph {et~al.} (\bibinfo {collaboration} {OPAL}),\ }\href
  {\doibase 10.1140/epjc/s2005-02120-6} {\bibfield  {journal} {\bibinfo
  {journal} {Eur. Phys. J. C}\ }\textbf {\bibinfo {volume} {40}},\ \bibinfo
  {pages} {287} (\bibinfo {year} {2005})},\ \Eprint
  {http://arxiv.org/abs/hep-ex/0503051} {arXiv:hep-ex/0503051} \BibitemShut
  {NoStop}%
\bibitem [{\citenamefont {Farhi}(1977)}]{Farhi:1977sg}%
  \BibitemOpen
  \bibfield  {author} {\bibinfo {author} {\bibfnamefont {E.}~\bibnamefont
  {Farhi}},\ }\href {\doibase 10.1103/PhysRevLett.39.1587} {\bibfield
  {journal} {\bibinfo  {journal} {Phys. Rev. Lett.}\ }\textbf {\bibinfo
  {volume} {39}},\ \bibinfo {pages} {1587} (\bibinfo {year}
  {1977})}\BibitemShut {NoStop}%
\bibitem [{\citenamefont {Nason}\ and\ \citenamefont
  {Zanderighi}(2023)}]{Nason:2023asn}%
  \BibitemOpen
  \bibfield  {author} {\bibinfo {author} {\bibfnamefont {P.}~\bibnamefont
  {Nason}}\ and\ \bibinfo {author} {\bibfnamefont {G.}~\bibnamefont
  {Zanderighi}},\ }\href {\doibase 10.1007/JHEP06(2023)058} {\bibfield
  {journal} {\bibinfo  {journal} {JHEP}\ }\textbf {\bibinfo {volume} {06}},\
  \bibinfo {pages} {058} (\bibinfo {year} {2023})},\ \Eprint
  {http://arxiv.org/abs/2301.03607} {arXiv:2301.03607 [hep-ph]} \BibitemShut
  {NoStop}%
\bibitem [{\citenamefont {Catani}\ and\ \citenamefont
  {Webber}(1997)}]{Catani:1997xc}%
  \BibitemOpen
  \bibfield  {author} {\bibinfo {author} {\bibfnamefont {S.}~\bibnamefont
  {Catani}}\ and\ \bibinfo {author} {\bibfnamefont {B.~R.}\ \bibnamefont
  {Webber}},\ }\href {\doibase 10.1088/1126-6708/1997/10/005} {\bibfield
  {journal} {\bibinfo  {journal} {JHEP}\ }\textbf {\bibinfo {volume} {10}},\
  \bibinfo {pages} {005} (\bibinfo {year} {1997})},\ \Eprint
  {http://arxiv.org/abs/hep-ph/9710333} {arXiv:hep-ph/9710333} \BibitemShut
  {NoStop}%
\bibitem [{\citenamefont {Abbiendi}\ \emph {et~al.}(2001)\citenamefont
  {Abbiendi} \emph {et~al.}}]{Abbiendi:2001qn}%
  \BibitemOpen
  \bibfield  {author} {\bibinfo {author} {\bibfnamefont {G.}~\bibnamefont
  {Abbiendi}} \emph {et~al.} (\bibinfo {collaboration} {OPAL}),\ }\href
  {\doibase 10.1007/s100520100699} {\bibfield  {journal} {\bibinfo  {journal}
  {Eur. Phys. J. C}\ }\textbf {\bibinfo {volume} {20}},\ \bibinfo {pages} {601}
  (\bibinfo {year} {2001})},\ \Eprint {http://arxiv.org/abs/hep-ex/0101044}
  {arXiv:hep-ex/0101044} \BibitemShut {NoStop}%
\bibitem [{\citenamefont {Frederix}\ \emph {et~al.}(2010)\citenamefont
  {Frederix}, \citenamefont {Frixione}, \citenamefont {Melnikov},\ and\
  \citenamefont {Zanderighi}}]{Frederix:2010ne}%
  \BibitemOpen
  \bibfield  {author} {\bibinfo {author} {\bibfnamefont {R.}~\bibnamefont
  {Frederix}}, \bibinfo {author} {\bibfnamefont {S.}~\bibnamefont {Frixione}},
  \bibinfo {author} {\bibfnamefont {K.}~\bibnamefont {Melnikov}}, \ and\
  \bibinfo {author} {\bibfnamefont {G.}~\bibnamefont {Zanderighi}},\ }\href
  {\doibase 10.1007/JHEP11(2010)050} {\bibfield  {journal} {\bibinfo  {journal}
  {JHEP}\ }\textbf {\bibinfo {volume} {11}},\ \bibinfo {pages} {050} (\bibinfo
  {year} {2010})},\ \Eprint {http://arxiv.org/abs/1008.5313} {arXiv:1008.5313
  [hep-ph]} \BibitemShut {NoStop}%
\end{thebibliography}%
\end{document}